\documentclass[pra,twocolumn,amsmath,amssymb,superscriptaddress,showpacs]{revtex4-2}
\usepackage[latin9]{inputenc}
\usepackage{hyperref}
\usepackage{ucs}
\usepackage{bbm}
\usepackage{epsfig}
\usepackage{subcaption} 
\usepackage{xcolor}
\usepackage[table]{xcolor}

\newcommand{\bra}[1]{\langle #1|}
\newcommand{\ket}[1]{|#1\rangle}
\newcommand{\ketbra}[1]{| #1\rangle \langle #1|}

\newcommand{\be}{\begin{equation}}
\newcommand{\ee}{\end{equation}}
\newcommand{\eea}{\end{eqnarray}}
\newcommand{\bea}{\begin{eqnarray}}

\newcommand{\va}[1]{\ensuremath{(\Delta#1)^2}}

\newcommand{\ex}[1]{\ensuremath{\langle{#1}\rangle}}

\newcommand{\qed}{\ensuremath{\hfill \blacksquare}}

\newcommand{\kommentar}[1]{}
\newcommand{\trace}{{\rm Tr}}

\newcommand{\forget}[1]{}

\newcommand{\EQ}[1]{Eq.~\eqref{#1}}
\newcommand{\EQS}[1]{Eqs.~\eqref{#1}}

\newcommand{\SEC}[1]{Sec.~\ref{#1}}
\newcommand{\FIG}[1]{Fig.~\ref{#1}}
\newcommand{\REF}[1]{Ref.~\cite{#1}}
\newcommand{\REFS}[1]{Refs.~\cite{#1}}

\newcommand{\APP}[1]{Appendix~\ref{#1}}

\newcommand{\OBS}[1]{Theorem~\ref{#1}}

\newcommand{\DEFOBS}[1]{{\bf Theorem \refstepcounter{observation}\theobservation\label{#1}.} }
\newcounter{observation}

\newcounter{example}

\newcounter{definition}

\begin{document}

\title{Quantum Wasserstein distance and its relation to several types of fidelities}
\author{G\'eza T\'oth}
\email{toth@alumni.nd.edu}
\affiliation{Theoretical Physics, University of the Basque Country UPV/EHU,  
48080 Bilbao, Spain}
\affiliation{EHU Quantum Center, University of the Basque Country UPV/EHU, 
48940 Leioa, 
Spain}
\affiliation{Donostia International Physics Center DIPC,  
20018 San Sebasti\'an, Spain}
\affiliation{IKERBASQUE, Basque Foundation for Science, 48009 Bilbao, Spain}
\affiliation{HUN-REN Wigner Research Centre for Physics,  
1525 Budapest, Hungary}

\author{J\'ozsef Pitrik}
\email{pitrik@math.bme.hu}
\affiliation{HUN-REN Wigner Research Centre for Physics,  
1525 Budapest, Hungary}
\affiliation{HUN-REN 
R\'enyi Institute of Mathematics, 
1053 Budapest, Hungary}
\affiliation{Department of Analysis and Operations Research, Institute of Mathematics, Budapest University of Technology and Economics, 
1111 Budapest, Hungary}

\begin{abstract}
We consider several definitions of the quantum Wasserstein distance based on an optimization over general bipartite quantum states with given marginals. Then, we examine the quantities obtained after the optimization is carried out over bipartite separable states instead. We prove that several of these quantities are equal to each other. Thus, we connect several approaches in the literature. We prove the triangle inequality for some of these quantities for the case of one of the three states being pure. As a byproduct, we show that  the square root of the Uhlmann-Jozsa quantum fidelity can also be written as an optimization over separable states with given marginals. We use this to prove that some of these quantities equal the Uhlmann-Jozsa quantum fidelity for qubits. We also find relations with the superfidelity.
\end{abstract}

\date{\today}

\maketitle

\section{Introduction}

Quantum optimal transport has been at the center of attention, as it led to the definition of several new and very useful notions in quantum physics. The first, semi-classical approach of \.Zyczkowski and S\l omczy\'nski, has been motivated by applications in quantum chaos \cite{Zyczkowski1998TheMonge,Zyczkowski2001TheMonge,Bengtsson2006Geometry}.  The method of Biane and Voiculescu is related to free probability \cite{Biane2011Free}, while the one of Carlen, Maas, Datta, and Rouz\'e \cite{CarlenMaas2014Analog,CarlenMaas2017Gradient,CarlenMaas2020Non-commutative,DattaRouze2019Concentration,DattaRouze2020Relating} is based on a dynamical interpretation. Caglioti, Golse, Mouhot, and Paul presented an approach based on a static interpretation \cite{Golse2016On,Golse2017The,Golse2018Wave,Golse2018TheQuantum,Caglioti2020Quantum,Caglioti2021Towards}, which has been connected to negative Sobolev norms \cite{lafleche2023quantumoptimaltransportweak}. Finally De Palma and Trevisan used quantum channels \cite{DePalma2021Quantum}, and De Palma, Marvian, Trevisan, and Lloyd defined the quantum earth mover's distance, i.~e., the quantum Wasserstein distance order 1 \cite{DePalma2021TheQuantum}, while Bistro\'n, Cole, Eckstein, Friedland, and \.Zyczkowski formulated a quantum Wasserstein distance based on an antisymmetric cost function and the SWAP-fidelity \cite{Friedland2022Quantum,Cole2023OnQuantum,Bistron2023Monotonicity}. 

One of the key results of quantum optimal transport is the definition of the quantum Wasserstein distance \cite{Zyczkowski1998TheMonge,Zyczkowski2001TheMonge,Bengtsson2006Geometry,Golse2016On,Golse2017The,Golse2018Wave,Golse2018TheQuantum,DePalma2021Quantum,DePalma2021TheQuantum,Caglioti2020Quantum,Caglioti2021Towards,Geher2023Quantum,Li2025Wasserstein}, for a review see \REF{beatty2025wassersteindistancesquantumstructures}. It has the often desirable feature that it is not necessarily maximal for two quantum states orthogonal to each other, which is beneficial, for instance, when  performing learning on quantum data \cite{Kiani2022Learning}. Some of the properties of the new quantities are puzzling, yet point to profound relations between seemingly unrelated fields of quantum physics. For instance, the quantum Wasserstein distance order 2 of the quantum state from itself can be nonzero, while in the classical case the self-distance is always zero. In particular, as we have mentioned, the quantum Wasserstein distance has been defined based on a quantum channel formalism for a given set of operators $\{H_n\}_{n=1}^N$ \cite{DePalma2021Quantum,DePalma2024QuantumOptimal}, and it has been shown that the square of the self-distance for a state $\varrho$ is equal to the sum of the Wigner-Yanase skew information $I_{\varrho}(H_n)$ of the quantum state \cite{Wigner1963INFORMATION}.  Such a quantum Wasserstein distance has been used to examine the distance of various ground state phases in spin systems \cite{camacho2025criticalscalingquantumwasserstein}. It has also been extended to define a $p$-Wasserstein distance \cite{bunth2025wassersteindistancesdivergencesorder}. Recently, the dual formalism has also been investigated \cite{bunth2025strongkantorovichdualityquantum}. A modified quantum Wasserstein distance has been defined connected to this approach, that does not have a self-distance \cite{DePalma2021Quantum} and  it fulfills the triangle inequalities \cite{Bunth2024MetricProperty,bunth2025wassersteindistancesdivergencesorder,wirth2025triangleinequalityquantumwasserstein}.

Recently, it has been found that if the optimization is taken over separable states then the self-distance equals the quantum Fisher information over four for when a single operator is given, i.~e., $N=1$ \cite{Toth2023QuantumWasserstein}. This result connects the quantum Wasserstein distance to quantum theory \cite{Horodecki2009Quantum,Guhne2009Entanglement,Friis2019}  and  quantum metrology \cite{Giovannetti2004Quantum-Enhanced,Paris2009QUANTUM,Demkowicz-Dobrzanski2014Quantum,Pezze2014Quantum,Toth2014Quantum,Pezze2018Quantum}.   In this approach, the formula containing an optimization over bipartite separable states with given marginals provides the distance for mixed states, where the cost function optimized is the distance based on the channel formalism between two pure states \cite{Toth2023QuantumWasserstein}. Another definition of the quantum Wasserstein distance based on an optimization over separable states has been considered in which the cost function is the trace distance, and for which the self-distance is zero  \cite{Beatty2025Order}. The approach has been generalized to convex sets different from separable states \cite{borsoni2026foldedoptimaltransportapplication}. 

Having several possible definitions of the quantum Wasserstein distance, the question arises: Is it possible to connect these approaches to each other, and possibly reduce them to few basic types?

In this paper, we answer this question affirmatively. We consider some of the definitions of the quantum Wasserstein distance, in which the optimization is taken over the set of all quantum states, and examine the quantities that arise when we carry out the optimization over separable states instead. We show that we obtain one of two cases. 

(i) In the first case, we obtain an optimization of a cost function given as the distance defined by the quantum channel formalism for pure states \cite{DePalma2024QuantumOptimal},  which is the case discussed in \REF{Toth2023QuantumWasserstein} and defines a quantity that can have a non-zero self-distance.

(ii) In the second case, we obtain an optimization of a cost function given as the {\it modified} distance defined by the quantum channel formalism  for pure states \cite{DePalma2024QuantumOptimal}, which defines a quantity that has a zero self-distance. In the case of a full set of observables $H_n$,  this quantity equals the definition based on the trace distance  \cite{Beatty2025Order}. It also equals the distance based on SWAP-fidelity, if we carry out an optimization over separable states rather than over general quantum states \cite{Friedland2022Quantum,Cole2023OnQuantum,Bistron2023Monotonicity}.

The paper is organized as follows. In \SEC{sec:Setting the stage}, we list some of the types of the quantum Wasserstein distance defined with an optimization over general states we consider in this article. In \SEC{sec:relations}, we calculate the above definitions for the case when we carry out the optimization over separable states. We also write down the quantum Fidelity and the Bures distance as an optimization over separable states and relate it to  the other distance definitions. In \SEC{sec:relations_pW}, we consider the case of the quantum $p$-Wasserstein distance.

\section{Setting the stage}
\label{sec:Setting the stage}

In this section, we list various definitions of the quantum Wasserstein distance. We consider the definitions that involve an optimization over all physical quantum states. 

\subsection{De Palma-Trevisan-type quantum Wasserstein distance}
\label{sec:De Palma-Trevisan-type quantum Wasserstein distance}

One of the definitions of the quantum Wasserstein distance has been given by De Palma and Trevisan as \cite{DePalma2021Quantum}
\begin{align}
\begin{split}
D_{\rm DPT}(\varrho,\sigma)^2\quad\\=\frac 1 2
\min_{ \varrho_{12}}\sum_{n=1}^N\;&
\trace[(H_n^T\otimes \openone-\openone\otimes H_n)^2  \varrho_{12} ],\\
\textrm{s.~t. }&
\varrho_{12}\in\mathcal D,\label{eq:DPT}  \\
& {\rm Tr}_1(\varrho_{12})=\varrho^T, \\
& {\rm Tr}_2(\varrho_{12})=\sigma,
\end{split}
\end{align}
where $H_n$ are Hermitian operators, and $\mathcal D$ is the set of physical states. 
Its self-distance is 
\be
D_{\rm DPT}(\varrho,\varrho)^2=\sum_n I_{\varrho}(H),\label{eq:self_Dsep}
\ee
where $I$ is the Wigner-Yanase skew information \cite{Wigner1963INFORMATION} defined as
\be
I_{\varrho}(H)={\rm Tr}(H^2\varrho )-{\rm Tr}( H \sqrt{\varrho} H\sqrt{\varrho}).
\ee
The advantage of the approach is that 

Golse, Mouhot, Paul and Caglioti  defined the square of the Wasserstein distance as \cite{Golse2016On,Caglioti2021Towards,Golse2018TheQuantum,Golse2017The,Golse2018Wave,Caglioti2020Quantum}
\begin{align}
D_{\rm GMPC}(\varrho,\sigma)^2\quad\quad\quad\nonumber\\=\frac 1 2
\min_{ \varrho_{12}} \sum_{n=1}^N\;&
\trace[(H_n\otimes \openone-\openone\otimes H_n)^2  \varrho_{12} ],\nonumber\\
\textrm{s.~t. }&
\varrho_{12}\in\mathcal D,\label{eq:GMPC_distance}  \nonumber\\
& {\rm Tr}_2(\varrho_{12})=\varrho, \nonumber\\
& {\rm Tr}_1(\varrho_{12})=\sigma.
\end{align}

The advantage of the formulation $D_{\rm DPT}(\varrho,\varrho)^2$ given in \EQ{eq:DPT} is that it has a formula for the self-distance given in \EQ{eq:self_Dsep}, and there is a transport map corresponding to 
all $\varrho_{12}$ couplings. Note that this is possible due to the specific form of the definition in \EQ{eq:DPT},
which includes several partial transposes.

\subsection{SWAP-fidelity and the corresponding quantum Wasserstein distance}
\label{sec:SWAP-fidelity and the corresponding quantum Wasserstein distance}

The SWAP-fidelity is defined as \cite{Friedland2022Quantum,Bistron2022Monotonicity,Cole2023OnQuantum}
\be
F_S(\varrho,\sigma)=\max_{\varrho_{12\in \mathcal D}} {\rm Tr}(\varrho_{12} S),\label{eq:FS}
\ee 
such that the marginals of $\varrho_{12}$ are $\varrho$ and $\sigma.$ Here $S$ is the SWAP operator defined as
\be
S\ket{\Psi}\otimes\ket{\Phi}=\ket{\Phi}\otimes\ket{\Psi}.
\ee
It is known that \cite{Friedland2022Quantum}
\be
F(\varrho,\sigma)\le  F_{S}(\varrho,\sigma) \le \sqrt{F}(\varrho,\sigma),\label{eq:FFF}
\ee
where the Uhlmann-Jozsa fidelity is defined as \cite{Uhlmann1976TheTransitionProbability,Jozsa1994Fidelity}
\be
F(\varrho,\sigma)={\rm Tr}\left(\sqrt{\sqrt{\varrho}\sigma\sqrt{\varrho}}\right)^2.\label{Fidelity_def}
\ee
Note that part of the literature defines fidelity as the square root of the right-hand side of \EQ{Fidelity_def}.
The corresponding distance is given as \cite{Friedland2022Quantum}
\be
D_S(\varrho,\sigma)^2=\frac1 2 [1-F_S(\varrho,\sigma)],\label{eq:DSFS}
\ee
which can be expressed as 
\be
D_S(\varrho,\sigma)^2=\min_{\varrho_{12}\in\mathcal D} {\rm Tr}(\varrho_{12} C_S) \label{eq:DSFS2}
\ee
such that the marginals of $\varrho_{12}$ are $\varrho$ and $\sigma,$ and the quantum cost matrix is
\be
C_S=\frac{\openone-S}{2},
\ee
with the property $(C_S)^2=C_S.$

\subsection{Modified De Palma-Trevisan-type quantum Wasserstein distance}
\label{sec:Modified De Palma-Trevisan-type quantum Wasserstein distance}

A modified De Palma-Trevisan distance $\tilde{D}_{\rm DPT}(\varrho,\sigma)$ has been considered that  does not have a nonzero self-distance  \cite{DePalma2024QuantumOptimal}
\begin{align}
\begin{split}
\tilde{D}_{\rm DPT}(\varrho,\sigma)^2&=D_{\rm DPT}(\varrho,\sigma)^2\\
&-[D_{\rm DPT}(\varrho,\varrho)^2+D_{\rm DPT}(\sigma,\sigma)^2]/2.\label{eq:Dmod}
\end{split}
\end{align}
The expression  fulfills the triangle inequality \cite{Bunth2024MetricProperty,bunth2025wassersteindistancesdivergencesorder,wirth2025triangleinequalityquantumwasserstein}. 

\section{Wasserstein distance based on an optimization over separable quantum states}
\label{sec:relations}

In this section, we consider the definitions given in \SEC{sec:Setting the stage}, and ask what happens if we optimize over separable quantum states instead of general quantum states. Separble quantum states are just mixtures of product states \cite{Werner1989Quantum}.  Any quantum state that is not separable is called entangled. (For a review on entanglement theory, see, e.~g., 
 \REFS{Horodecki2009Quantum,Guhne2009Entanglement,Friis2019}).
 
\subsection{De Palma-Trevisan-type quantum Wasserstein distance optimized over separable states}
\label{sec:sep_De Palma-Trevisan-type quantum Wasserstein distance}

It has been found that a quantum Wasserstein distance between quantum states can be defined similarly to \EQ{eq:DPT}, 
\begin{align}
\begin{split}
D_{\rm DPT, sep}(\varrho,\sigma)^2\quad\\=\frac 1 2
\min_{ \varrho_{12}\in\mathcal D}\sum_{n=1}^N\;&
\trace[(H_n\otimes \openone-\openone\otimes H_n)^2  \varrho_{12} ],\\
\textrm{s.~t. }&
\varrho_{12}\in\mathrm{Sep},\label{eq:GMPC_distance_sep}  \\
& {\rm Tr}_1(\varrho_{12})=\varrho, \\
& {\rm Tr}_2(\varrho_{12})=\sigma,
\end{split}
\end{align}
where the optimization is over separable quantum states \cite{Toth2023QuantumWasserstein}. Note that when considering the optimization over separable states, a simpler form  has been obtained without  transposes \cite{Toth2023QuantumWasserstein}. While 
\be
D_{\rm DPT, sep}(\varrho,\sigma)^2\ge D_{\rm DPT}(\varrho,\sigma)^2,\label{eq:sep_notsep_ineq}
\ee
its self-distance for $N=1$ is 
\be
D_{\rm DPT, sep}(\varrho,\varrho)^2=\frac 1 4 F_Q[\varrho,H_1],\label{eq:selfd}
\ee
where $F_Q$ is the quantum Fisher information defined as \cite{Helstrom1976Quantum,Holevo1982Probabilistic,Braunstein1994Statistical,Braunstein1996Generalized,Petz2008Quantum}
\begin{equation}
\label{eq:FQ}
{\mathcal F}_Q[\varrho,H]=2\sum_{k,l}\frac{(\lambda_{k}-\lambda_{l})^{2}}{\lambda_{k}+\lambda_{l}}\vert \langle k \vert H \vert l \rangle \vert^{2},
\end{equation}
where the density matrix has the eigendecomposition
\begin{equation}\label{eq:rho_eigdecomp}
\varrho=\sum_{k}\lambda_k \ketbra{k}.
\end{equation}
The definition of $D_{\rm DPT, sep}(\varrho,\sigma)^2$ was motivated by the fact that the  quantum Fisher information is four times the convex roof of the variance \cite{Toth2013Extremal,Yu2013Quantum}, and this statement can be reformulated as an optimization over separable states \cite{Toth2015Evaluating}, which can be used to estimate the  quantum Fisher information based on measurements \cite{Apellaniz2017Optimal}, and to formulate new uncertainty relations \cite{Toth2022Uncertainty,Chiew2022Improving}. In the more general case of non-unitary dynamics, the quantum Fisher information is not four times the convex roof of the variance any more, but it can be obtained via  semidefinite programming \cite{MullerRigat2023CertifyingQuantum}. In \EQ{eq:GMPC_distance_sep},  the coupling $\varrho_{12}$ is a separable state \cite{Werner1989Quantum}
\be
\varrho_{12}=\sum_k p_k \ketbra{\Psi_k} \otimes \ketbra{\Phi_k}.\label{eq:sep}
\ee
The marginals are
\begin{subequations}
\begin{align}
\varrho&=\sum_k p_k  \ketbra{\Psi_k},\label{eq:marginalcond_a}\\
\sigma&=\sum_k p_k  \ketbra{\Phi_k}.\label{eq:marginalcond_b}
\end{align}\label{eq:marginalcond}
\end{subequations}
The self-distance for $N>1$ is given as  \cite{Toth2023QuantumWasserstein}
\be
D_{\rm DPT, sep}(\varrho,\varrho)^2=\min_{\{p_k,\ket{\Psi_k}\}}\sum_{n=1}^N \sum_k p_k \va{H_n}_{\Psi_k}, \label{eq:selfd_Ngreaterthan1}
\ee
where $\varrho$ is decomposed as in \EQ{eq:marginalcond_a}.

Let us consider the case when $\varrho=\ketbra{\Psi}$ is a pure state of any dimension and $\sigma$ is an arbitrary density matrix of the same dimension. Then, when computing the various quantum Wasserstein distance measures between $\varrho$ and $\sigma,$ the state $\varrho_{12}$ in the optimization is constrained to be the tensor product of the two density matrices.  Let us define the auxiliary quantity
\begin{align}
\begin{split}
&\Delta(\varrho,\sigma)^2=\frac1 2 \sum_{n=1}^N\bigg[\va{H_n}_{\varrho}+\va{H_n}_{\sigma}\\
&\quad\quad\quad\quad\quad\quad+(\ex{H_n}_{\varrho}-\ex{H_n}_{\sigma})^2\bigg].\label{eq:Delta}
\end{split}
\end{align}
Hence, for the distance from a pure state \cite{Toth2023QuantumWasserstein}
\begin{align}
\begin{split}
D_{\rm DPT}(\ketbra{\Psi},\sigma)^2&=D_{\rm DPT, sep}(\ketbra{\Psi},\sigma)^2\\
&=\Delta(\ketbra{\Psi},\sigma)^2
\label{eq:D2_pure_mixed}
\end{split}
\end{align}
holds. 

The optimization given in \EQ{eq:GMPC_distance_sep} can be carried out numerically exactly using semidefinite programming for the case of qubits, while we can obtain vary good lower bounds for large systems, if we carry out the minimization over quantum states with a positive partial transpose \cite{Toth2023QuantumWasserstein}. The Wasserstein distance above is known to fulfill the condition (see Eq.~(126) in \REF{Toth2023QuantumWasserstein})
\begin{align}\label{eq:GMPC_distance_sep_recursive}
D_{\rm DPT, sep}(\varrho,\sigma)^2\quad\quad\quad\nonumber\\=
\min_{\{p_k,\ket{\Psi_k},\ket{\Phi_k}\}}\;&
\sum_k p_k D_{\rm DPT}(\ket{\Psi_k},\ket{\Phi_k})^2,
\end{align}
where for pure states we have 
\begin{align}
\begin{split}
D_{\rm DPT}(\ket{\Psi},\ket{\Phi})^2&=D_{\rm DPT, sep}(\ket{\Psi},\ket{\Phi})^2\\
&=\Delta(\ket{\Psi},\ket{\Phi})^2,
\label{eq:eq:D2_pure_mixedB}
\end{split}
\end{align}
where the conditions for the marginals given in \EQ{eq:marginalcond} are fulfilled and $\Delta(\varrho,\sigma)^2$ is defined in \EQ{eq:Delta}. Note that here we used the notation $D_{\rm DPT}(\ket{\Psi},\ket{\Phi})^2$ instead of $D_{\rm DPT}(\ketbra{\Psi},\ketbra{\Phi})^2$ for shortness, and we will follow this convention in the rest of the paper.

Thus, we  can define $D_{\rm DPT, sep}(\varrho,\sigma)^2$  based on optimizing a quantity that is given for pure states as in \EQ{eq:GMPC_distance_sep_recursive}. This is an advantage, since we can start from a quantity defined for pure states only, and then define a quantity that is also valid for mixed states. However, we also know how to compute it numerically using semidefinite programming based on \EQ{eq:GMPC_distance_sep}.

An upper bound can be found based on an optimization only over one of the two states as
\be
D_{\rm DPT, sep}(\varrho,\sigma)^2\le \min_{\{p_k,\Psi_k\}}\sum_k p_k D_{\rm DPT, sep}(\ket{\Psi_k},\sigma)^2.\label{eq:upperbound}
\ee
where for $\varrho$ \EQ{eq:marginalcond_a} holds.

Note that we could also write down an optimization over mixed separable states as 
\begin{align}\label{eq:GMPC_distance_sep_recursive2b}
D_{\rm DPT, sep}(\varrho,\sigma)^2\quad\quad\quad\nonumber\\=
\min_{\{p_k, \varrho_k, \sigma_k\}}\;&
\sum_k p_k D_{\rm DPT}(\varrho_k,\sigma_k)^2,
\end{align}
where the separable state is given as 
\be
\varrho_{12}=\sum_k p_k \varrho_k \otimes \sigma_k.\label{eq:sep2}
\ee
The marginals are
\begin{align}
\begin{split}
\varrho&=\sum_k p_k  \varrho_k,\\
\sigma&=\sum_k p_k  \sigma_k.\label{eq:marginalcond2}
\end{split}
\end{align}
Since in \EQ{eq:GMPC_distance_sep} we optimize an expression linear in expectation values, the optimization in \EQ{eq:GMPC_distance_sep_recursive2b} would lead to the same result as in \EQ{eq:GMPC_distance_sep_recursive}.

For $N>1,$ for the self-distance 
\begin{align}
&\frac 1 4 \sum_{n=1}^N F_Q[\varrho,H_n] \nonumber\\
&\quad\quad\le D_{\rm DPT, sep}(\varrho,\varrho)^2=\min_{\{p_k,\ket{\Psi_k}\}}&
\sum_{n=1}^N\sum_k p_k \va{H_n}_{\Psi_k}\nonumber\\
&\quad\quad\le \sum_{n=1}^N \va{H_n}_{\varrho}\label{eq:selfd2}
\end{align}
holds, since the variance is the concave roof of itself \cite{Toth2013Extremal}. Here $\varrho$ is decomposed as  in \EQ{eq:marginalcond_a}  \cite{Toth2023QuantumWasserstein}. Note the important relation between the  the Wigner-Yanase skew information and the quantum Fisher information given as (e.~g., see \REF{Toth2013Extremal})
\be
I_{\varrho}(H)\le\frac1 4 F_Q[\varrho,H].\label{eq:FQI}
\ee

The quantum Wasserstein distance can be bounded from below with the self-distances as 
\be
D_{\rm DPT, sep}(\varrho,\sigma)^2\ge \frac1 2 [D_{\rm DPT, sep}(\varrho,\varrho)^2 + D_{\rm DPT, sep}(\sigma,\sigma)^2],\label{eq:selfdbound}
\ee
see Eq.~(133) in  \REF{Toth2023QuantumWasserstein}.

Note that even for  $D_{\rm DPT, sep}(\varrho,\sigma)^2$ there is a transport map corresponding to all separable $\varrho_{12}$ couplings, see Eq.~(83) in \REF{Toth2023QuantumWasserstein}. 

Finally, it is possible to define the Golse-Mouhot-Paul-Caglioti distance given in \EQ{eq:GMPC_distance} for the case that we optimize over separable states. It has been proven that \cite{Toth2023QuantumWasserstein}
\be
D_{\rm GMPC, sep}(\varrho,\sigma)^2=D_{\rm DPT, sep}(\varrho,\sigma)^2
\ee
holds.

\subsection{SWAP-fidelity and the corresponding quantum Wasserstein distance optimized over separable states}
\label{sec:sep_SWAP-fidelity and the corresponding quantum Wasserstein distance}

Before turning to the SWAP-fidelity, let us obtain a useful relation for the Uhlmann-Jozsa fidelity, which we will use later. 

The Uhlmann-Jozsa fidelity can be given as  \cite{Uhlmann1976TheTransitionProbability}
\be
F(\varrho,\sigma)=\max_{\Psi_\sigma,\Phi_\sigma} |\langle \Psi_\varrho | \Phi_\sigma\rangle|^2,\label{eq:overlap2}
\ee
where $\ket{\Psi_\varrho}$ and $\ket{\Phi_\sigma}$ are the purifications of $\varrho$ and $\sigma,$ respectively. 
In fact, it has also been shown that it is sufficient to optimize over one of the purifications
\be
F(\varrho,\sigma)=\max_{\Phi_\sigma} |\langle \Psi_\varrho | \Phi_\sigma\rangle|^2.\label{eq:overlap2b}
\ee
The measurement of quantum fidelity and the characterization of its properties is at the center of attention in quantum information science \cite{BHATIA2019165,Cerezo2020variationalquantum,afham2022quantummeanstatesnicer,gilyen2022improvedquantumalgorithmsfidelity}.

We can reformulate \EQ{eq:overlap2} as an optimization over separable states as follows.

\DEFOBS{thm:UhlmannJzsaFidelity}For the Uhlmann-Jozsa fidelity the relation 
\begin{align}
F(\varrho,\sigma)=\left(\max_{\{p_k,\Psi_k,\Phi_k\}}\sum_k p_k  |\bra{\Psi_k} \Phi_k\rangle |\right)^2,\label{eq:Fidelity}
\end{align}
holds such that the conditions for the marginals given in \EQ{eq:marginalcond} are fulfilled. In other words,
$\sqrt{F}(\varrho,\sigma)$ can be given by an optimization over separable states as 
\be
\sqrt{F}(\varrho,\sigma)=\max_{\{p_k,\Psi_k,\Phi_k\}}\sum_k p_k \sqrt{F}(\ket{\Psi_k},\ket{\Phi_k}).\label{eq:sqrtFoptim}
\ee

{\it Proof.}  Let us consider the purifications
\begin{align}
\begin{split}
\ket{\Psi_\varrho}&=\sum_k \sqrt{p_k} \ket{\Psi_k}_S\otimes \ket{k}_A,\\
\ket{\Phi_\sigma}&=\sum_k \sqrt{q_k} \ket{\Phi_k}_S\otimes \ket{k}_A.
\end{split}
\end{align}
where $S$ refers to the system and  $A$ refers to the ancilla. Then, if we measure the ancilla in the $z$-basis, we obtain decompositons of $\varrho$ and $\sigma$ as
\begin{align}
\begin{split}
\varrho=\sum_k p_k \ketbra{\Psi_k},\\
\sigma=\sum_k q_k \ketbra{\Phi_k}.
\end{split}
\end{align}
Note that if we measure the two ancillas in the $\{\ket{k}\}$-basis then we obtain the probabilities as
\begin{align}
\begin{split}
\bra{k} {\rm Tr}_{S}(\ketbra{\Psi_\varrho})\ket{k}=p_k,\\
\bra{k} {\rm Tr}_{S}(\ketbra{\Phi_\sigma})\ket{k}=q_k.
\end{split}
\end{align}
The reduced state of ancilla is not necessarily diagonal in the $\{\ket k\}$-basis. With these, we have (e. g., Equation~(23) in \REF{Streltsov2010Linking})
\begin{align}
\begin{split}
\langle \Psi_\varrho | \Phi_\sigma\rangle=\sum_k \sqrt{p_kq_k}\bra{\Psi_k}\Phi_k\rangle.\label{eq:optim}
\end{split}
\end{align}
For an optimization over all purifications, we obtain
\begin{align}
\begin{split}
\max_{\{\Psi_\varrho,\Phi_\sigma\}} |\langle \Psi_\varrho | \Phi_\sigma\rangle|=\max_{\{p_k,q_k,\Psi_k,\Phi_k\}} \sum_k \sqrt{p_kq_k}|\bra{\Psi_k}\Phi_k\rangle|.\label{eq:optim2}
\end{split}
\end{align}
Here, since the maximum is taken when $\langle  \Psi_k| \Phi_k\rangle$ are all real and positive 
we replaced $\langle \Psi_\varrho | \Phi_\sigma\rangle$ with $|\langle \Psi_\varrho | \Phi_\sigma\rangle|$ in the maximization \cite{Streltsov2010Linking}.

Let us see now how to construct the various purifications from each other. Other purifications of $\varrho$ and $\sigma$ can be obtained as
\begin{align}
\begin{split}
\ket{\Psi'_\varrho}=(\openone_S \otimes U_A)\ket{\Psi_\varrho}&=\sum_k \sqrt{p'_k} \ket{\Psi'_k}_S\otimes \ket{k}_A,\\
\ket{\Phi'_\sigma}=(\openone_S \otimes V_A)\ket{\Phi_\sigma}&=\sum_k \sqrt{q'_k} \ket{\Phi'_k}_S\otimes \ket{k}_A,
\end{split}
\end{align}
where $U_A$ and $V_A$ are unitaries. 
After measuring in the $\{\ket k\}$-basis again we find further decompositions of $\varrho$ and $\sigma$ as
\begin{align}
\begin{split}
\varrho=\sum_k p'_k \ketbra{\Psi_k'},\\
\sigma=\sum_k q'_k \ketbra{\Phi_k'}.
\end{split}
\end{align}

It is well known that for any traceless matrix there is a basis in which all diagonal elements of the matrix are zero (see Example 2.2.3 in \REF{Horn_Johnson_2012}).
Hence, there is a basis $\{\ket{e_k}\}$ such that \be\langle e_k | \left[{\rm Tr}_{S}(\ketbra{\Psi_\varrho})- {\rm Tr}_{S}(\ketbra{\Phi_\sigma})\right]|e_k\rangle =0.\ee 
 Then, if we employ $U_A=V_A=\ket{k}\bra{e_k}$ and then we measure in the $\{\ket k\}$-basis, we get decompositions with $q'_k=p'_k.$
Thus, for any $\ket{\Psi_\varrho}$ and $\ket{\Phi_\sigma}$ there is $\ket{\Psi'_\varrho}$ and $\ket{\Phi_\sigma'}$ such that 
$p_k'=q_k'$ for all $k,$ while $|\langle \Psi_\varrho | \Phi_\sigma\rangle|=|\langle \Psi'_\varrho | \Phi'_\sigma\rangle|.$ Thus,  in the optimization in \EQ{eq:optim2}, it is sufficient to consider the purifications of $\varrho$ and $\sigma$ for which $p_k=q_k$ holds, and
we arrive at \EQ{eq:Fidelity}. $\qed$

From \OBS{thm:UhlmannJzsaFidelity}, the joint concavity of $\sqrt{F}(\varrho,\sigma)$ can be seen immediately, that is, for any $\varrho, \sigma$ and $0\le p\le1$
\be
\sqrt{F}(\varrho,\sigma) \ge p \sqrt{F}(\varrho_1,\sigma_1) + (1-p) \sqrt{F}(\varrho_2,\sigma_2)
\ee
holds, where $\varrho=p\varrho_1+(1-p)\varrho_2$ and $\sigma=p\sigma_1+(1-p)\sigma_2.$
 
 Let us rederive some known relations using the definition of the fidelity via an optimization given in \EQ{eq:sqrtFoptim}. Let us consider the dynamics
\be
\varrho_\theta=e^{-iH\theta}\varrho e^{+iH\theta}\label{eq:varrho_theta}
\ee
for small $\theta.$ Clearly, if $\varrho$ is a pure state, we have
\be
\sqrt{F}(\varrho,\varrho_{\theta}) = 1-\frac {\theta^2} 2 \va{H}_{\varrho}+O(\theta^3), 
\ee
where $O(x)$ is the usual asymptotic notation. For a mixed $\varrho,$ we have
\begin{align}
\begin{split}
\sqrt{F}(\varrho,\varrho_\theta) &\ge \max_{\{p_k,\Psi_k\}}\sum_k p_k \sqrt{F}(\ket{\Psi_k},e^{-iH\theta}\ket{\Psi_k})\\
&=1-\frac {\theta^2} 2 \min_{\{p_k,\Psi_k\}} \sum_k p_k \va{H}_{\Psi_k}+O(\theta^3),
\end{split}\label{eq:ineqF}
\end{align}
where we have an inequality due to the fact that we optimize such that $\ket{\Phi_k}$ depend on  $\ket{\Psi_k},$ 
while the optimization in \EQ{eq:sqrtFoptim} is over a larger set of decompositions.

For pure states the two sides of the inequality in \EQ{eq:ineqF} are equal to each other.
Based on \EQ{eq:Fidelity}, the left-hand side is a concave roof.
A concave roof is the smallest concave function that has given values for pure states \cite{Hill1997Entanglement,Wootters1998Entanglement,Bennett1996Mixed-state,Uhlmann2010Roofs}.
The right-hand side is also a concave roof. 
Thus, there must be an equality in \EQ{eq:ineqF}, and hence 
\begin{align}
\begin{split}
\frac{\partial^2\sqrt{F}(\varrho,\varrho_\theta)}{\partial\theta^2}\bigg|_{\theta=0} &= - \min_{\{p_k,\Psi_k\}} \sum_k p_k \va{H}_{\Psi_k}\\
&=- \frac 1 4 F_Q[\varrho,H],
\end{split}\label{eq:ineqF2}
\end{align}
where we used that the quantum Fisher information is the convex roof of the variance over four \cite{Toth2013Extremal,Yu2013Quantum}.
For the second derivative of the fidelity we have 
\begin{align}
\begin{split}
\frac{\partial^2F(\varrho,\varrho_\theta)}{\partial\theta^2}\bigg|_{\theta=0} &= - 2\min_{\{p_k,\Psi_k\}} \sum_k p_k \va{H}_{\Psi_k}\\
&=- \frac 1 2 F_Q[\varrho,H].
\end{split}\label{eq:ineqF3}
\end{align}

We can also make the following statement.

\DEFOBS{thm:Bures}We can give the Bures metric with an optimization over separable states as
\be
D_{\rm Bures}(\varrho,\sigma)^2=2\min_{\{p_k,\Psi_k,\Phi_k\}} \sum_k p_k (1-\left| \langle \Psi_k| \Phi_k\rangle\right|).\label{eq:Brues_opt}
\ee
Thus, we can say that the square of the Bures metric  is the convex roof of itself
\be
D_{\rm Bures}(\varrho,\sigma)^2=\min_{\{p_k,\Psi_k,\Phi_k\}} \sum_k p_k D_{\rm Bures}(\ket{\Phi_k},\ket{\Psi_k})^2.\label{eq:Brues_opt2}
\ee
{\it Proof.} We used that the  Bures distance can be given with the fidelity as \cite{Uhlmann1976TheTransitionProbability,Uhlmann1995Geometric}
\be
D_{\rm Bures}(\varrho,\sigma)^2=2\left[1-\sqrt{F(\varrho,\sigma)}\right].\label{eq:Bures}
\ee
$\qed$

After discussing the  Uhlmann-Jozsa fidelity and the Bures metric, let us consider now the SWAP-fidelity based on an optimization over separable states as
\be\label{eq:Fssep_S}
F_{S,{\rm sep}}(\varrho,\sigma)=\max_{\varrho_{12}\in \mathrm{Sep}} {\rm Tr}(\varrho_{12} S),
\ee 
where $\mathrm{Sep}$ is the set of separable states. We can formulate the following theorem.

\DEFOBS{thm:FSsep}For the maximum over separable states
\be
F_{S,{\rm sep}}(\varrho,\sigma)=\max_{\{p_k,\Phi_k,\Psi_k\}}\sum_k p_k | \bra{\Phi_k} \Psi_k\rangle |^2\label{eq:Fssep}
\ee 
holds,  and the conditions for the marginals given in \EQ{eq:marginalcond} are fulfilled. It is instructive to compare \EQ{eq:Fssep} to \EQ{eq:Fidelity}.

{\it Proof.}  We have to use that the trace of the product of the Hermitian matrices $A$ and $B$ can be expressed with their tensor product and the SWAP operator as 
\be
{\rm Tr}(AB)={\rm Tr}[(A\otimes B) S].\label{eq:ABswapAotimesB}
\ee
Then, for a separable $\varrho_{12}$ given in \EQ{eq:sep} 
\be
{\rm Tr}(\varrho_{12} S)=\sum_k p_k | \bra{\Phi_k} \Psi_k\rangle |^2\label{eq_sepineq}
\ee
holds. $\qed$

Let us make a comment concerning the relation of this finding to entanglement theory. The relation in \EQ{eq_sepineq} indicates that for any separable state $\varrho_{12}$
\be
{\rm Tr}(\varrho_{12} S)\ge 0, \label{eq_sepineq1}
\ee
holds, and any state that violates the inequality in \EQ{eq_sepineq1} is entangled. Note that some of the eigenvalues of the SWAP operator are negative. For the two-qubit singlet state, $(\ket{01}-\ket{10})/\sqrt{2},$ its expectation value is $-1.$ It is also known that any bipartite quantum state in the symmetric subspace violating  \EQ{eq_sepineq1}  has a non-positive partial transpose \cite{Toth2009Entanglement}. 

Let us now compare the various quantities discussed so far. We find that the inequalities
\be
F(\varrho,\sigma)\le F_{S,{\rm sep}}(\varrho,\sigma)\le F_{S}(\varrho,\sigma)\label{eq:FFFlele}
\ee 
hold. In particular, the first inequality holds since for any $p_k$ probabilities and $\left| \langle \Psi_k| \Phi_k\rangle\right|$  real numbers we have
\be
\sum_k p_k \left| \langle \Psi_k| \Phi_k\rangle\right|^2 \ge \left(\sum_k p_k \left| \langle \Psi_k| \Phi_k\rangle\right|\right)^2, \label{eq:xkxk}
\ee
which is due to the convexity of the function $f(x)=x^2.$  The second inequality holds since  on the left-hand side we are maximizing a quantity over a subset of physical states, while on the right-hand side we are maximizing over the entire set of physical states. 

Note that for pure states we have
\be
F_S(\ket{\Psi},\ket{\Phi})=|\bra{\Phi} \Psi\rangle|^2,
\ee
thus $F_{S,{\rm sep}}(\varrho,\sigma)$ can be given by an optimization over separable states as 
\be
F_{S,{\rm sep}}(\varrho,\sigma)=\max_{\{p_k,\Phi_k,\Psi_k\}}\sum_k p_k F_S(\ket{\Psi_k},\ket{\Phi_k})
\ee 
such that the conditions for the marginals given in \EQ{eq:marginalcond} are fulfilled. The distance based on an optimization over separable states is defined similarly to \EQ{eq:DSFS} as
\be
D_{S,{\rm sep}}(\varrho,\sigma)^2=\frac1 2[1-F_{S,{\rm sep}}(\varrho,\sigma)].  \label{eq:Dssep}
\ee
Then, analogously to the statements about the SWAP-fidelity, we have that
\be
D_{S,{\rm sep}}(\varrho,\sigma)^2=\min_{\{p_k,\Phi_k,\Psi_k\}}\sum_k p_k D_S(\ket{\Psi_k},\ket{\Phi_k})^2 \label{eq:Dssep2}
\ee 
holds such that the conditions for the marginals given in \EQ{eq:marginalcond} are fulfilled.

Moreover, for a pure state $\varrho=\ketbra{\Psi}$ and a mixed state $\sigma$ we have all the inequalities in \EQ{eq:FFFlele} saturated and hence
\be
F(\ketbra{\Psi},\sigma)= F_{S,{\rm sep}}(\ketbra{\Psi},\sigma)= F_{S}(\ketbra{\Psi},\sigma).\label{eq:FFFlele2}
\ee

Next, let us consider another related quantity, the superfidelity defined as  \cite{Miszczak2009Sub}
\be
F_{\rm super}(\varrho,\sigma)={\rm Tr}(\varrho\sigma)+\sqrt{[1-{\rm Tr}(\varrho^2)][1-{\rm Tr}(\sigma^2)]}.\label{eq:Fsuper}
\ee
We know that the fidelity is bounded from above as 
\be
F(\varrho,\sigma) \le F_{\rm super}(\varrho,\sigma).
\ee
For $d=2$ there is an equality \cite{Miszczak2009Sub,Hubner1992Explicit}. 
\be
F(\varrho,\sigma) = F_{\rm super}(\varrho,\sigma). \label{eq:FFsuper}
\ee
We can now formulate the following bound.

\DEFOBS{thm:swapfidelitysuperfidelity}The SWAP-fidelity for separable states is bounded from above as 
\begin{align}
\begin{split}
&F_{S,{\rm sep}}(\varrho,\sigma)\le F_{\rm super}(\varrho,\sigma)\label{eq:Fsuperbound}
\end{split}
\end{align}
As a consequence, for $d=2$ (i.~e., for qubits), we have 
\be
F_{S,{\rm sep}}(\varrho,\sigma)=F(\varrho,\sigma).\label{eq:FseqF}
\ee

{\it Proof.} Let us consider a separable state $\varrho_{12}$ given in \EQ{eq:sep}, with the marginals $\varrho$ and $\sigma$ as in \EQ{eq:marginalcond}. Let us introduce the traceless Hermitian matrices characterizing the deviation of the pure subensembles form the ensemble average as
\begin{align}
\begin{split}
\varrho_k&=\ketbra{\Psi_k}-\varrho,\\
\sigma_k&=\ketbra{\Phi_k}-\sigma.\label{eq:rhok_sigmak}
\end{split}
\end{align}
Then, for any decomposition of $\varrho$ and $\sigma$ we have 
\begin{align}
\begin{split}
&{\rm Tr}(\varrho_{12}S)\\
&\quad\quad=\sum_k p_k | \bra{\Phi_k} \Psi_k\rangle |^2\\
&\quad\quad=\sum_k p_k {\rm Tr}(\ketbra{\Psi_k} \Phi_k\rangle\langle \Phi_k|)\\
&\quad\quad= \sum_k p_k {\rm Tr}[(\varrho+\varrho_k)(\sigma+\sigma_k)]\\
&\quad\quad=  {\rm Tr}(\varrho\sigma) + {\rm Tr}\left( \sum_k p_k \varrho_k\sigma_k\right) \\
&\quad\quad\le {\rm Tr}(\varrho\sigma) + \sqrt{  {\rm Tr}\left( \sum_k p_k \varrho_k^2\right) {\rm Tr}\left(\sum_k p_k \sigma_k^2\right) },\label{eq:bigineq}
\end{split}
\end{align}
where we used in the fourth equality that
\be
\sum_k p_k \varrho_k \sigma=\sum_k p_k \sigma_k \varrho=0.
\ee
In  the inequality in \EQ{eq:bigineq}, we used the Cauchy-Schwarz inequality in the following form
\be
{\rm Tr}\left( A^\dagger B \right)\le \sqrt{{\rm Tr}\left( A^\dagger A \right)}  \sqrt{{\rm Tr}\left( B^\dagger B\right)},
\ee
where we define the block diagonal matrices 
\begin{align}
\begin{split}
A&={\rm diag}(\sqrt{p_1}\varrho_1,\sqrt{p_2}\varrho_2,\sqrt{p_3}\varrho_3,...),\\
B&={\rm diag}(\sqrt{p_1}\sigma_1,\sqrt{p_2}\sigma_2,\sqrt{p_3}\sigma_3,...).\\
\end{split}
\end{align}
Then, we need the following identities 
\begin{align}
\begin{split}
{\rm Tr}\left(\sum_k p_k  \varrho_k^2\right)&=1-{\rm Tr}(\varrho^2),\\
{\rm Tr}\left(\sum_k p_k  \sigma_k^2\right)&=1-{\rm Tr}(\sigma^2),
\end{split}
\end{align}
where we used the definitions in \EQ{eq:rhok_sigmak}, taking into account
\be
\sum_k p_k \varrho_k \varrho=\sum_k p_k \sigma_k \sigma=0.
\ee
Hence, we arrive at \EQ{eq:Fsuperbound}.

\begin{figure}[t!]
\begin{center}
\begin{subfigure}[b]{0.42\textwidth}  
        \centering
        \includegraphics[width=\textwidth]{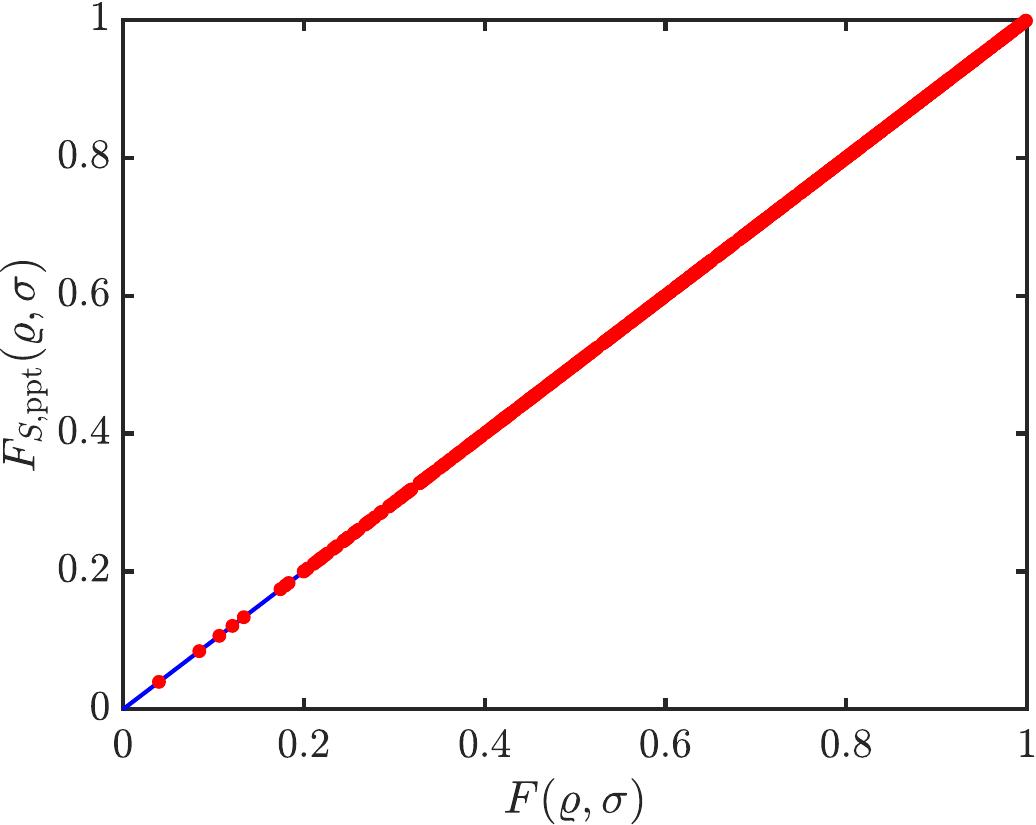}  
        \caption{}  
        \label{subfig:a}
    \end{subfigure}
    
      \vspace{0.20cm}
    \begin{subfigure}[b]{0.42\textwidth}  
        \centering
        \includegraphics[width=\textwidth]{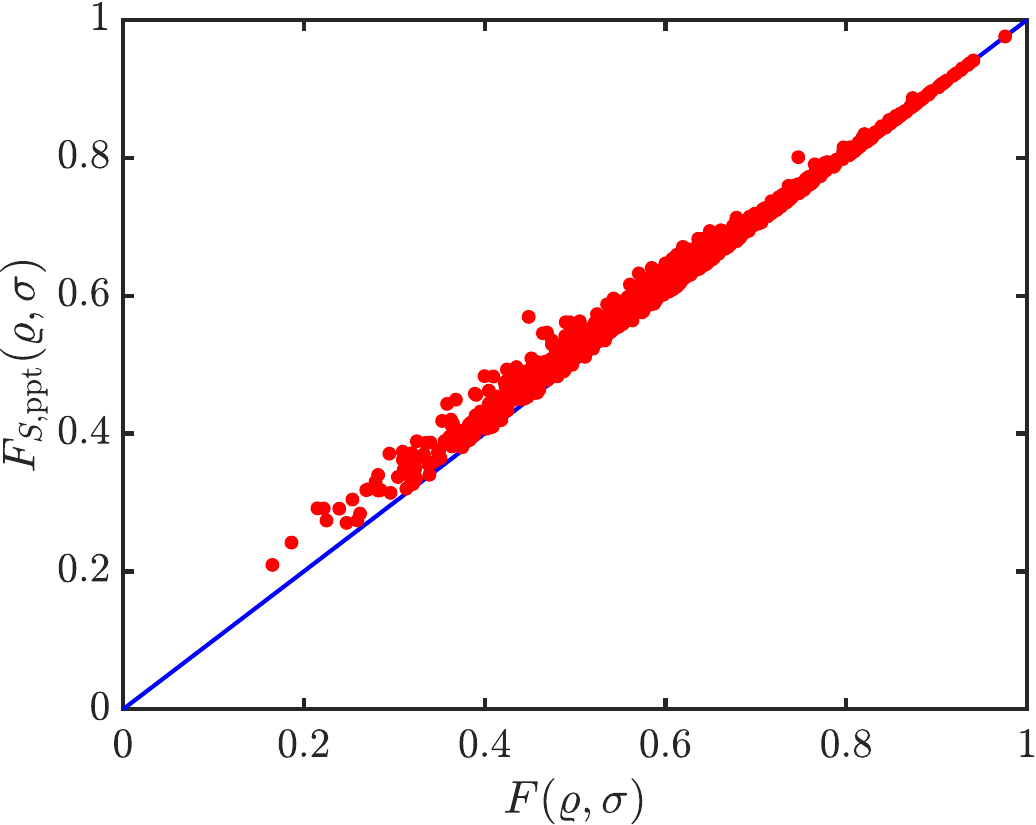}  
        \caption{}  
        \label{subfig:b}
            \end{subfigure}
            
       \vspace{0.20cm}
      \begin{subfigure}[b]{0.42\textwidth}  
        \centering
        \includegraphics[width=\textwidth]{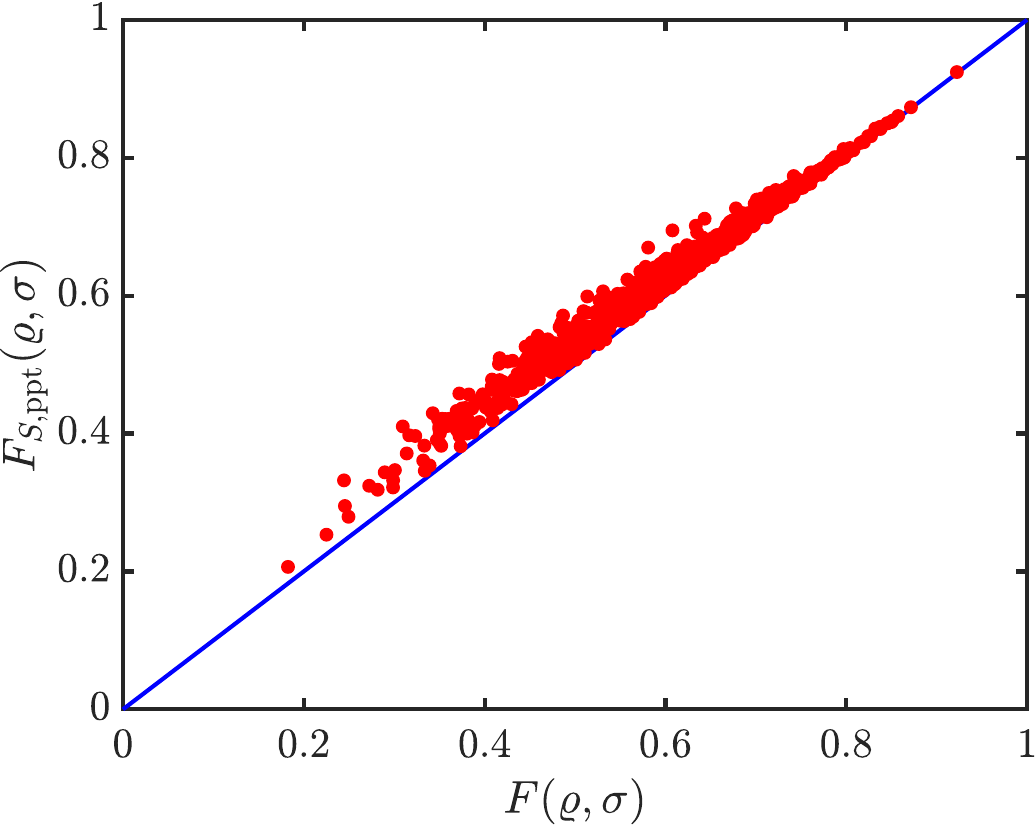}  
        \caption{}  
        \label{subfig:c}
    \end{subfigure}
    \end{center}
\caption{ $F_{S,{\rm ppt}}(\varrho,\sigma)$ vs. $F(\varrho,\sigma)$ for 1000 random pairs of states for (a) $d=2$, (b) $d=3$, and (c) $d=4.$ In the $d=2$ case the two quantities are equal to each other, see \OBS{thm:swapfidelitysuperfidelity}.
For $d>2,$ the inequality in \EQ{eq:FFFlele} holds.} \label{fig:Fig_F_Fs_ppt}
\end{figure}

For the $d=2$ case, we know that \EQ{eq:FFsuper} holds, and also the SWAP-fidelity computed over separable states is bounded as in \EQ{eq:FFFlele}. Hence, \EQ{eq:FseqF} follows. $\qed$

The relation in \EQ{eq:FseqF} can hold only if the inequality in \EQ{eq:xkxk} is saturated. Thus, for all $k$
\be
 \left| \langle \Psi_k| \Phi_k\rangle\right|^2=F(\varrho,\sigma)=F_{S,{\rm sep}}(\varrho,\sigma)
\ee
holds.


Note that \EQ{eq:FseqF}  also holds for two rank-2 states of a system with a dimension larger than two living in the same subspace.

Let us know verify some of the bounds we found numerically. We generated 1000 random pairs of density matrices and plotted  $F_{S,{\rm ppt}}(\varrho,\sigma)$ and $F(\varrho,\sigma)$ in \FIG{fig:Fig_F_Fs_ppt}. 
$F_{S,{\rm ppt}}(\varrho,\varrho_\theta)$ obtained via an optimization over PPT states instead of  $F_{S,{\rm sep}}(\varrho,\varrho_\theta),$ since the former can be computed with semidefinite programming and it is a very good upper bound on the latter.  
For qubits, the two quantities are equal to each other. We used a method that provides random density matrices uniformly distributed according to the Hilbert-Schmidt norm \cite{Sommers2004Statistical}.

Let us consider the unitary dynamics given in \EQ{eq:varrho_theta}. Then, following similar steps as in the case of the quantum fidelity, for the second derivative we have
\begin{align}
\begin{split}
\frac{\partial^2F_{S,{\rm sep}}(\varrho,\varrho_\theta)}{\partial\theta^2}\bigg|_{\theta=0} &= -2 \min_{\{p_k,\Psi_k\}} \sum_k p_k \va{H}_{\Psi_k}\\
&=- \frac 1 2 F_Q[\varrho,H],
\end{split}\label{eq:ineqF3b}
\end{align}
where we used that the quantum Fisher information is the convex roof of the variance over four \cite{Toth2013Extremal,Yu2013Quantum}, c.~f. \EQ{eq:ineqF3}.

\begin{figure}[t!]
\includegraphics[width=0.43\textwidth]{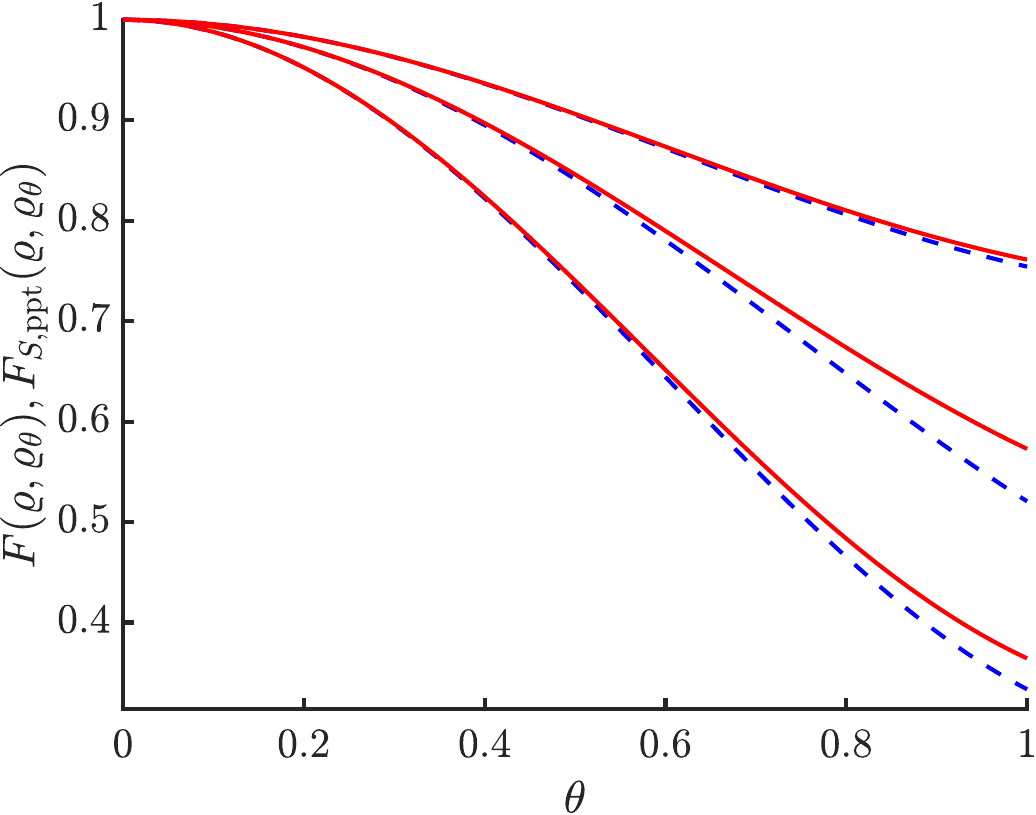}
\caption{Dynamics of (solid) $F_{S,{\rm ppt}}(\varrho,\varrho_\theta)$ and (dashed) $F(\varrho,\varrho_\theta)$ as a function of $\theta$ for a quantum state of dimension $d=4$.
We considered three randomly chosen $\varrho.$ The unitary dynamics is given in \EQ{eq:varrho_theta}, where the Hamiltonian is given in \EQ{eq:HJz}.
} \label{fig:Fig_F_Fs_small_angle}
\end{figure}

In \FIG{fig:Fig_F_Fs_small_angle}, we considered systems of dimension $d=4,$ and the Hamiltonian is the $z$-component of the collective angular momentum
\be
H=J_z={\rm diag}\left(-\frac 3 2,-\frac 1 2, \frac 1 2, \frac 3 2\right). \label{eq:HJz}
\ee
The two curves coincide with each other for small $\theta$ values, which is consistent with our findings that their second order derivatives at $\theta=0$ are equal.
We plotted $F_{S,{\rm ppt}}(\varrho,\varrho_\theta)$ obtained via an optimization over PPT states instead of  $F_{S,{\rm sep}}(\varrho,\varrho_\theta),$ since the former can be computed with semidefinite programming and it is a very good upper bound on the latter.  

\subsection{Modified De Palma-Trevisan-type quantum Wasserstein distance optimized over separable states}
\label{sec:sep_Modified De Palma-Trevisan-type quantum Wasserstein distance}

\subsubsection{Two possible generalizations for the optimization over separable states}

In \EQ{eq:GMPC_distance_sep_recursive}, we optimized the square of the De Palma-Trevisan-type quantum Wasserstein distance. Next, we examine what happens if we replace  $D_{\rm DPT}(\ket{\Psi_k},\ket{\Phi_k})^2$ in \EQ{eq:GMPC_distance_sep_recursive} with the square of the modified distance $\tilde{D}_{\rm DPT}(\ket{\Psi_k},\ket{\Phi_k})^2$  given in  \EQ{eq:HHH}. 
Let us consider now the quantity defining the Euclidean distance squared with the expectation values of $H_n$ as
\be
E(\varrho,\sigma)^2=\frac 1 2 \sum_{n=1}^N (\ex{H_n}_{\varrho}-\ex{H_n}_{\sigma})^2.\label{eq:Erho}
\ee
For simplifying our calculations, we also included the constant $1/2$  in \EQ{eq:Erho}. For pure states, the modified De~Palma-Trevisan-type quantum Wasserstein distance is given as 
\be
\tilde{D}_{\rm DPT}(\ket{\Psi},\ket{\Phi})^2=E(\ket{\Psi},\ket{\Phi})^2.\label{eq:HHH}
\ee
Clearly, if $\ket{\Psi}=\ket{\Phi}$ then the above expression is zero. Then, we consider the quantity
\begin{align}
\begin{split}
\label{eq:GMPC_distance_sep_recursive2}
&\tilde D_{\rm DPT, decomp}(\varrho,\sigma)^2\\
&\quad=\min_{\{p_k,\ket{\Psi_k},\ket{\Phi_k}\}}
\sum_k p_k \tilde{D}_{\rm DPT}(\ket{\Psi_k},\ket{\Phi_k})^2,
\end{split}
\end{align}
such that the conditions for the marginals given in \EQ{eq:marginalcond} are fulfilled. Here the subscript "decomp"  in \EQ{eq:GMPC_distance_sep_recursive2} indicates that the quantity is defined with an optimization over decompositions to pure product states. Based on the definition given in \EQ{eq:GMPC_distance_sep_recursive2}, we can see that
\begin{align}
\begin{split}
&\tilde D_{\rm DPT, decomp}(\varrho,\sigma)^2\\
&\quad\le\min_{\{p_k,\ket{\Psi_k}\}}  \sum_k p_k \tilde D_{\rm DPT, decomp}(\ket{\Psi_k},\sigma)^2,\label{eq:decomp_single_optmization}
\end{split}
\end{align}
where $\varrho$ is decomposed as in \EQ{eq:marginalcond_a}. 

Next, we will find a relevant inequality for the quantity we have just introduced.

\DEFOBS{thm:decomp}The expression based on a  minimization over separable decompositions given in \EQ{eq:GMPC_distance_sep_recursive2} is bounded from below with the distance based on the expectation values  as 
\begin{align}
\tilde D_{\rm DPT, decomp}(\varrho,\sigma)^2&\ge
\min_{\{p_k,\varrho_k,\sigma_k\}}\;
\sum_k p_k E(\varrho_k,\sigma_k)^2\nonumber\\
& = E(\varrho,\sigma)^2, \label{eq:decomp}
\end{align}
where the optimization is over separable states of the form given in \EQ{eq:sep2}, the marginals fulfill \EQ{eq:marginalcond2}, and $E(\varrho,\sigma)^2$ is defined in \EQ{eq:Erho}.

{\it Proof.} The inequality  in \EQ{eq:decomp} is true since the minimization is over a larger set on the right-hand side than on the left-hand side. 
We will now prove that result of the minimization equals $E(\varrho,\sigma),$ thus the equality in \EQ{eq:decomp}  is true.
Let us define
\begin{align}
\begin{split}
Q_{n,k}&=\langle H_n\rangle_{\varrho_k}-\langle H_n\rangle_{\sigma_k},\\
W_n&=\langle H_n\rangle_{\varrho}-\langle H_n\rangle_{\sigma}.
\end{split}
\end{align}
Then, let us consider the non-negative quantity
\begin{align}
Z^2&=\sum_{n=1}^N  \sum _k p_k(Q_{n,k}-W_n)^2\nonumber\\
&=\sum_{n=1}^N  \left(\sum _k p_kQ_{n,k}^2\right)+W_n^2-2\left(\sum _k p_kQ_{n,k}\right)W_n\nonumber\\
&=\sum_{n=1}^N  \sum _k p_k Q_{n,k}^2-W_n^2,
\end{align}
where we use that
\be
\sum_k p_k Q_{n,k}=W_n.
\ee
From these, it follows that
\begin{equation}
\sum_{n=1}^N \sum _k p_kQ_{n,k}^2 =Z^2+ \sum_{n=1}^N W_n^2\ge \sum_{n=1}^N W_n^2=2E(\varrho,\sigma)^2.\label{eq:ineqHnHn}
\end{equation}
So far we have proved that the result of the minimization in  in \EQ{eq:decomp} is larger than or equal to $E(\varrho,\sigma).$
The equality in \EQ{eq:decomp} holds, since $p_1=1, \varrho_1=\varrho,$ and $\sigma_1=\sigma$ minimizes the sum, and the value of the minimum is $E(\varrho,\sigma).$  $\qed$

If $N=1$ then the inequality in \EQ{eq:decomp} is saturated. The reason is that for all $\varrho$ there is a decomposition of the type \EQ{eq:marginalcond_a} such that $\ex{H_1}_{\Psi_k}=\ex{H_1}_{\varrho}$ for all $k.$ For all $\sigma$ there is a decomposition of the type \EQ{eq:marginalcond_b} such that $\ex{H_1}_{\Phi_k}=\ex{H_1}_{\sigma}$ for all $k$  \cite{Toth2013Extremal}.

If $N=2$ then the inequality in \EQ{eq:decomp} is saturated, since a similar statement holds for the expectation values of two operators, $H_1$ and $H_2$ \cite{Leka2013Some,Petz2014}. 

Following the idea for the definition of the modified quantum Wasserstein distance in \EQ{eq:Dmod}, let us now define an analogous quantity for the 
quantum Wasserstein distance  $D_{\rm DPT, sep}(\varrho,\sigma)$ as
\begin{align}
\begin{split}
&\tilde D_{\rm DPT, sep}(\varrho,\sigma)^2=D_{\rm DPT, sep}(\varrho,\sigma)^2\\
&\quad-[D_{\rm DPT,sep}(\varrho,\varrho)^2+D_{\rm DPT, sep}(\sigma,\sigma)^2]/2.\label{eq:tildeDPTsep}
\end{split}
\end{align}
This is a non-negative quantity due to the inequality in \EQ{eq:selfdbound}. Clearly, for pure states 
\be
\tilde D_{\rm DPT, decomp}(\varrho,\sigma)^2=\tilde D_{\rm DPT}(\varrho,\sigma)^2
\ee
holds. Then, we can relate $\tilde D_{\rm DPT, decomp}(\varrho,\sigma)^2$ and $\tilde D_{\rm DPT, sep}(\varrho,\sigma)^2$ to each other as follows.

\DEFOBS{thm:decomp_sep}For the modified quantum Wasserstein distance based on an optimization over separable states
\begin{align}
&\tilde D_{\rm DPT, decomp}(\varrho,\sigma)^2\le \tilde D_{\rm DPT, sep}(\varrho,\sigma)^2\label{eq:tildeDineq}
\end{align}
holds. For the $N=1$ case, this means 
\begin{align}
\begin{split}
&\tilde D_{\rm DPT, decomp}(\varrho,\sigma)^2\le D_{\rm DPT, sep}(\varrho,\sigma)^2\\
&\quad-(F_Q[\varrho,H_1]+F_Q[\sigma,H_1])/8.\label{eq:tildeDineq3}
\end{split}
\end{align}
{\it Proof.} Based on \EQ{eq:GMPC_distance_sep_recursive} and \EQ{eq:eq:D2_pure_mixedB}, we can write that 
\begin{align}
\begin{split}
&D_{\rm DPT, sep}(\varrho,\sigma)^2\quad\quad\quad\\
&\quad=\min_{\{p_k,\ket{\Psi_k},\ket{\Phi_k}\}}\;\sum_k p_k D_{\rm DPT}(\ket{\Psi_k},\ket{\Phi_k})^2\\
&\quad\ge [D_{\rm DPT,sep}(\varrho,\varrho)^2+D_{\rm DPT, sep}(\sigma,\sigma)^2]/2\\
&\quad\quad+\min_{\{p_k,\ket{\Psi_k},\ket{\Phi_k}\}}\;\sum_k p_k \tilde D_{\rm DPT}(\ket{\Psi_k},\ket{\Phi_k})^2.
\end{split}\label{eq:tildeDineq2b}
\end{align}
For $N=1$, the self-distance is given in \EQ{eq:selfd}. $\qed$

Next, we relate $\tilde D_{\rm DPT, sep}(\varrho,\sigma)^2$ and $\tilde D_{\rm DPT}(\varrho,\sigma)^2$ to each other. Since the inequality in \EQ{eq:sep_notsep_ineq} holds, 
we would expect an analogous relation for the modified quantities. Unexpectedly, we find an inequality in the opposite direction. 

\DEFOBS{thm:tide_ineq} The following inequality
\be
\tilde D_{\rm DPT, sep}(\varrho,\sigma)^2 \le \tilde D_{\rm DPT}(\varrho,\sigma)^2\label{eq:DPTsepDPT}
\ee
holds if $\varrho$  is pure. 

{\it Proof.}  The relation given in \EQ{eq:D2_pure_mixed} holds for the case when $\varrho=\ketbra{\Psi}$ is a pure state and $\sigma$ is an arbitrary density matrix.
Hence, the difference between the two quantities is just the difference of the two self-distances for $\sigma$ over two
\begin{align}
\begin{split}
&\tilde D_{\rm DPT}(\varrho,\sigma)^2-\tilde D_{\rm DPT, sep}(\varrho,\sigma)^2\\
&=
\frac 1 2 \left[\min_{\{p_k,\ket{\Phi_k}\}}\sum_{n=1}^N \sum_k p_k \va{H_n}_{\Phi_k}-\sum_{n=1}^N I_{\sigma}(H_n)\right]\\
&\ge\frac  1 2 \left[ \sum_{n=1}^N\frac1 4 F_Q[\sigma,H_n]-\sum_{n=1}^N I_{\sigma}(H_n)\right] \\
&\ge0,
\end{split}
\end{align}
where $\sigma$ is decomposed as in \EQ{eq:marginalcond_b}, for the first inequality we used \EQ{eq:selfd2}, which puts a lower bound on the self-distance with the quantum Fisher information. For the second inequality we used \EQ{eq:FQI}, which is an inequality between the quantum Fisher information over four and the Wigner-Yanase skew information. The self-distances are given in \EQS{eq:self_Dsep} and \eqref{eq:selfd_Ngreaterthan1}. Note that for $N=1,$ we have  
\begin{align}
\begin{split}
&\tilde D_{\rm DPT}(\varrho,\sigma)^2-\tilde D_{\rm DPT, sep}(\varrho,\sigma)^2\\
&\quad\quad=\frac 1 2 \left[\frac 1 4 F_Q[\sigma,H_1]-I_{\sigma}(H_1)\right]\\
&\quad\quad\ge0.
\end{split}
\end{align}
This finishes our proof. $\qed$

\begin{figure}[t!]
\begin{center}
\begin{subfigure}[b]{0.42\textwidth}  
        \centering
        \includegraphics[width=\textwidth]{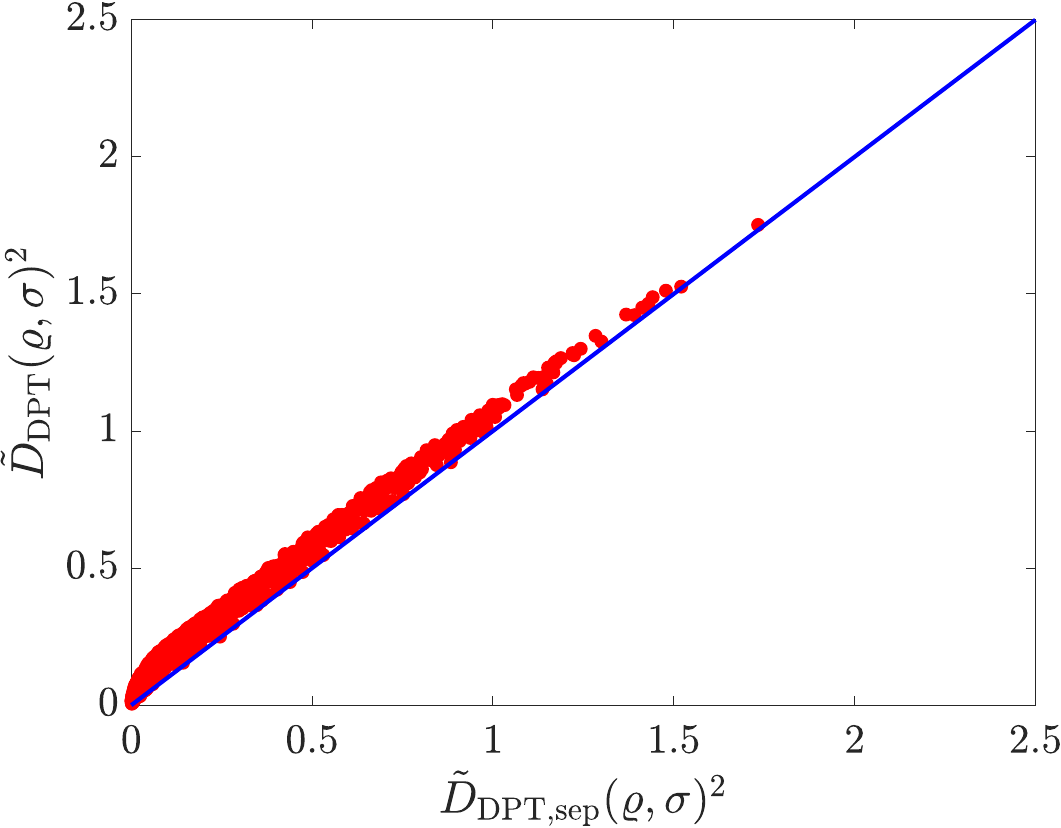}  
        \caption{}  
        \label{subfig:a}
    \end{subfigure}
    \vspace{0.5cm}
    
    \begin{subfigure}[b]{0.42\textwidth}  
        \centering
        \includegraphics[width=\textwidth]{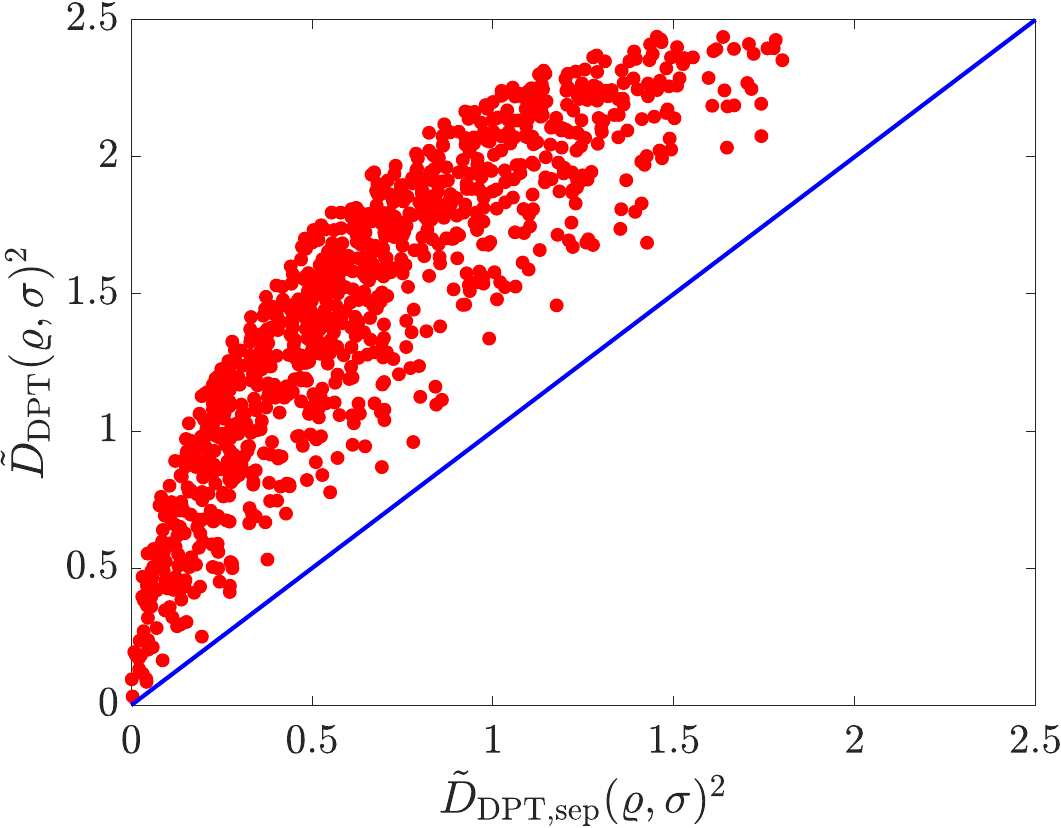}  
        \caption{}  
        \label{subfig:b}
          
    \end{subfigure}
\end{center}
\caption{ $\tilde D_{\rm DPT}(\varrho,\sigma)^2$ and $\tilde D_{\rm DPT, sep}(\varrho,\sigma)^2$  for random states for $d=2$ for (a) $H_1=\sigma_z$ and (b) $\{H_n\}_{n=1}^3=\{\sigma_x,\sigma_y,\sigma_z\}$.
The numerical results are consistent with the conjecture given in \EQ{eq:DPTsepDPTConj}.
} \label{fig:Fig_tilde_Dsep_tildeD}
\end{figure}

In \OBS{thm:tide_ineq} , the inequality in \EQ{eq:DPTsepDPT} has been proven if one of the states is pure. We also have numerical evidence for the case of two mixed states.
In \FIG{fig:Fig_tilde_Dsep_tildeD}, 1000 random state pairs are generated, and $\tilde D_{\rm DPT}(\varrho,\sigma)^2$ and $\tilde D_{\rm DPT, sep}(\varrho,\sigma)^2$ have been compared. Thus, we can make the following statement.

{\bf Conjecture 1.} The following inequality seems to hold
\begin{align}
 \tilde D_{\rm DPT, sep}(\varrho,\sigma)^2  \le \tilde D_{\rm DPT}(\varrho,\sigma)^2\label{eq:DPTsepDPTConj}
\end{align}
if $\varrho$ and $\sigma$ are arbitrary mixed states.

\subsubsection{The special case of a full set of $H_n$ operators}

So far we considered an arbitrary set of $H_n$ operators. This makes it possible to choose operators that are important for the problem considered. Moreover, two orthogonal states do not have necessarily a maximal distance. Next, we consider the opposite situation, when we have a full set of $(d^2-1)$ traceless pairwise orthogonal $H_n$ operator, i.~e., $SU(d)$ generators (see \REF{Vitagliano2011Spin}, and Observation 10 in \REF{Toth2023QuantumWasserstein}). The operators must satisfy
\be
{\rm Tr}(H_n H_{n'})=2\delta_{nn'}.\label{eq:orthog}
\ee
In this case, we have a  unitarily invariant distance and two orthogonal pure states do have a maximal distance.
It has been also shown that (Observation 10 in \REF{Toth2023QuantumWasserstein}).
\be
D_{\rm DPT, sep}(\varrho,\sigma)^2 \ge 2(d-1) \label{eq:DPTseplowerbound}
\ee
holds in this case.

For this choice of $H_n,$ we can state the following.

{\bf Lemma 1.} The Euclidean distance between pure states $\ket{\Psi}$ and $\ket{\Phi}$ in the space of $\ex{H_n}$ equals constant times $(1-|\langle \Psi | \Phi \rangle|^2),$ which equals also the trace distance, i. e., 
\begin{align}
&\frac 1 2 \sum_{n=1}^{d^2-1} (\ex{H_n}_{\Psi}-\ex{H_n}_{\Phi})^2=2(1-|\langle \Psi | \Phi \rangle|^2).\label{eq:HnHnS}
\end{align}
{\it Proof.}  For such operators, for pure states for the sum of expectation value squares (e.g., see \REFS{Vitagliano2011Spin,Toth2023QuantumWasserstein})
\be
\sum_{n=1}^{d^2-1}  \ex{H_n}^2=2(1-1/d)
\ee
holds, while for the sum of the tensor products we have 
\be
\sum_{n=1}^{d^2-1}  H_n \otimes H_n = 2(S-\openone/d).
\ee
Then, for the sum of the squares of the differences we have 
\begin{align}
&\frac 1 2 \sum_{n=1}^{d^2-1} (\ex{H_n}_{\Psi}-\ex{H_n}_{\Phi})^2\nonumber\\
&\quad\quad=\frac 1 2  \sum_{n=1}^{d^2-1} \ex{H_n}_{\Psi}^2+\ex{H_n}_{\Phi}^2-\sum_{n=1}^{d^2-1}  \ex{ H_n\otimes H_n}_{\Psi\otimes\Phi}\nonumber\\
&\quad\quad=2(1-\ex{S}_{\Psi\otimes\Phi})=2(1-|\langle \Psi | \Phi \rangle|^2),\label{eq:HnHnS1}
\end{align}
where $S$ is again the SWAP operator.  $\qed$
 
 Hence, we find the following.

\DEFOBS{thm:fullsetHn}If we consider a full set of $H_n$ operators, then the following two quantities are equal to each 
\begin{align}
\begin{split}
\tilde D_{\rm DPT, decomp}(\varrho,\sigma)^2&=2\left[1-F_{S,\rm sep}(\varrho,\sigma)\right]\\
&=4 D_{S,\rm sep}(\varrho,\sigma)^2.\label{eq:decomp_Ssep}
\end{split}
\end{align}

From \EQ{eq:decomp_Ssep} follows immediately that, for the case of a full set of $H_n$ operators, there is a simple tight upper bound on $\tilde D_{\rm DPT, decomp}(\varrho,\sigma)^2,$ i.e., 
\be
\tilde D_{\rm DPT, decomp}(\varrho,\sigma)^2 \le 2\label{eq:decomp_Ssep2}
\ee
holds, since $F_{S,\rm sep}(\varrho,\sigma)\ge 0.$

We will now show that the inequality in \EQ{eq:tildeDineq}  can be saturated. We will use that for all pure states \cite{Toth2023QuantumWasserstein}
\be
\sum_{n=1}^{d^2-1} (\Delta H_n)^2=2(d-1)\label{eq:sumnvarHn}
\ee
holds, hence the self-distance is  independent from the state and it is
\be
D_{\rm DPT, sep}(\varrho,\varrho)^2=2(d-1).\label{DPSsepselfdist2dm1}
\ee
Based on these, we find the following.

\DEFOBS{thm:fdecomp_sep} The inequality in \EQ{eq:tildeDineq} is saturated and thus
\be
\tilde D_{\rm DPT, decomp}(\varrho,\sigma)^2=\tilde D_{\rm DPT, sep}(\varrho,\sigma)^2\label{eq:tildeDineq2}
\ee
holds,  if we consider a full set of $H_n$ operators fulfilling \EQ{eq:orthog}. 

For the proof, see \APP{app:SAT}. We also present an example with $N=1,$ when the inequality is not saturated.

Due to \OBS{thm:fdecomp_sep}  and the definition of $\tilde D_{\rm DPT, decomp}(\varrho,\sigma)^2$ in \EQ{eq:GMPC_distance_sep_recursive2}, in the special case of having a full set of $H_n$ operators, 
\begin{align}
\begin{split}\label{eq:GMPC_distance_sep_recursive2_sep}
&\tilde D_{\rm DPT, sep}(\varrho,\sigma)^2\\
&\quad=\min_{\{p_k,\ket{\Psi_k},\ket{\Phi_k}\}}
\sum_k p_k \tilde{D}_{\rm DPT,sep}(\ket{\Psi_k},\ket{\Phi_k})^2
\end{split}
\end{align}
holds, while it is not true in general. Here we used that for pure states we have $D_{\rm DPT, sep}(\varrho,\sigma)^2=D_{\rm DPT}(\varrho,\sigma)^2.$ 

For the case of having a  full set of $(d^2-1)$ operators $H_n$ fulfilling \EQ{eq:orthog}, we can find an upper bound on $D_{\rm DPT, sep}(\varrho,\sigma)^2$ as
\be
D_{\rm DPT, sep}(\varrho,\sigma)^2\le 2d.
\ee
Here the upper bound is based on \EQS{eq:tildeDPTsep},  \eqref{eq:decomp_Ssep2}, \eqref{DPSsepselfdist2dm1}, and \eqref{eq:tildeDineq2}.

Due to \OBS{thm:swapfidelitysuperfidelity}, \OBS{thm:fullsetHn} and \OBS{thm:fdecomp_sep},  for the qubit case, if we choose $H_1=\sigma_x, H_2=\sigma_y, H_3=\sigma_z,$ we have
\be
\tilde D_{\rm DPT, sep}(\varrho,\sigma)^2=D_{\rm DPT, sep}(\varrho,\sigma)^2-2=2-2F(\varrho,\sigma).
\ee
Thus, in this case we know $\tilde D_{\rm DPT, sep}(\varrho,\sigma)^2$ with an explicit formula.
Here we used the self-distance given in \EQ{DPSsepselfdist2dm1} to express $\tilde D_{\rm DPT, sep}(\varrho,\sigma)^2$ with $D_{\rm DPT, sep}(\varrho,\sigma)^2.$

Let us again consider the case of having a  full set of $(d^2-1)$ operators $H_n$ fulfilling \EQ{eq:orthog}. Let us assume that $\varrho=\ketbra{\Psi}$ is pure. Then, we find that several quantites are equal to each other 
\begin{align}
\begin{split}
&\tilde D_{\rm DPT, decomp}(\ketbra{\Psi},\sigma)^2=\tilde D_{\rm DPT, sep}(\ketbra{\Psi},\sigma)^2\\
&\quad\quad=2[1-F_{\rm S, sep}(\ketbra{\Psi},\sigma)]=2[1-F(\ketbra{\Psi},\sigma)]\\
&\quad\quad=2[1-{\rm Tr}(\ketbra{\Psi}\sigma)]. \label{eq:decompF}
\end{split}
\end{align}
Here, the first equality is due to \EQ{eq:tildeDineq2}.
The second equality is due to \EQS{eq:tildeDineq2} and  \eqref{eq:decomp_Ssep}.
The third equality is due to \EQ{eq:FFFlele2}.
Finally, the last equality is due to a trivial property of the quantum fidelity.

\subsubsection{Triangle inequality}

For using the quantities based on an optimization over separable states defined above to measure how far a quantum state is from another one, it would be important to show that these quantities fulfill the triangle inequality. Let us now examine this in the case of $\tilde D_{\rm DPT, decomp}(\varrho,\sigma).$ We will take advantage of the fact that it is defined as an optimization of an expression given for two pure states and the optimization is taken over all possible separable decompositions of a bipartite state with given marginals.

\DEFOBS{thm:triangle}The inequality 
\begin{align}
\begin{split}
&\tilde D_{\rm DPT, decomp}(\varrho,\omega)+\tilde D_{\rm DPT, decomp}(\omega,\tau)\\
&\quad\quad\quad\quad\quad\quad\quad\quad\quad\quad-\tilde D_{\rm DPT, decomp}(\varrho,\tau)\ge 0,\label{eq:trianglesep}
\end{split}
\end{align}
where $\varrho, \omega$ and $\tau$ are density matrices, is fulfilled for the case of a pure $\omega.$

{\it Proof.} We know that $\tilde D_{\rm DPT}(\varrho,\omega)$ fulfills an analogous inequality for $\omega$ being pure or both $\varrho$ and $\tau$ being pure  \cite{Bunth2024MetricProperty}, or one of the three states being pure \cite{bunth2025wassersteindistancesdivergencesorder}. From this, it follows trivially that \EQ{eq:trianglesep} is true for pure states, since for pure states $\tilde D_{\rm DPT,decomp}(\varrho,\omega)$=$\tilde D_{\rm DPT}(\varrho,\omega)$ holds. However, the triangle inequality can also be proven easily for pure states directly.

Let us assume that $w=\ketbra{\Omega}$ is a pure state. Let us consider the optimal decompositions
\begin{subequations}
\begin{align}
&\tilde D_{\rm DPT, decomp}(\varrho,\omega)^2= \sum_k p_k \tilde{D}_{\rm DPT}(\ket{\Psi_k},\ket{\Omega})^2,\\
&\tilde D_{\rm DPT, decomp}(\omega,\tau)^2= \sum_k p_k \tilde{D}_{\rm DPT}(\ket{\Omega},\ket{\Upsilon_k})^2,\\
&\tilde D_{\rm DPT, decomp}(\varrho,\tau)^2= \sum_k p_k \tilde{D}_{\rm DPT}(\ket{\Psi_k'},\ket{\Upsilon_k'})^2,\label{eq:cond_c}
\end{align}
\end{subequations}
where
\begin{align}
\begin{split}
&\varrho=\sum_k p_k \ketbra{\Psi_k}=\sum_k p_k \ketbra{\Psi_k'},\\
&\tau=\sum_k p_k \ketbra{\Upsilon_k}=\sum_k p_k \ketbra{\Upsilon_k'}.\\
\end{split}\label{eq:cond_cc}
\end{align}
In these decompositions states with different indices are not necessarily different, that is why we can use common coefficients $p_k.$  Then, 
\begin{align}
\begin{split}
&[\tilde D_{\rm DPT, decomp}(\varrho,\omega)+\tilde D_{\rm DPT, decomp}(\omega,\tau)]^2\\
&\quad\quad=\sum_k p_k \bigg[\tilde{D}_{\rm DPT}(\ket{\Psi_k},\ket{\Omega})^2+\tilde{D}_{\rm DPT}(\ket{\Omega},\ket{\Upsilon_k})^2\bigg]\\
&\quad\quad+2\sqrt{\sum_k p_k \tilde{D}_{\rm DPT}(\ket{\Psi_k},\ket{\Omega})^2}\sqrt{\sum_k p_k \tilde{D}_{\rm DPT}(\ket{\Omega},\ket{\Upsilon_k})^2}\\
&\quad\quad\ge \sum_k p_k \bigg[\tilde{D}_{\rm DPT}(\ket{\Psi_k},\ket{\Omega})^2+\tilde{D}_{\rm DPT}(\ket{\Omega},\ket{\Upsilon_k})^2\\
&\quad\quad+2\tilde{D}_{\rm DPT}(\ket{\Psi_k},\ket{\Omega})\tilde{D}_{\rm DPT}(\ket{\Omega},\ket{\Upsilon_k})\bigg]\\
&\quad\quad=\sum_k p_k [\tilde{D}_{\rm DPT}(\ket{\Psi_k},\ket{\Omega})+\tilde{D}_{\rm DPT}(\ket{\Omega},\ket{\Upsilon_k})]^2 \\
&\quad\quad\ge\sum_k p_k \tilde{D}_{\rm DPT}(\ket{\Psi_k},\ket{\Upsilon_k})^2\\
&\quad\quad\ge\sum_k p_k \tilde{D}_{\rm DPT}(\ket{\Psi_k'},\ket{\Upsilon_k'})^2=\tilde D_{\rm DPT, decomp}(\varrho,\tau)\label{eq:PsiOmegaUpsilon}^2.
\end{split}
\end{align}
In the first inequality, we used that for any $x_k$ and $y_k$ non-negative real numbers and $p_k$ probabilities we have the Cauchy-Schwarz inequality 
\begin{align}\begin{split}&\sum_k p_kx_ky_k=\sum_k (\sqrt{p_k} x_k)(\sqrt{p_k} y_k)\\&\quad\quad\quad\quad\quad\quad\le\sqrt{\sum_k p_k x_k^2}\sqrt{\sum_k p_k y_k^2}.\end{split}\label{eq:xyineq}\end{align}
In the second inequality we used that the triangle inequality is valid for pure states. In the third inequality, we used that the optimal decomposition for $\tilde D_{\rm DPT, decomp}(\varrho,\tau)^2$ is given by  \EQ{eq:cond_c}. $\qed$

\subsection{Summary of the methods based on an optimization over separable states presented in \SEC{sec:relations}}
 
After considering several possibilities of defining the quantum Wasserstein distance based on an optimization over separable states, let us summarize our findings. In many of the examples in \SEC{sec:relations}, we have one of the two possibilities. We had the square of the De Palma-Trevisan distance, $D_{\rm DPT}(\varrho,\sigma)^2$  optimized over separable states 
\begin{align}
\min_{\{p_k,\ket{\Psi_k},\ket{\Phi_k}\}}\;&
\sum_k p_k D_{\rm DPT}(\ket{\Psi_k},\ket{\Phi_k})^2,
\end{align}
where $D_{\rm DPT}(\ket{\Psi_k},\ket{\Phi_k})^2=\Delta(\ket{\Psi_k},\ket{\Phi_k})^2$ and $\Delta$ is given in \EQ{eq:Delta}.
The other possibility was the square of the {\it modified} De Palma-Trevisan distance, $\tilde D_{\rm DPT}(\varrho,\sigma)^2$ optimized over separable states 
\begin{align}
\min_{\{p_k,\ket{\Psi_k},\ket{\Phi_k}\}}\;&
\sum_k p_k \tilde{D}_{\rm DPT}(\ket{\Psi_k},\ket{\Phi_k})^2,
\end{align}
where $\tilde D_{\rm DPT}(\ket{\Psi_k},\ket{\Phi_k})^2=E(\ket{\Psi_k},\ket{\Phi_k})^2$ and $E$ is given in \EQ{eq:Erho}.
We assumed that that the conditions for the marginals given in \EQ{eq:marginalcond} are fulfilled in both cases. In the first case, we have a quantity that might have a nonzero self-distance. In the second case, the quantify given by the optimization does not have a self-distance.

\newlength{\mywidtha}
\settowidth{\mywidtha}{13}
\newlength{\mywidthb}
\settowidth{\mywidthb}{13.6421}
\newcommand{\minuswithspace}{\makebox[\mywidthb][c]{-}}

{\renewcommand{\arraystretch}{1.25}

\begin{table*}[t!]
\begin{tabular}{|l|l|}
\hline
\text{Notation}&\text{Name and equation number}\\
\hline
$F_{S}(\varrho,\sigma)$                                                & SWAP-fidelity, \EQ{eq:FS}\\
 $F(\varrho,\sigma)$                                                 & Uhlmann-Jozsa fidelity, \EQ{Fidelity_def} and \EQ{eq:Fidelity}\\

$F_{S,{\rm sep}}(\varrho,\sigma)$                                                & SWAP-fidelity based on an optimization \\ & over separable states, \EQ{eq:Fssep_S} and \EQ{eq:Fssep}\\
$F_{\rm super}(\varrho,\sigma)$                                                 & Superfidelity, \EQ{eq:Fsuper}\\
\hline
\end{tabular}
\caption{Fidelities in this article}
\label{tab:quantities_in_the_paper}
\end{table*}

{\renewcommand{\arraystretch}{1.25}

\begin{table*}[t!]
\begin{tabular}{|l|l|}
\hline
\text{Notation}&\text{Name, Section and equation number}\\
\hline
$D_{\rm DPT}(\varrho,\sigma)^2$                                                  & De Palma-Trevisan Wasserstein distance, \SEC{sec:De Palma-Trevisan-type quantum Wasserstein distance}, \EQ{eq:DPT}\\
$D_{\rm GMPC}(\varrho,\sigma)^2$                                                   & Golse-Mouhot-Paul-Caglioti Wasserstein distance, \SEC{sec:De Palma-Trevisan-type quantum Wasserstein distance}, \EQ{eq:GMPC_distance}\\
$D_S(\varrho,\sigma)^2=[1-F_S(\varrho,\sigma)]/2$ & Distance based on the SWAP-fidelity, \SEC{sec:SWAP-fidelity and the corresponding quantum Wasserstein distance}, \EQ{eq:DSFS} and \EQ{eq:DSFS2} \\
$\tilde D_{\rm DPT}(\varrho,\sigma)^2$                                                  &Modified  De Palma-Trevisan Wasserstein distance, \SEC{sec:Modified De Palma-Trevisan-type quantum Wasserstein distance}, \EQ{eq:Dmod}\\
\hline
$D_{\rm DPT, sep}(\varrho,\sigma)^2=D_{\rm GMPC, sep}(\varrho,\sigma)^2$                                                 & De Palma-Trevisan Wasserstein distance\\
& based on an optimization over separable states, \SEC{sec:sep_De Palma-Trevisan-type quantum Wasserstein distance}, \EQ{eq:GMPC_distance_sep} \\
$D_{S,{\rm sep}}(\varrho,\sigma)^2=[1-F_{S,{\rm sep}}(\varrho,\sigma)]/2$ & Distance based on the SWAP-fidelity and an optimization \\& over  separable states, \SEC{sec:sep_SWAP-fidelity and the corresponding quantum Wasserstein distance}, \EQ{eq:Dssep} and  \EQ{eq:Dssep2}\\
$\tilde D_{\rm DPT, sep}(\varrho,\sigma)^2$                                                 & Modified De Palma-Trevisan Wasserstein distance\\
& based on an optimization over separable states, \SEC{sec:sep_Modified De Palma-Trevisan-type quantum Wasserstein distance},  \EQ{eq:tildeDPTsep}\\
 $\tilde D_{\rm DPT, decomp}(\varrho,\sigma)^2$                                                    & Alternative generalization for separable states, \SEC{sec:sep_Modified De Palma-Trevisan-type quantum Wasserstein distance}, \EQ{eq:GMPC_distance_sep_recursive2}\\
\hline
$D_{{\rm BSF},p}(\varrho,\sigma)^2$                                                    & Beatty-Stilck Fran\c{c}a Wasserstein distance, \\& \SEC{sec:relations_pW}, \EQ{eq:tracedist} and \EQ{eq:tracedist_p}\\
\hline
\end{tabular}
\caption{Various types of quantum Wasserstein distance in this article}
\label{tab:quantities_in_the_paper2}
\end{table*}

{\renewcommand{\arraystretch}{1.25}
\begin{table*}[t!]
\begin{tabular}{|l|l|}
\hline
\text{Notation}&\text{Equation number}\\
\hline
$D_{\rm DPT, sep}(\varrho,\sigma)^2\ge D_{\rm DPT}(\varrho,\sigma)^2$&\EQ{eq:sep_notsep_ineq}\\
$F(\varrho,\sigma)\le F_{S,{\rm sep}}(\varrho,\sigma)\le F_{S}(\varrho,\sigma)$&\EQ{eq:FFFlele}\\
$F_{S,{\rm sep}}(\varrho,\sigma)\le F_{\rm super}(\varrho,\sigma)$&\EQ{eq:Fsuperbound}\\
$\tilde D_{\rm DPT, decomp}(\varrho,\sigma)^2\ge
 E(\varrho,\sigma)^2, $ & \EQ{eq:decomp}\\
$\tilde D_{\rm DPT, decomp}(\varrho,\sigma)^2\le \tilde D_{\rm DPT, sep}(\varrho,\sigma)^2$&\EQ{eq:tildeDineq}\\
$\tilde D_{\rm DPT, sep}(\varrho,\sigma)^2 \le \tilde D_{\rm DPT}(\varrho,\sigma)^2$&\EQ{eq:DPTsepDPT}, pure $\varrho$\\
$D_{\rm BSF,2}(\varrho,\sigma)^2=1-F_{S,\rm sep}(\varrho,\sigma)$&\EQ{eq:DFS}\\
$D_{\rm DPT,sep}(\varrho,\sigma)^2-2(d-1)=\tilde D_{\rm DPT,sep}(\varrho,\sigma)^2$&\EQ{eq:manyeq},  \\
$\quad\quad=\tilde D_{\rm DPT,decomp}(\varrho,\sigma)^2=4D_{\rm S,sep}(\varrho,\sigma)^2$&for a full set of $H_n$ \\
$\quad\quad=2(1-F_{\rm S,sep}(\varrho,\sigma)^2)=2D_{\rm BSF,2}(\varrho,\sigma)^2$&\\
\hline
\end{tabular}
\caption{Some of the most relevant relations from this publication}
\label{tab:quantities_in_the_paper3}
\end{table*}

\section{Quantum $p$-Wasserstein distance based on an optimization over separable states}
\label{sec:relations_pW}

In this section, we will examine quantum $p$-SWAP-fidelity and quantum $p$-Wasserstein distance definitions based on an optimization over separable states.

Beatty and Stilck Fran\c{c}a considered  the optimization of several distances for pure states, and defined quantum $p$-Wasserstein distances with them \cite{Beatty2025Order}.  In particular, they considered for $p=2$ with the trace distance
\begin{align}\label{eq:tracedist}
\begin{split}
&D_{\rm BSF,2}(\varrho,\sigma)^2\\
&\quad\quad=\min_{\{p_k,\ket{\Psi_k},\ket{\Phi_k}\}}\;\sum_k p_k  (1-|\langle \Psi_k | \Phi_k\rangle|^2),
\end{split}
\end{align}
such that the conditions for the marginals given in \EQ{eq:marginalcond} are fulfilled. They also defined a quantum $p$-Wasserstein distance as 
\begin{align}
\begin{split}\label{eq:tracedist_p}
&D_{{\rm BSF},p}(\varrho,\sigma)^2\\
&\quad\quad=\left[\min_{\{p_k,\ket{\Psi_k},\ket{\Phi_k}\}}\;\sum_k p_k  (1-|\langle \Psi_k | \Phi_k\rangle|^2)^{\frac p 2}\right]^{\frac 2 p},
\end{split}
\end{align}
such that for the marginals  \EQ{eq:marginalcond} are fulfilled. An advantage of the definition is that we can define an expression for general mixed states using a definition given for pure states. Moreover,  for pure states the distance gives back the underlying metric, which is the trace distance.  For a given $p,$ we have just the $p^{th}$ power of the trace distance \cite{Beatty2025Order}. 

Hence, we can see the relation to the quantities we have considered before
\be
D_{\rm BSF,2}(\varrho,\sigma)^2=1-F_{S,\rm sep}(\varrho,\sigma)=2D_{S,\rm sep}(\varrho,\sigma)^2.\label{eq:DFS}
\ee
These ideas can be used to calculate $D_{\rm BSF,2}^2(\varrho,\sigma)$ numerically based on the methods explained in \REFS{Toth2015Evaluating,Toth2023QuantumWasserstein}.

We now arrive at an important theorem.

\DEFOBS{thm:bsf}If we have a full set of $(d^2-1)$ operators $H_n$ fulfilling \EQ{eq:orthog} then 
\begin{align}
\begin{split}
2D_{\rm BSF,2}(\varrho,\sigma)^2&=\tilde D_{\rm DPT, sep}(\varrho,\sigma)^2\\
&=D_{\rm DPT, sep}(\varrho,\sigma)^2-2(d-1).\label{eq:BSF2tildeDPTsep}
\end{split}
\end{align}
holds. 

{\it Proof.} Here, the first equality is from \OBS{thm:fdecomp_sep},  \EQS{eq:decomp_Ssep}, and \eqref{eq:DFS}.
The second equality is from \EQS{eq:tildeDPTsep} and \eqref{DPSsepselfdist2dm1}.$\qed$

Thus, the definitions of $D_{\rm BSF,2}(\varrho,\sigma)^2$ given in \REF{Beatty2025Order} and $D_{\rm DPT, sep}(\varrho,\sigma)^2$ given in \REF{Toth2023QuantumWasserstein} are unexpectedly linear functions of each other.

We can summarize some relevant relations bounding $D_{\rm BSF,2}(\varrho,\sigma)^2$  as 
\begin{align}
\begin{split}
&D_{\rm Bures}(\varrho,\sigma)^2/2=1-\sqrt{F(\varrho,\sigma)}\\
&\quad \quad\quad\quad \le 2D_{S}(\varrho,\sigma)^2= 1-F_S(\varrho,\sigma)\\
&\quad \quad\quad\quad \le D_{\rm BSF,2}(\varrho,\sigma)^2=2D_{S,\rm sep}(\varrho,\sigma)^2\\
&\quad\quad\quad\quad  \le 1-F(\varrho,\sigma).\label{eq:ineqs}
\end{split}
\end{align}
The first equality is from \EQ{eq:Bures}.
The first inequality and the second equality are from  \EQS{eq:FFF} and \eqref{eq:DSFS}.
The second inequality together with the third equality is from  \EQS{eq:FFFlele} and \eqref{eq:DFS}.
The last inequality is also due to \EQS{eq:FFFlele} and \eqref{eq:DFS}.
Note that for qubits the last inequality is saturated due to \OBS{thm:swapfidelitysuperfidelity}. Thus, for qubits
\be
D_{\rm BSF,2}(\varrho,\sigma)^2=1-F(\varrho,\sigma)\label{eq:ineqs2}
\ee
holds. The last inequality in \EQ{eq:ineqs} is also saturated and \EQ{eq:ineqs2} holds, if one of the two states is pure due to \EQ{eq:decompF}. Finally, due to \OBS{thm:swapfidelitysuperfidelity} and \EQ{eq:DFS}, we have
\be
1-F_{\rm super}(\varrho,\sigma)\le D_{\rm BSF,2}(\varrho,\sigma)^2.
\ee

Next we will define an optimization of the powers of the overlap knowing that an optimization of some power of an expectation value over a quantum state can be expressed an an optimization over a separable state with a given marginal \cite{Toth2015Evaluating}. Let us define the $p$-SWAP-fidelity based on an optimization over separable states as
\be
F^{(p)}_{S,{\rm sep}}(\varrho,\sigma)=\max_{\{p_k,\Phi_k,\Psi_k\}}\sum_k p_k | \bra{\Phi_k} \Psi_k\rangle |^{2p},
\ee
which can also be rewritten as 
\be
F^{(p)}_{S,{\rm sep}}(\varrho,\sigma)=\max_{\varrho_{123...2p}\in \mathrm{Sep}_p} {\rm Tr}(\varrho_{123..2p} S_p),\label{eq:Fp}
\ee 
where $\mathrm{Sep}_p$ is the set of $2p$-particle separable states, ${\rm Tr}_{I\backslash\{n\}}(\varrho_{123...2p})=\varrho$ if $n\le p$ and ${\rm Tr}_{I\backslash\{n\}}(\varrho_{123...2p})=\sigma$  if $n\ge p+1.$ Here the set of all indices is denoted by $I=\{1,2,..,2p\}.$ The SWAP operator $S_p$ acts between the particle groups $1,2,..,p$ and $p+1,p+2,...,2p.$ For $p\ge 2,$ we require that the separable state is symmetric within the $1,2,...,p$ particle unit and also  within the $p+1,...,2p$ unit, i.~e.,
\begin{align}
\begin{split}
{\rm Tr}[(P_{\rm sym} \otimes \openone)\varrho_{123...2p})]&=1,\\
{\rm Tr}[( \openone\otimes P_{\rm sym})\varrho_{123...2p})]&=1,
\end{split}
\end{align}
where $P_{\rm sym}$ is the projector to the symmetric subspace of the quantum states of $p$ particles, and  $\openone$ acts also on $p$ particles. It is instructive to write down the $\varrho_{123...2p}$ consistent with these conditions as
\be
\sum_k p_k (\ketbra{\Psi_k})^{\otimes p} \otimes (\ketbra{\Phi_k})^{\otimes p}.\label{eq:optmimumsym}
\ee

The value of the maximum will remain the same, even if the conditions with the projector to the symmetric subspace are neglected (see Observations 1 and 4 in \REF{Toth2023QuantumWasserstein} for a similar problem), see also \APP{app:symopt}.

The quantum $p$-Wasserstein distance can be calculated based on similar ideas.  We find that 
\be
D_{{\rm BSF},p}(\varrho,\sigma)^2=\left[\min_{\varrho_{123...2p}\in \mathrm{Sep}_p} {\rm Tr}\left(\varrho_{123..2p} O\right)\right]^{\frac 2 p},
\ee
where the operator for the optimization is 
\be
O=\otimes_{k=1}^p (\openone-S^{(k,k+p)}),
\ee
and the conditions for the optimization are the same. Here, $S^{(k,l)}$ is the SWAP operator acting between particle $k$ and $l.$ For pure states, we have
\begin{align}
\begin{split}
&{\rm Tr}(\ketbra{\Phi}^{\otimes p} \ketbra{\Psi}^{\otimes p} O)\\
&\quad\quad\quad\quad\quad\propto\left[\sum_{n=1}^{d^2-1} (\ex{H_n}_{\Psi}-\ex{H_n}_{\Phi})^{2}\right]^{\frac p 2},\label{eq:HHH1}
\end{split}
\end{align}
where $O$ acts on a $2^p \times 2^p$ bipartite state, and $H_n$ is a full set of $(d^2-1)$ operators fulfilling \EQ{eq:orthog} mentioned before.
These quantities can be calculated numerically based on the methods explained in \REFS{Toth2015Evaluating,Toth2023QuantumWasserstein}, e.~g., by approximating the separable states using the separability criterion based on symmetric extensions \cite{Doherty2002Distinguishing, Doherty2004Complete,Doherty2005Detecting}. 
We add that quantum $p$-Wasserstein distances have been considered for the case of optimization over all quantum states, rather than
separable states \cite{bunth2025wassersteindistancesdivergencesorder}.

Let us summarize our findings for the case of if we consider a full set of $H_n$ matrices. Then, several of the quantities considered in this paper are equal to each other
\begin{align}
\begin{split}
&D_{\rm DPT,sep}(\varrho,\sigma)^2-2(d-1)=\tilde D_{\rm DPT,sep}(\varrho,\sigma)^2\\
&\quad\quad=\tilde D_{\rm DPT,decomp}(\varrho,\sigma)^2=4D_{\rm S,sep}(\varrho,\sigma)^2\\
&\quad\quad=2(1-F_{\rm S,sep}(\varrho,\sigma)^2)=2D_{\rm BSF,2}(\varrho,\sigma)^2.\label{eq:manyeq}
\end{split}
\end{align}
Here, the first equality is from \EQS{eq:tildeDPTsep} and \eqref{DPSsepselfdist2dm1}. The second equality is due to \OBS{thm:fdecomp_sep} .
The third and fourth equalities are due to \EQ{eq:decomp_Ssep}.
The last equality is due to \EQ{eq:DFS}.
Thus, in \OBS{thm:triangle} we proved the 
triangle inequality also for all the quantities above.

The main quantities and relations discussed in this article are given in Tables~\ref{tab:quantities_in_the_paper}, \ref{tab:quantities_in_the_paper2}, and \ref{tab:quantities_in_the_paper3}.

\section{Further directions}


{\renewcommand{\arraystretch}{1.25}

\begin{table*}[t!]
\begin{tabular}{|l|l|}
\hline
\text{Notation}&\text{Name and equation number}\\
\hline
$D_{\rm DPT, sep}(\varrho,\sigma)^2$&De Palma-Trevisan Wasserstein distance square defined in \REF{Toth2023QuantumWasserstein}\\
& based on an optimization over separable states, \SEC{sec:sep_De Palma-Trevisan-type quantum Wasserstein distance}, \EQ{eq:GMPC_distance_sep_recursive}\\
 $\sqrt F(\varrho,\sigma)$                                                 & Square root of the Uhlmann-Jozsa fidelity,   \SEC{sec:sep_SWAP-fidelity and the corresponding quantum Wasserstein distance}, \EQ{eq:sqrtFoptim}\\
$D_{\rm Bures}(\varrho,\sigma)^2$ &Square of the Bures distance,  \SEC{sec:sep_SWAP-fidelity and the corresponding quantum Wasserstein distance}, \EQ{eq:Brues_opt} \\
$D_{S,{\rm sep}}(\varrho,\sigma)^2=[1-F_{S,{\rm sep}}(\varrho,\sigma)]/2$ & Square of the distance based on the SWAP-fidelity and an optimization \\& over  separable states, \SEC{sec:sep_SWAP-fidelity and the corresponding quantum Wasserstein distance}, \EQ{eq:Dssep2}\\
 $\tilde D_{\rm DPT, decomp}(\varrho,\sigma)^2$                                                    & Alternative generalization of the De Palma-Trevisan Wasserstein distance square\\
 & for an optimization over separable states, \SEC{sec:sep_Modified De Palma-Trevisan-type quantum Wasserstein distance}, \EQ{eq:GMPC_distance_sep_recursive2}\\
$D_{{\rm BSF},p}(\varrho,\sigma)^2$                                                    & Square of the Beatty-Stilck Fran\c{c}a Wasserstein distance defined in \REF{Beatty2025Order}, \\& \SEC{sec:relations_pW}, \EQ{eq:tracedist} and \EQ{eq:tracedist_p}\\
\hline
\end{tabular}
\caption{Quantities defined as an optimization over separable decompositions of a function given for pure states. If the optimization is a minimization, then our quantity is jointly convex. If it is a maximization, then it is jointly concave.}
\label{tab:quantities_in_the_paper_opt}
\end{table*}

There are quantities defined as a function of a density matrix that can be written as a convex roof, such as the entanglement of formation \cite{Hill1997Entanglement,Wootters1998Entanglement}, and the quantum Fisher information \cite{Toth2013Extremal,Yu2013Quantum}. Clearly, a quantity defined as a convex roof is a convex function of the state. Any convex function is bounded from above by its convex roof. For this reason, the 
Wigner-Yanase skew information defined in is smaller than the quantum Fisher information over four as given in \EQ{eq:FQI} \cite{Toth2013Extremal}. 
The variance can be given as a concave roof  of itself \cite{Toth2013Extremal}.  Any concave function  is bounded from below by its concave roof. 

In this paper we considered quantities that are the function of two density matrices.
We obtained several quantities that are a convex roof of an expression with pure state components, see Table~\ref{tab:quantities_in_the_paper_opt}.
Even if  a jointly convex function $D(\varrho,\sigma)^2$ cannot be written as its own convex roof,  we can always obtain an upper bound on the function
using its convex roof as 
\be
D(\varrho,\sigma)^2 \le \min_{\{p_k,\Psi_k,\Phi_k\}} \sum_k p_k D(\ket{\Phi_k},\ket{\Psi_k})^2.\label{eq:Brues_opt2_bound}
\ee

We now consider a number of relevant quantities in quantum information science. They are not distances in the sense that they do not fulfill all the required properties of a distance.
The Jensen divergence is defined as \cite{VIROSZTEK201967}
 \be
 J_{f,\lambda}(A,B)=(1 - \lambda){\rm Tr} f(A)+ \lambda {\rm Tr} f(B) -{\rm Tr} f ((1 - \lambda)A+ \lambda B) .
\ee
It is jointly convex if $ f$ is a convex function from the matrix entropy class \cite{VIROSZTEK201967}. Then, we obtain the bounds
\bea
 &&J_{f,\lambda}(\varrho,\sigma) \nonumber\\
 &&\quad\quad\le \min_{p_k,\Psi_k,\Phi_k} \sum_k p_k \bigg[f(1)+f(0)(d-1) \nonumber\\
 &&\quad\quad-{\rm Tr} f ((1 - \lambda)\ketbra{\Psi_k}+ \lambda \ketbra{\Phi_k})\bigg]\nonumber\\
  &&\quad\quad\le f(0)+ f(1)-f(\lambda)-f(1-\lambda).
\eea
The upper bound comes from assuming that $\ket{\Psi_k}$ and $\ket{\Phi_k}$ are pairwise orthogonal.
The quantum Jensen-Shannon divergence is a special case \cite{VIROSZTEK2021107595}.
Then, we have $f(0)=f(1)=0$ and $f(x)= x \log(x).$

The quantum Bregman divergence is defined as 
\be
H_f (X, Y )={\rm Tr} [f (X)- f (Y )-f'(Y )(X -Y )].
\ee
It is jointly convex if $ f$ is a convex function from the matrix entropy class \cite{Pitrik2015JointConvexityBregmanDivergence}.
For this case, we obtain
\bea
H_{f} (\varrho, \sigma) &&\le (d-1)f(0) + [f'(1)-f'(0)]\nonumber\\
&&- \max_{\{p_k,\Psi_k,\Phi_k\}}  \sum_k p_k\langle \Psi_k| f'(\ketbra{\Phi_k})   |\Psi_k\rangle\nonumber\\
&&= (d-1)f(0)\nonumber\\
&&+ [ f'(1)-f'(0) ] \min_{\{p_k,\Psi_k,\Phi_k\}}  \sum_k p_k (1-|\langle \Psi_k | \Phi_k\rangle |^2).\nonumber\\
\eea
Then, a lower bound can be obtained as 
\begin{align}
\begin{split}
&\max_{f(x)}\frac{H_{f} (\varrho, \sigma) - (d-1)f(0) }{[ f'(1)-f'(0) ] }\\
&\quad\quad\le  1-F_{S,\rm sep}(\varrho,\sigma)=2D_{S,\rm sep}(\varrho,\sigma)^2\\
&\quad\quad=D_{\rm BSF,2}(\varrho,\sigma)^2 .
\end{split}
\end{align}
Here the optimization is carried over all possible allowed $f(x)$ functions.

Let us consdier some concrete $f(x)$ functions.
Here, $f(x)=x^2$ is for the Hilbert-Schmidt distance.
The Tsallis divergence is another special case of the Bregman divergence given  as 
\be
H_{f_q} (A, B)={\rm Tr}(B^q) + \frac{1}{q -1} [{\rm Tr}(A^q) -q{\rm Tr}(AB^{q-1})]
\ee
for $q\ne1.$ Here $f_q(x)=x\ln_q(x)$ where $\ln_q(x)$ is the $q$-deformed logarithm. Then, we obtain the upper bound as 
\be
H_{f_q} (\varrho, \sigma) \le \frac{q}{q-1} \min_{\{p_k,\Psi_k,\Phi_k\}}  \sum_k p_k (1- |\langle \Psi_k|\Phi_k\rangle|^2).
\ee
$H_{f_q} (\varrho, \sigma)$ is jointly convex if $1< q\le 2$ \cite{Pitrik2015JointConvexityBregmanDivergence}.
It is possible to obtain a lower bound as
\begin{align}
\begin{split}
\max_{1<q\le2}\frac{q-1}{q} H_{f_q} &\le  1-F_{S,\rm sep}(\varrho,\sigma)=2D_{S,\rm sep}(\varrho,\sigma)^2.\\
&=D_{\rm BSF,2}(\varrho,\sigma)^2.
\end{split}
\end{align}

It is an intriguing question under what conditions these inequalities can be saturated.

\section*{Conclusions}
We studied several possible definitions of the quantum Wasserstein distance. We considered some that are based on an optimization over general bipartite states. Then, we considered reducing the optimization to separable states, and determined the corresponding quantities. We studied the relations between these quantities. We proved a triangle inequality for some of these quantities for the case when of the states is pure. The details of the numerical methods we have been using are in \APP{app:Numerics}. 

\acknowledgments

We thank I.~Apellaniz, T. ~Borsoni, M.~Eckstein, F. Fr\"owis, I.~L.~Egusquiza, C.~Klempt, J.~Ko\l ody\'nski, M.~W.~Mitchell, M.~Mosonyi, G.~Muga, J.~Siewert, Sz.~Szalay, K. \.Zyczkowski, T.~V\'ertesi, G. Vitagliano, and D. Virosztek  for discussions. We acknowledge the support of the  EU (QuantERA MENTA, QuantERA QuSiED, COST Action CA23115),
the Spanish MCIU (Grant No.~PCI2022-132947), the Basque Government (Grant No. IT1470-22), and the National Research, Development and Innovation Office of Hungary (NKFIH) (Grant No. 2019-2.1.7-ERA-NET-2021-00036, Advanced Grant No. 152794). We thank the National Research, Development and Innovation Office of Hungary (NKFIH) within the Quantum Information National Laboratory of Hungary.   We acknowledge the support of the Grant No.~PID2021-126273NB-I00 funded by MCIN/AEI/10.13039/501100011033 and by ``ERDF A way of making Europe''.  We thank the ``Frontline'' Research Excellence Programme of the NKFIH (Grant No. KKP133827). We thank Project no. TKP2021-NVA-04, which has been implemented with the support provided by the Ministry of Innovation and Technology of Hungary from the National Research, Development and Innovation Fund, financed under the TKP2021-NVA funding scheme. G.~T. is thankful for a  Bessel Research Award from the Humboldt Foundation.

\appendix

\section{Proof of \OBS{thm:fdecomp_sep} }
\label{app:SAT}

We show that the inequality in \EQ{eq:tildeDineq} is saturated if we consider a full set of $(d^2-1)$ operators $H_n$ fulfilling \EQ{eq:orthog}.

Let us look at the inequality given in \EQ{eq:tildeDineq2b}. In general, the right-hand side can be smaller than the left-hand side since on the left-hand side we minimize the sum of several terms, while 
on the right-hand side we compute the minimum of the these terms independently and sum them.
Now, we need to use that for all pure states the equality in \EQ{eq:sumnvarHn} holds.
Hence for all $\{p_k,\Psi_k\}$ decompositions we have 
\be
\sum_k p_k  \sum_{n=1}^{d^2-1}\va{H_n}_{\Psi_k}=2(d-1).
\ee
Similarly, for all $\{p_k,\Phi_k\}$ decompositions we have 
\be
\sum_k p_k  \sum_{n=1}^{d^2-1}\va{H_n}_{\Phi_k}=2(d-1).
\ee
Hence, for the self-distance [c.~f., \EQ{DPSsepselfdist2dm1}]
\be
D_{\rm DPT, sep}(\varrho,\varrho)^2=D_{\rm DPT, sep}(\sigma,\sigma)^2=2(d-1)
\ee
holds.
Thus, the self-distances are constant independent from the quantum state, and in order to obtain them, we do not even need a minimization.
The decomposition $\{p_k,\Psi_k,\Phi_k\}$ that minimizes 
\be
\min_{\{p_k,\ket{\Psi_k},\ket{\Phi_k}\}}\;(\ex{H_n}_{\Psi_k}-\ex{H_n}_{\Phi_k})^2,
\ee
will saturate the inequality in \EQ{eq:tildeDineq2b} and consequently, also the inequality in \EQ{eq:tildeDineq}. 

We show a simple example with $N=1$ where the two sides in \EQ{eq:tildeDineq} are not equal to each other. For $N=1$,  \EQ{eq:tildeDineq}  can be rewritten as \EQ{eq:tildeDineq3}. Let us now consider $H_1=\sigma_z$ and
\begin{align}
\begin{split}
\varrho&=\openone/2,\\
\sigma&=\ketbra{0}.
\end{split}
\end{align}
For these, for the quantum Fisher information
\be
F_Q[\varrho,H_1]=F_Q[\sigma,H_1]=0
\ee
hold, thus the self distances are zero.
Then, based on \EQS{eq:GMPC_distance_sep}   and \eqref{eq:GMPC_distance_sep_recursive2}, we have 
\begin{subequations}
\begin{align}
D_{\rm DPT, sep}(\varrho,\sigma)^2=\tilde D_{\rm DPT, sep}(\varrho,\sigma)^2&=1 \label{eq:eq1},\\
\tilde D_{\rm DPT, decomp}(\varrho,\sigma)^2&=1/2.\label{eq:eq2}
\end{align}
\end{subequations}
An optimal coupling for \EQ{eq:eq1} is
\be
\varrho\otimes\sigma=\frac{1}{2}(\ketbra{0}\otimes\ketbra{0}+\ketbra{1}\otimes\ketbra{0}).
\ee
For calculating \EQ{eq:eq2}, we need the decomposition
\be
\varrho\otimes\sigma=\frac{1}{2}(\ketbra{0}_x\otimes\ketbra{0}+\ketbra{1}_x\otimes\ketbra{0}),
\ee
where $\ket{.}_x$ refers to the $\sigma_x$ basis.

\section{Symmetry properties of the optimal state}
\label{app:symopt}

In this Appendix, we prove that there is an optimal state for the maximization in \EQ{eq:Fp} that is of the form \EQ{eq:optmimumsym}.

From the inequality between the arithmetic and geometric mean follows that
\be
\frac 1 p \sum_{n=1}^p {\rm Tr}(A_n) \ge {\rm Tr}\left(\bigotimes_{n=1}^p A_n\right)^{\frac 1 p}, 
\ee
where $A_n$ are positive-semidefinite Hermitian matrices, which can be rewritten as 
\be
{\rm Tr}\left(\overline{A}^{\otimes p}\right)\ge {\rm Tr}\left(A_1\otimes A_2 \otimes ... \otimes A_p\right),\label{eq:geomarit}
\ee
where the average is defined as
\be
\overline{A}=\frac 1 p \sum_{n=1}^p A_n.
\ee

Let us consider two positive-semidefinite Hermitian matrices $A$ and $B.$ Substituting $B_n^{\frac 1 2} A_n B_n^{\frac 1 2} $ into $A_n$ in \EQ{eq:geomarit}, we arrive at 
\begin{align}
\begin{split}
&{\rm Tr}\left[\left(\frac 1 p \sum_{n=1}^p B_n^{\frac 1 2} A_n B_n^{\frac 1 2} \right)^{\otimes p}\right]\nonumber\\
&\quad\ge{\rm Tr}\left[(B_1^{\frac 1 2} A_1 B_1^{\frac 1 2})\otimes (B_2^{\frac 1 2} A_2 B_2{\frac 1 2}) \otimes ... \otimes (B_p^{\frac 1 2} A_p B_p^{\frac 1 2})\right]
\end{split}
\end{align}
Hence, based on straightforward algebra using the properties of trace follows  that 
\begin{align}
\begin{split}
&{\rm Tr}\left[\left(\frac 1 p \sum_{n=1}^p A_n B_n\right)^{\otimes p}\right].\nonumber\\
&\quad\ge {\rm Tr}\left[(A_1  B_1)\otimes (A_2B_2) \otimes ... \otimes (A_pB_p)\right] 
\end{split}
\end{align}

We will also use that for $A$ and $B$ matrices \EQ{eq:ABswapAotimesB} holds. Then, we can write an inequality relevant for our problem as
\begin{align}
\begin{split}
&{\rm Tr}\left(\sum_k {\tilde p_k}\bigotimes_{n=1}^p \varrho_k^{(n,n+p)} S_p\right)\\
&\quad\quad\quad\quad\ge  {\rm Tr}\left(\sum_k {\tilde p_k}\bigotimes_{n=1}^p \varrho^{(n,n+p)}_{k,n} S_p\right),\label{eq:ineqSp}
\end{split}
\end{align}
where we define the two-particle product density matrices as 
\be
\varrho_{k,n}= \ketbra{\Psi_k^{(n)}}\otimes \ketbra{\Phi_k^{(n)}},
\ee
where for all $n=1,2,...,p$ we have
\begin{subequations}
\begin{align}
\varrho&=\sum_k {\tilde p_k}  \ketbra{\Psi_k^{(n)}},\label{eq:marginalcond_a1}\\
\sigma&=\sum_k {\tilde p_k}  \ketbra{\Phi_k^{(n)}},\label{eq:marginalcond_b1}
\end{align}\label{eq:marginalcond1}
\end{subequations}
and the average two-particle separable density matrix is
\be
\varrho_k=\frac1 p \sum_{n=1}^p \varrho_{k,n}.
\ee
The superscript $(n,n+p)$ indicates that the two-particle density matrix corresponds to spins $n$ and $(n+p).$ However, note that the two-body matrix $\varrho_k^{(n,n+p)} $ does not depend on $n,$ while 
$\varrho^{(n,n+p)}_{k,n} $ depends on $n.$

Let us consider the state
\be
\varrho_{12}=\sum_k {\tilde p_k} \varrho_k,
\ee
which can be written as in \EQ{eq:sep}. Note that the reduced states are the following
\be
{\rm Tr}_2\left(\varrho_{12}\right)=\varrho,\quad {\rm Tr}_1\left(\varrho_{12}\right)=\sigma. 
\ee
The inequality in \EQ{eq:ineqSp} is saturated if 
\be
\varrho_{k,n}=\varrho_{k,1}
\ee
for all $k$ and $n.$ 

Now we have to use that  
\be
{\rm Tr}\left(\bigotimes_{n=1}^p \varrho_k^{(n,n+p)} S_p\right)={\rm Tr}(\varrho_k S)^p\label{eq:nnpSp}
\ee
is a convex function of the two-body density matrix $\varrho.$ This statement is true for general quantum states for an even $p,$ however, it is also true for an odd $p$ for separable states due to the fact that for a separable state  \EQ{eq_sepineq1} holds. Thus, when maximized over a convex set, the expression in \EQ{eq:nnpSp} takes its maximum at the extreme points of the set.

Hence, we can see that that there is a $\{p_k,\tau_k\}$ decomposition such that
\begin{align}
\begin{split}
& {\rm Tr}\left(\sum_k {p_k}\bigotimes_{n=1}^p \tau_{k}^{(n,n+p)} S_p\right)\\
&\quad\quad\quad\quad\ge {\rm Tr}\left(\sum_k \tilde p_k\bigotimes_{n=1}^p \varrho_k^{(n,n+p)} S_p\right),\label{eq:ineqSp2}
\end{split}
\end{align}
where the two-particle pure state is given as 
\be
\tau_k=\ketbra{\Psi_k}\otimes\ketbra{\Phi_k},
\ee
and $\varrho$ and $\sigma$ is decomposed as in \EQ{eq:marginalcond}. Together with \EQ{eq:ineqSp} this indicates that there is an optimal state of the form \EQ{eq:optmimumsym}. The quantum state appearing on the left-hand side of \EQ{eq:ineqSp2} is of that form.

\section{Details of the numerical calculations}
\label{app:Numerics}

We used MATLAB \cite{MATLAB2020} for numerical calculations. We used the semidefinite solver MOSEK \cite{MOSEK} and the front-end YALMIP \cite{Lofberg2004Yalmip}. We also used the QUBIT4MATLAB package \cite{Toth2008QUBIT4MATLAB,QUBIT4MATLAB_actual_note_href}. $D_{\rm DPT}(\varrho,\sigma)^2$ given in \EQ{eq:DPT} can be obtained using semidefinite programming.   $D_{\rm DPT, sep}(\varrho,\sigma)^2$ in \EQ{eq:GMPC_distance_sep} need an optimization over separable states.  Instead of an optmization over separable states, we can carry out numerically an optimization over the set of PPT states (i.~e., states with a positive partial transpose). Since it is a superset of separable states, if we maximize, we obtain an upper bound on the maximum over separable states, and if we minimize, we obtain a lower bound on the minimum over separable states. However, for two qubits we will get the true optimal value, since in that case the set of PPT states  equals the set of separable states. For a detailed discussion about how to compute numerically the quantum Wasserstein distance via an optimization over separable states see \REF{Toth2023QuantumWasserstein}.

We have already included the routines computing the various quantum Wasserstein distance measures for \REF{Toth2023QuantumWasserstein}, such as 
{\tt wdistsquare\_GMPC.m},
{\tt wdistsquare\_DPT\_ppt.m}, and
{\tt wdistsquare\_DPT.m}.
We included the various types of the variance-like quantities as
{\tt wvar\_GMPC\_ppt.m},
{\tt wvar\_GMPC.m},
{\tt wvar\_DPT\_ppt.m}, and
{\tt wvar\_DPT.m}.
The usage of these routines is demonstrated in {\tt example\_wdistsquare.m}. 

For this publication, we added
{\tt swapfidelity.m},
{\tt swapfidelity\_ppt.m}, and
{\tt example\_swapfidelity.m}. Instead of the command {\tt swapfidelity\_ppt.m}, the old command {\tt wdistsquare\_DPT\_ppt.m} could also be used with an appropriate choice of the $H_n$ operators. This is due to the relation given in \EQ{eq:manyeq} for the case when we consider a full set of $(d^2-1)$ traceless operators $H_n,$  which are the SU(d) generators.
We included the new command for convenience.

\bibliography{Bibliography2}

\begin{thebibliography}{89}%
\makeatletter
\providecommand \@ifxundefined [1]{%
 \@ifx{#1\undefined}
}%
\providecommand \@ifnum [1]{%
 \ifnum #1\expandafter \@firstoftwo
 \else \expandafter \@secondoftwo
 \fi
}%
\providecommand \@ifx [1]{%
 \ifx #1\expandafter \@firstoftwo
 \else \expandafter \@secondoftwo
 \fi
}%
\providecommand \natexlab [1]{#1}%
\providecommand \enquote  [1]{``#1''}%
\providecommand \bibnamefont  [1]{#1}%
\providecommand \bibfnamefont [1]{#1}%
\providecommand \citenamefont [1]{#1}%
\providecommand \href@noop [0]{\@secondoftwo}%
\providecommand \href [0]{\begingroup \@sanitize@url \@href}%
\providecommand \@href[1]{\@@startlink{#1}\@@href}%
\providecommand \@@href[1]{\endgroup#1\@@endlink}%
\providecommand \@sanitize@url [0]{\catcode `\\12\catcode `\$12\catcode
  `\&12\catcode `\#12\catcode `\^12\catcode `\_12\catcode `\%12\relax}%
\providecommand \@@startlink[1]{}%
\providecommand \@@endlink[0]{}%
\providecommand \url  [0]{\begingroup\@sanitize@url \@url }%
\providecommand \@url [1]{\endgroup\@href {#1}{\urlprefix }}%
\providecommand \urlprefix  [0]{URL }%
\providecommand \Eprint [0]{\href }%
\providecommand \doibase [0]{https://doi.org/}%
\providecommand \selectlanguage [0]{\@gobble}%
\providecommand \bibinfo  [0]{\@secondoftwo}%
\providecommand \bibfield  [0]{\@secondoftwo}%
\providecommand \translation [1]{[#1]}%
\providecommand \BibitemOpen [0]{}%
\providecommand \bibitemStop [0]{}%
\providecommand \bibitemNoStop [0]{.\EOS\space}%
\providecommand \EOS [0]{\spacefactor3000\relax}%
\providecommand \BibitemShut  [1]{\csname bibitem#1\endcsname}%
\let\auto@bib@innerbib\@empty
\bibitem [{\citenamefont {\.Zyczkowski}\ and\ \citenamefont
  {Slomczynski}(1998)}]{Zyczkowski1998TheMonge}%
  \BibitemOpen
  \bibfield  {author} {\bibinfo {author} {\bibfnamefont {K.}~\bibnamefont
  {\.Zyczkowski}}\ and\ \bibinfo {author} {\bibfnamefont {W.}~\bibnamefont
  {Slomczynski}},\ }\bibfield  {title} {\bibinfo {title} {The {Monge} distance
  between quantum states},\ }\href
  {https://doi.org/10.1088/0305-4470/31/45/009} {\bibfield  {journal} {\bibinfo
   {journal} {J. Phys. A: Math. Gen.}\ }\textbf {\bibinfo {volume} {31}},\
  \bibinfo {pages} {9095} (\bibinfo {year} {1998})}\BibitemShut {NoStop}%
\bibitem [{\citenamefont {\.Zyczkowski}\ and\ \citenamefont
  {Slomczynski}(2001)}]{Zyczkowski2001TheMonge}%
  \BibitemOpen
  \bibfield  {author} {\bibinfo {author} {\bibfnamefont {K.}~\bibnamefont
  {\.Zyczkowski}}\ and\ \bibinfo {author} {\bibfnamefont {W.}~\bibnamefont
  {Slomczynski}},\ }\bibfield  {title} {\bibinfo {title} {The {Monge} metric on
  the sphere and geometry of quantum states},\ }\href
  {https://doi.org/10.1088/0305-4470/34/34/311} {\bibfield  {journal} {\bibinfo
   {journal} {J. Phys. A: Math. Gen.}\ }\textbf {\bibinfo {volume} {34}},\
  \bibinfo {pages} {6689} (\bibinfo {year} {2001})}\BibitemShut {NoStop}%
\bibitem [{\citenamefont {Bengtsson}\ and\ \citenamefont
  {\.Zyczkowski}(2006)}]{Bengtsson2006Geometry}%
  \BibitemOpen
  \bibfield  {author} {\bibinfo {author} {\bibfnamefont {I.}~\bibnamefont
  {Bengtsson}}\ and\ \bibinfo {author} {\bibfnamefont {K.}~\bibnamefont
  {\.Zyczkowski}},\ }\href {https://doi.org/10.1017/CBO9780511535048} {\emph
  {\bibinfo {title} {Geometry of Quantum States: An Introduction to Quantum
  Entanglement}}}\ (\bibinfo  {publisher} {Cambridge University Press},\
  \bibinfo {year} {2006})\BibitemShut {NoStop}%
\bibitem [{\citenamefont {{P. Biane and D. Voiculescu}}(2001)}]{Biane2011Free}%
  \BibitemOpen
  \bibfield  {author} {\bibinfo {author} {\bibnamefont {{P. Biane and D.
  Voiculescu}}},\ }\bibfield  {title} {\bibinfo {title} {{A free probability
  analogue of the Wasserstein metric on the trace-state space}},\ }\href
  {https://doi.org/10.1007/s00039-001-8226-4} {\bibfield  {journal} {\bibinfo
  {journal} {GAFA, Geom. Funct. Anal.}\ }\textbf {\bibinfo {volume} {11}},\
  \bibinfo {pages} {1125} (\bibinfo {year} {2001})}\BibitemShut {NoStop}%
\bibitem [{\citenamefont {Carlen}\ and\ \citenamefont
  {Maas}(2014)}]{CarlenMaas2014Analog}%
  \BibitemOpen
  \bibfield  {author} {\bibinfo {author} {\bibfnamefont {E.}~\bibnamefont
  {Carlen}}\ and\ \bibinfo {author} {\bibfnamefont {J.}~\bibnamefont {Maas}},\
  }\bibfield  {title} {\bibinfo {title} {{An Analog of the 2-Wasserstein Metric
  in Non-Commutative Probability Under Which the Fermionic Fokker-Planck
  Equation is Gradient Flow for the Entropy}},\ }\href
  {https://doi.org/10.1007/s00220-014-2124-8} {\bibfield  {journal} {\bibinfo
  {journal} {Commun. Math. Phys.}\ }\textbf {\bibinfo {volume} {331}},\
  \bibinfo {pages} {887} (\bibinfo {year} {2014})}\BibitemShut {NoStop}%
\bibitem [{\citenamefont {Carlen}\ and\ \citenamefont
  {Maas}(2017)}]{CarlenMaas2017Gradient}%
  \BibitemOpen
  \bibfield  {author} {\bibinfo {author} {\bibfnamefont {E.~A.}\ \bibnamefont
  {Carlen}}\ and\ \bibinfo {author} {\bibfnamefont {J.}~\bibnamefont {Maas}},\
  }\bibfield  {title} {\bibinfo {title} {{Gradient flow and entropy
  inequalities for quantum Markov semigroups with detailed balance}},\ }\href
  {https://doi.org/10.1016/j.jfa.2017.05.003} {\bibfield  {journal} {\bibinfo
  {journal} {J. Funct. Anal.}\ }\textbf {\bibinfo {volume} {273}},\ \bibinfo
  {pages} {1810} (\bibinfo {year} {2017})}\BibitemShut {NoStop}%
\bibitem [{\citenamefont {Carlen}\ and\ \citenamefont
  {Maas}(2020)}]{CarlenMaas2020Non-commutative}%
  \BibitemOpen
  \bibfield  {author} {\bibinfo {author} {\bibfnamefont {E.~A.}\ \bibnamefont
  {Carlen}}\ and\ \bibinfo {author} {\bibfnamefont {J.}~\bibnamefont {Maas}},\
  }\bibfield  {title} {\bibinfo {title} {Non-commutative calculus, optimal
  transport and functional inequalities in dissipative quantum systems},\
  }\href {https://doi.org/10.1007/s10955-019-02434-w} {\bibfield  {journal}
  {\bibinfo  {journal} {J. Stat. Phys.}\ }\textbf {\bibinfo {volume} {178}},\
  \bibinfo {pages} {319} (\bibinfo {year} {2020})}\BibitemShut {NoStop}%
\bibitem [{\citenamefont {Datta}\ and\ \citenamefont
  {Rouz\'e}(2019)}]{DattaRouze2019Concentration}%
  \BibitemOpen
  \bibfield  {author} {\bibinfo {author} {\bibfnamefont {N.}~\bibnamefont
  {Datta}}\ and\ \bibinfo {author} {\bibfnamefont {C.}~\bibnamefont
  {Rouz\'e}},\ }\bibfield  {title} {\bibinfo {title} {Concentration of quantum
  states from quantum functional and transportation cost inequalities},\ }\href
  {https://doi.org/10.1063/1.5023210} {\bibfield  {journal} {\bibinfo
  {journal} {J. Math. Phys.}\ }\textbf {\bibinfo {volume} {60}},\ \bibinfo
  {pages} {012202} (\bibinfo {year} {2019})}\BibitemShut {NoStop}%
\bibitem [{\citenamefont {Datta}\ and\ \citenamefont
  {Rouz\'e}(2020)}]{DattaRouze2020Relating}%
  \BibitemOpen
  \bibfield  {author} {\bibinfo {author} {\bibfnamefont {N.}~\bibnamefont
  {Datta}}\ and\ \bibinfo {author} {\bibfnamefont {C.}~\bibnamefont
  {Rouz\'e}},\ }\bibfield  {title} {\bibinfo {title} {{Relating relative
  entropy, optimal transport and Fisher information: A quantum HWI
  inequality}},\ }\href {https://doi.org/10.1007/s00023-020-00891-8} {\bibfield
   {journal} {\bibinfo  {journal} {Ann. Henri Poincar\'e}\ }\textbf {\bibinfo
  {volume} {21}},\ \bibinfo {pages} {2115} (\bibinfo {year}
  {2020})}\BibitemShut {NoStop}%
\bibitem [{\citenamefont {Golse}\ \emph {et~al.}(2016)\citenamefont {Golse},
  \citenamefont {Mouhot},\ and\ \citenamefont {Paul}}]{Golse2016On}%
  \BibitemOpen
  \bibfield  {author} {\bibinfo {author} {\bibfnamefont {F.}~\bibnamefont
  {Golse}}, \bibinfo {author} {\bibfnamefont {C.}~\bibnamefont {Mouhot}},\ and\
  \bibinfo {author} {\bibfnamefont {T.}~\bibnamefont {Paul}},\ }\bibfield
  {title} {\bibinfo {title} {On the mean field and classical limits of quantum
  mechanics},\ }\href {https://doi.org/10.1007/s00220-015-2485-7} {\bibfield
  {journal} {\bibinfo  {journal} {Commun. Math. Phys.}\ }\textbf {\bibinfo
  {volume} {343}},\ \bibinfo {pages} {165} (\bibinfo {year}
  {2016})}\BibitemShut {NoStop}%
\bibitem [{\citenamefont {Golse}\ and\ \citenamefont
  {Paul}(2017)}]{Golse2017The}%
  \BibitemOpen
  \bibfield  {author} {\bibinfo {author} {\bibfnamefont {F.}~\bibnamefont
  {Golse}}\ and\ \bibinfo {author} {\bibfnamefont {T.}~\bibnamefont {Paul}},\
  }\bibfield  {title} {\bibinfo {title} {The {Schr{\"o}dinger} equation in the
  mean-field and semiclassical regime},\ }\href
  {https://doi.org/10.1007/s00205-016-1031-x} {\bibfield  {journal} {\bibinfo
  {journal} {Arch. Ration. Mech. Anal.}\ }\textbf {\bibinfo {volume} {223}},\
  \bibinfo {pages} {57} (\bibinfo {year} {2017})}\BibitemShut {NoStop}%
\bibitem [{\citenamefont {Golse}\ and\ \citenamefont
  {Paul}(2018)}]{Golse2018Wave}%
  \BibitemOpen
  \bibfield  {author} {\bibinfo {author} {\bibfnamefont {F.}~\bibnamefont
  {Golse}}\ and\ \bibinfo {author} {\bibfnamefont {T.}~\bibnamefont {Paul}},\
  }\bibfield  {title} {\bibinfo {title} {Wave packets and the quadratic
  {Monge}-{Kantorovich} distance in quantum mechanics},\ }\href
  {https://doi.org/https://doi.org/10.1016/j.crma.2017.12.007} {\bibfield
  {journal} {\bibinfo  {journal} {Comptes Rendus Math.}\ }\textbf {\bibinfo
  {volume} {356}},\ \bibinfo {pages} {177} (\bibinfo {year}
  {2018})}\BibitemShut {NoStop}%
\bibitem [{\citenamefont {Golse}(2018)}]{Golse2018TheQuantum}%
  \BibitemOpen
  \bibfield  {author} {\bibinfo {author} {\bibfnamefont {F.}~\bibnamefont
  {Golse}},\ }\bibfield  {title} {\bibinfo {title} {The quantum {$N$}-body
  problem in the mean-field and semiclassical regime},\ }\href
  {https://doi.org/10.1098/rsta.2017.0229} {\bibfield  {journal} {\bibinfo
  {journal} {Phil. Trans. R. Soc. A}\ }\textbf {\bibinfo {volume} {376}},\
  \bibinfo {pages} {20170229} (\bibinfo {year} {2018})}\BibitemShut {NoStop}%
\bibitem [{\citenamefont {Caglioti}\ \emph {et~al.}(2020)\citenamefont
  {Caglioti}, \citenamefont {Golse},\ and\ \citenamefont
  {Paul}}]{Caglioti2020Quantum}%
  \BibitemOpen
  \bibfield  {author} {\bibinfo {author} {\bibfnamefont {E.}~\bibnamefont
  {Caglioti}}, \bibinfo {author} {\bibfnamefont {F.}~\bibnamefont {Golse}},\
  and\ \bibinfo {author} {\bibfnamefont {T.}~\bibnamefont {Paul}},\ }\bibfield
  {title} {\bibinfo {title} {Quantum optimal transport is cheaper},\ }\href
  {https://doi.org/10.1007/s10955-020-02571-7} {\bibfield  {journal} {\bibinfo
  {journal} {J. Stat. Phys.}\ }\textbf {\bibinfo {volume} {181}},\ \bibinfo
  {pages} {149} (\bibinfo {year} {2020})}\BibitemShut {NoStop}%
\bibitem [{\citenamefont {Caglioti}\ \emph {et~al.}(2021)\citenamefont
  {Caglioti}, \citenamefont {Golse},\ and\ \citenamefont
  {Paul}}]{Caglioti2021Towards}%
  \BibitemOpen
  \bibfield  {author} {\bibinfo {author} {\bibfnamefont {E.}~\bibnamefont
  {Caglioti}}, \bibinfo {author} {\bibfnamefont {F.}~\bibnamefont {Golse}},\
  and\ \bibinfo {author} {\bibfnamefont {T.}~\bibnamefont {Paul}},\ }\bibfield
  {title} {\bibinfo {title} {Towards optimal transport for quantum densities},\
  }\href {https://arxiv.org/abs/2101.03256} {\bibfield  {journal} {\bibinfo
  {journal} {arXiv:2101.03256}\ } (\bibinfo {year} {2021})}\BibitemShut
  {NoStop}%
\bibitem [{\citenamefont
  {Lafleche}(2023)}]{lafleche2023quantumoptimaltransportweak}%
  \BibitemOpen
  \bibfield  {author} {\bibinfo {author} {\bibfnamefont {L.}~\bibnamefont
  {Lafleche}},\ }\href {https://arxiv.org/abs/2306.12944} {\bibinfo {title}
  {Quantum optimal transport and weak topologies}} (\bibinfo {year} {2023}),\
  \Eprint {https://arxiv.org/abs/2306.12944} {arXiv:2306.12944 [math.AP]}
  \BibitemShut {NoStop}%
\bibitem [{\citenamefont {De~Palma}\ and\ \citenamefont
  {Trevisan}(2021)}]{DePalma2021Quantum}%
  \BibitemOpen
  \bibfield  {author} {\bibinfo {author} {\bibfnamefont {G.}~\bibnamefont
  {De~Palma}}\ and\ \bibinfo {author} {\bibfnamefont {D.}~\bibnamefont
  {Trevisan}},\ }\bibfield  {title} {\bibinfo {title} {Quantum optimal
  transport with quantum channels},\ }\href
  {https://doi.org/10.1007/s00023-021-01042-3} {\bibfield  {journal} {\bibinfo
  {journal} {Ann. Henri Poincar{\'e}}\ }\textbf {\bibinfo {volume} {22}},\
  \bibinfo {pages} {3199} (\bibinfo {year} {2021})}\BibitemShut {NoStop}%
\bibitem [{\citenamefont {De~Palma}\ \emph {et~al.}(2021)\citenamefont
  {De~Palma}, \citenamefont {Marvian}, \citenamefont {Trevisan},\ and\
  \citenamefont {Lloyd}}]{DePalma2021TheQuantum}%
  \BibitemOpen
  \bibfield  {author} {\bibinfo {author} {\bibfnamefont {G.}~\bibnamefont
  {De~Palma}}, \bibinfo {author} {\bibfnamefont {M.}~\bibnamefont {Marvian}},
  \bibinfo {author} {\bibfnamefont {D.}~\bibnamefont {Trevisan}},\ and\
  \bibinfo {author} {\bibfnamefont {S.}~\bibnamefont {Lloyd}},\ }\bibfield
  {title} {\bibinfo {title} {The quantum {Wasserstein} distance of order 1},\
  }\href {https://doi.org/10.1109/TIT.2021.3076442} {\bibfield  {journal}
  {\bibinfo  {journal} {IEEE Trans. Inf. Theory}\ }\textbf {\bibinfo {volume}
  {67}},\ \bibinfo {pages} {6627} (\bibinfo {year} {2021})}\BibitemShut
  {NoStop}%
\bibitem [{\citenamefont {{S. Friedland, M. Eckstein, S. Cole, and K.
  \.Zyczkowski}}(2022)}]{Friedland2022Quantum}%
  \BibitemOpen
  \bibfield  {author} {\bibinfo {author} {\bibnamefont {{S. Friedland, M.
  Eckstein, S. Cole, and K. \.Zyczkowski}}},\ }\bibfield  {title} {\bibinfo
  {title} {{Quantum Monge--Kantorovich problem and transport distance between
  density matrices}},\ }\href {https://doi.org/10.1103/PhysRevLett.129.110402}
  {\bibfield  {journal} {\bibinfo  {journal} {{Phys. Rev. Lett.}}\ }\textbf
  {\bibinfo {volume} {129}},\ \bibinfo {pages} {110402} (\bibinfo {year}
  {2022})}\BibitemShut {NoStop}%
\bibitem [{\citenamefont {Cole}\ \emph {et~al.}(2023)\citenamefont {Cole},
  \citenamefont {Eckstein}, \citenamefont {Friedland},\ and\ \citenamefont
  {{\.{Z}}yczkowski}}]{Cole2023OnQuantum}%
  \BibitemOpen
  \bibfield  {author} {\bibinfo {author} {\bibfnamefont {S.}~\bibnamefont
  {Cole}}, \bibinfo {author} {\bibfnamefont {M.}~\bibnamefont {Eckstein}},
  \bibinfo {author} {\bibfnamefont {S.}~\bibnamefont {Friedland}},\ and\
  \bibinfo {author} {\bibfnamefont {K.}~\bibnamefont {{\.{Z}}yczkowski}},\
  }\bibfield  {title} {\bibinfo {title} {On quantum optimal transport},\ }\href
  {https://doi.org/10.1007/s11040-023-09456-7} {\bibfield  {journal} {\bibinfo
  {journal} {Mathematical Physics, Analysis and Geometry}\ }\textbf {\bibinfo
  {volume} {26}},\ \bibinfo {pages} {14} (\bibinfo {year} {2023})}\BibitemShut
  {NoStop}%
\bibitem [{\citenamefont {Bistro\'n}\ \emph {et~al.}(2023)\citenamefont
  {Bistro\'n}, \citenamefont {Eckstein},\ and\ \citenamefont
  {\.Zyczkowski}}]{Bistron2023Monotonicity}%
  \BibitemOpen
  \bibfield  {author} {\bibinfo {author} {\bibfnamefont {R.}~\bibnamefont
  {Bistro\'n}}, \bibinfo {author} {\bibfnamefont {M.}~\bibnamefont
  {Eckstein}},\ and\ \bibinfo {author} {\bibfnamefont {K.}~\bibnamefont
  {\.Zyczkowski}},\ }\bibfield  {title} {\bibinfo {title} {{Monotonicity of a
  quantum 2-Wasserstein distance}},\ }\href
  {https://doi.org/10.1088/1751-8121/acb9c8} {\bibfield  {journal} {\bibinfo
  {journal} {J. Phys. A: Math. Theor.}\ }\textbf {\bibinfo {volume} {56}},\
  \bibinfo {pages} {095301} (\bibinfo {year} {2023})}\BibitemShut {NoStop}%
\bibitem [{\citenamefont {Geh\'er}\ \emph {et~al.}(2023)\citenamefont
  {Geh\'er}, \citenamefont {Pitrik}, \citenamefont {Titkos},\ and\
  \citenamefont {Virosztek}}]{Geher2023Quantum}%
  \BibitemOpen
  \bibfield  {author} {\bibinfo {author} {\bibfnamefont {G.~P.}\ \bibnamefont
  {Geh\'er}}, \bibinfo {author} {\bibfnamefont {J.}~\bibnamefont {Pitrik}},
  \bibinfo {author} {\bibfnamefont {T.}~\bibnamefont {Titkos}},\ and\ \bibinfo
  {author} {\bibfnamefont {D.}~\bibnamefont {Virosztek}},\ }\bibfield  {title}
  {\bibinfo {title} {{Quantum Wasserstein isometries on the qubit state
  space}},\ }\href {https://doi.org/https://doi.org/10.1016/j.jmaa.2022.126955}
  {\bibfield  {journal} {\bibinfo  {journal} {J. Math. Anal. Appl.}\ }\textbf
  {\bibinfo {volume} {522}},\ \bibinfo {pages} {126955} (\bibinfo {year}
  {2023})}\BibitemShut {NoStop}%
\bibitem [{\citenamefont {Li}\ \emph {et~al.}(2025)\citenamefont {Li},
  \citenamefont {Bu}, \citenamefont {Enshan~Koh}, \citenamefont {Jaffe},\ and\
  \citenamefont {Lloyd}}]{Li2025Wasserstein}%
  \BibitemOpen
  \bibfield  {author} {\bibinfo {author} {\bibfnamefont {L.}~\bibnamefont
  {Li}}, \bibinfo {author} {\bibfnamefont {K.}~\bibnamefont {Bu}}, \bibinfo
  {author} {\bibfnamefont {D.}~\bibnamefont {Enshan~Koh}}, \bibinfo {author}
  {\bibfnamefont {A.}~\bibnamefont {Jaffe}},\ and\ \bibinfo {author}
  {\bibfnamefont {S.}~\bibnamefont {Lloyd}},\ }\bibfield  {title} {\bibinfo
  {title} {Wasserstein complexity of quantum circuits},\ }\href
  {https://doi.org/10.1088/1751-8121/ade381} {\bibfield  {journal} {\bibinfo
  {journal} {J. Phys. A: Math. Theor.}\ }\textbf {\bibinfo {volume} {58}},\
  \bibinfo {pages} {265302} (\bibinfo {year} {2025})}\BibitemShut {NoStop}%
\bibitem [{\citenamefont
  {Beatty}(2025)}]{beatty2025wassersteindistancesquantumstructures}%
  \BibitemOpen
  \bibfield  {author} {\bibinfo {author} {\bibfnamefont {E.}~\bibnamefont
  {Beatty}},\ }\href {https://arxiv.org/abs/2506.09794} {\bibinfo {title}
  {Wasserstein distances on quantum structures: an overview}} (\bibinfo {year}
  {2025}),\ \Eprint {https://arxiv.org/abs/2506.09794} {arXiv:2506.09794
  [quant-ph]} \BibitemShut {NoStop}%
\bibitem [{\citenamefont {Kiani}\ \emph {et~al.}(2022)\citenamefont {Kiani},
  \citenamefont {Palma}, \citenamefont {Marvian}, \citenamefont {Liu},\ and\
  \citenamefont {Lloyd}}]{Kiani2022Learning}%
  \BibitemOpen
  \bibfield  {author} {\bibinfo {author} {\bibfnamefont {B.~T.}\ \bibnamefont
  {Kiani}}, \bibinfo {author} {\bibfnamefont {G.~D.}\ \bibnamefont {Palma}},
  \bibinfo {author} {\bibfnamefont {M.}~\bibnamefont {Marvian}}, \bibinfo
  {author} {\bibfnamefont {Z.-W.}\ \bibnamefont {Liu}},\ and\ \bibinfo {author}
  {\bibfnamefont {S.}~\bibnamefont {Lloyd}},\ }\bibfield  {title} {\bibinfo
  {title} {Learning quantum data with the quantum earth mover's distance},\
  }\href {https://doi.org/10.1088/2058-9565/ac79c9} {\bibfield  {journal}
  {\bibinfo  {journal} {Quantum Sci. Technol.}\ }\textbf {\bibinfo {volume}
  {7}},\ \bibinfo {pages} {045002} (\bibinfo {year} {2022})}\BibitemShut
  {NoStop}%
\bibitem [{\citenamefont {De~Palma}\ and\ \citenamefont
  {Trevisan}(2024)}]{DePalma2024QuantumOptimal}%
  \BibitemOpen
  \bibfield  {author} {\bibinfo {author} {\bibfnamefont {G.}~\bibnamefont
  {De~Palma}}\ and\ \bibinfo {author} {\bibfnamefont {D.}~\bibnamefont
  {Trevisan}},\ }\bibinfo {title} {Quantum optimal transport: Quantum channels
  and qubits},\ in\ \href {https://doi.org/10.1007/978-3-031-50466-2_4} {\emph
  {\bibinfo {booktitle} {Optimal Transport on Quantum Structures}}},\ \bibinfo
  {editor} {edited by\ \bibinfo {editor} {\bibfnamefont {J.}~\bibnamefont
  {Maas}}, \bibinfo {editor} {\bibfnamefont {S.}~\bibnamefont {Rademacher}},
  \bibinfo {editor} {\bibfnamefont {T.}~\bibnamefont {Titkos}},\ and\ \bibinfo
  {editor} {\bibfnamefont {D.}~\bibnamefont {Virosztek}}}\ (\bibinfo
  {publisher} {Springer Nature Switzerland},\ \bibinfo {address} {Cham},\
  \bibinfo {year} {2024})\ pp.\ \bibinfo {pages} {203--239}\BibitemShut
  {NoStop}%
\bibitem [{\citenamefont {Wigner}\ and\ \citenamefont
  {Yanase}(1963)}]{Wigner1963INFORMATION}%
  \BibitemOpen
  \bibfield  {author} {\bibinfo {author} {\bibfnamefont {E.~P.}\ \bibnamefont
  {Wigner}}\ and\ \bibinfo {author} {\bibfnamefont {M.~M.}\ \bibnamefont
  {Yanase}},\ }\bibfield  {title} {\bibinfo {title} {Information contents of
  distributions},\ }\href {https://doi.org/10.1073/pnas.49.6.910} {\bibfield
  {journal} {\bibinfo  {journal} {Proc. Natl. Acad. Sci. U.S.A.}\ }\textbf
  {\bibinfo {volume} {49}},\ \bibinfo {pages} {910} (\bibinfo {year}
  {1963})}\BibitemShut {NoStop}%
\bibitem [{\citenamefont {Camacho}\ and\ \citenamefont
  {Fauseweh}(2025)}]{camacho2025criticalscalingquantumwasserstein}%
  \BibitemOpen
  \bibfield  {author} {\bibinfo {author} {\bibfnamefont {G.}~\bibnamefont
  {Camacho}}\ and\ \bibinfo {author} {\bibfnamefont {B.}~\bibnamefont
  {Fauseweh}},\ }\href {https://arxiv.org/abs/2504.02709} {\bibinfo {title}
  {{Critical Scaling of the Quantum Wasserstein Distance}}} (\bibinfo {year}
  {2025}),\ \Eprint {https://arxiv.org/abs/2504.02709} {arXiv:2504.02709
  [quant-ph]} \BibitemShut {NoStop}%
\bibitem [{\citenamefont {Bunth}\ \emph
  {et~al.}(2025{\natexlab{a}})\citenamefont {Bunth}, \citenamefont {Pitrik},
  \citenamefont {Titkos},\ and\ \citenamefont
  {Virosztek}}]{bunth2025wassersteindistancesdivergencesorder}%
  \BibitemOpen
  \bibfield  {author} {\bibinfo {author} {\bibfnamefont {G.}~\bibnamefont
  {Bunth}}, \bibinfo {author} {\bibfnamefont {J.}~\bibnamefont {Pitrik}},
  \bibinfo {author} {\bibfnamefont {T.}~\bibnamefont {Titkos}},\ and\ \bibinfo
  {author} {\bibfnamefont {D.}~\bibnamefont {Virosztek}},\ }\href
  {https://arxiv.org/abs/2501.08066} {\bibinfo {title} {Wasserstein distances
  and divergences of order $p$ by quantum channels}} (\bibinfo {year}
  {2025}{\natexlab{a}}),\ \Eprint {https://arxiv.org/abs/2501.08066}
  {arXiv:2501.08066 [math-ph]} \BibitemShut {NoStop}%
\bibitem [{\citenamefont {Bunth}\ \emph
  {et~al.}(2025{\natexlab{b}})\citenamefont {Bunth}, \citenamefont {Pitrik},
  \citenamefont {Titkos},\ and\ \citenamefont
  {Virosztek}}]{bunth2025strongkantorovichdualityquantum}%
  \BibitemOpen
  \bibfield  {author} {\bibinfo {author} {\bibfnamefont {G.}~\bibnamefont
  {Bunth}}, \bibinfo {author} {\bibfnamefont {J.}~\bibnamefont {Pitrik}},
  \bibinfo {author} {\bibfnamefont {T.}~\bibnamefont {Titkos}},\ and\ \bibinfo
  {author} {\bibfnamefont {D.}~\bibnamefont {Virosztek}},\ }\href
  {https://arxiv.org/abs/2510.26326} {\bibinfo {title} {Strong {Kantorovich}
  duality for quantum optimal transport with generic cost and optimal couplings
  on quantum bits}} (\bibinfo {year} {2025}{\natexlab{b}}),\ \Eprint
  {https://arxiv.org/abs/2510.26326} {arXiv:2510.26326 [math-ph]} \BibitemShut
  {NoStop}%
\bibitem [{\citenamefont {Bunth}\ \emph {et~al.}(2024)\citenamefont {Bunth},
  \citenamefont {Pitrik}, \citenamefont {Titkos},\ and\ \citenamefont
  {Virosztek}}]{Bunth2024MetricProperty}%
  \BibitemOpen
  \bibfield  {author} {\bibinfo {author} {\bibfnamefont {G.}~\bibnamefont
  {Bunth}}, \bibinfo {author} {\bibfnamefont {J.}~\bibnamefont {Pitrik}},
  \bibinfo {author} {\bibfnamefont {T.}~\bibnamefont {Titkos}},\ and\ \bibinfo
  {author} {\bibfnamefont {D.}~\bibnamefont {Virosztek}},\ }\bibfield  {title}
  {\bibinfo {title} {Metric property of quantum wasserstein divergences},\
  }\href {https://doi.org/10.1103/PhysRevA.110.022211} {\bibfield  {journal}
  {\bibinfo  {journal} {Phys. Rev. A}\ }\textbf {\bibinfo {volume} {110}},\
  \bibinfo {pages} {022211} (\bibinfo {year} {2024})}\BibitemShut {NoStop}%
\bibitem [{\citenamefont
  {Wirth}(2025)}]{wirth2025triangleinequalityquantumwasserstein}%
  \BibitemOpen
  \bibfield  {author} {\bibinfo {author} {\bibfnamefont {M.}~\bibnamefont
  {Wirth}},\ }\href {https://arxiv.org/abs/2511.20450} {\bibinfo {title}
  {Triangle inequality for a quantum {Wasserstein} divergence}} (\bibinfo
  {year} {2025}),\ \Eprint {https://arxiv.org/abs/2511.20450} {arXiv:2511.20450
  [math-ph]} \BibitemShut {NoStop}%
\bibitem [{\citenamefont {T{\'{o}}th}\ and\ \citenamefont
  {Pitrik}(2023)}]{Toth2023QuantumWasserstein}%
  \BibitemOpen
  \bibfield  {author} {\bibinfo {author} {\bibfnamefont {G.}~\bibnamefont
  {T{\'{o}}th}}\ and\ \bibinfo {author} {\bibfnamefont {J.}~\bibnamefont
  {Pitrik}},\ }\bibfield  {title} {\bibinfo {title} {Quantum {W}asserstein
  distance based on an optimization over separable states},\ }\href
  {https://doi.org/10.22331/q-2023-10-16-1143} {\bibfield  {journal} {\bibinfo
  {journal} {{Quantum}}\ }\textbf {\bibinfo {volume} {7}},\ \bibinfo {pages}
  {1143} (\bibinfo {year} {2023})}\BibitemShut {NoStop}%
\bibitem [{\citenamefont {Horodecki}\ \emph {et~al.}(2009)\citenamefont
  {Horodecki}, \citenamefont {Horodecki}, \citenamefont {Horodecki},\ and\
  \citenamefont {Horodecki}}]{Horodecki2009Quantum}%
  \BibitemOpen
  \bibfield  {author} {\bibinfo {author} {\bibfnamefont {R.}~\bibnamefont
  {Horodecki}}, \bibinfo {author} {\bibfnamefont {P.}~\bibnamefont
  {Horodecki}}, \bibinfo {author} {\bibfnamefont {M.}~\bibnamefont
  {Horodecki}},\ and\ \bibinfo {author} {\bibfnamefont {K.}~\bibnamefont
  {Horodecki}},\ }\bibfield  {title} {\bibinfo {title} {Quantum entanglement},\
  }\href {https://doi.org/10.1103/RevModPhys.81.865} {\bibfield  {journal}
  {\bibinfo  {journal} {Rev. Mod. Phys.}\ }\textbf {\bibinfo {volume} {81}},\
  \bibinfo {pages} {865} (\bibinfo {year} {2009})}\BibitemShut {NoStop}%
\bibitem [{\citenamefont {G{\"u}hne}\ and\ \citenamefont
  {T{\'o}th}(2009)}]{Guhne2009Entanglement}%
  \BibitemOpen
  \bibfield  {author} {\bibinfo {author} {\bibfnamefont {O.}~\bibnamefont
  {G{\"u}hne}}\ and\ \bibinfo {author} {\bibfnamefont {G.}~\bibnamefont
  {T{\'o}th}},\ }\bibfield  {title} {\bibinfo {title} {Entanglement
  detection},\ }\href
  {https://doi.org/https://doi.org/10.1016/j.physrep.2009.02.004} {\bibfield
  {journal} {\bibinfo  {journal} {Phys. Rep.}\ }\textbf {\bibinfo {volume}
  {474}},\ \bibinfo {pages} {1} (\bibinfo {year} {2009})}\BibitemShut {NoStop}%
\bibitem [{\citenamefont {Friis}\ \emph {et~al.}(2019)\citenamefont {Friis},
  \citenamefont {Vitagliano}, \citenamefont {Malik},\ and\ \citenamefont
  {Huber}}]{Friis2019}%
  \BibitemOpen
  \bibfield  {author} {\bibinfo {author} {\bibfnamefont {N.}~\bibnamefont
  {Friis}}, \bibinfo {author} {\bibfnamefont {G.}~\bibnamefont {Vitagliano}},
  \bibinfo {author} {\bibfnamefont {M.}~\bibnamefont {Malik}},\ and\ \bibinfo
  {author} {\bibfnamefont {M.}~\bibnamefont {Huber}},\ }\bibfield  {title}
  {\bibinfo {title} {Entanglement certification from theory to experiment},\
  }\href {https://doi.org/10.1038/s42254-018-0003-5} {\bibfield  {journal}
  {\bibinfo  {journal} {Nat. Rev. Phys.}\ }\textbf {\bibinfo {volume} {1}},\
  \bibinfo {pages} {72} (\bibinfo {year} {2019})}\BibitemShut {NoStop}%
\bibitem [{\citenamefont {Giovannetti}\ \emph {et~al.}(2004)\citenamefont
  {Giovannetti}, \citenamefont {Lloyd},\ and\ \citenamefont
  {Maccone}}]{Giovannetti2004Quantum-Enhanced}%
  \BibitemOpen
  \bibfield  {author} {\bibinfo {author} {\bibfnamefont {V.}~\bibnamefont
  {Giovannetti}}, \bibinfo {author} {\bibfnamefont {S.}~\bibnamefont {Lloyd}},\
  and\ \bibinfo {author} {\bibfnamefont {L.}~\bibnamefont {Maccone}},\
  }\bibfield  {title} {\bibinfo {title} {Quantum-enhanced measurements: Beating
  the standard quantum limit},\ }\href
  {https://doi.org/10.1126/science.1104149} {\bibfield  {journal} {\bibinfo
  {journal} {Science}\ }\textbf {\bibinfo {volume} {306}},\ \bibinfo {pages}
  {1330} (\bibinfo {year} {2004})}\BibitemShut {NoStop}%
\bibitem [{\citenamefont {Paris}(2009)}]{Paris2009QUANTUM}%
  \BibitemOpen
  \bibfield  {author} {\bibinfo {author} {\bibfnamefont {M.~G.~A.}\
  \bibnamefont {Paris}},\ }\bibfield  {title} {\bibinfo {title} {Quantum
  estimation for quantum technology},\ }\href
  {https://doi.org/10.1142/S0219749909004839} {\bibfield  {journal} {\bibinfo
  {journal} {Int. J. Quant. Inf.}\ }\textbf {\bibinfo {volume} {07}},\ \bibinfo
  {pages} {125} (\bibinfo {year} {2009})}\BibitemShut {NoStop}%
\bibitem [{\citenamefont {Demkowicz-Dobrzanski}\ \emph
  {et~al.}(2015)\citenamefont {Demkowicz-Dobrzanski}, \citenamefont {Jarzyna},\
  and\ \citenamefont {Kolodynski}}]{Demkowicz-Dobrzanski2014Quantum}%
  \BibitemOpen
  \bibfield  {author} {\bibinfo {author} {\bibfnamefont {R.}~\bibnamefont
  {Demkowicz-Dobrzanski}}, \bibinfo {author} {\bibfnamefont {M.}~\bibnamefont
  {Jarzyna}},\ and\ \bibinfo {author} {\bibfnamefont {J.}~\bibnamefont
  {Kolodynski}},\ }\bibfield  {title} {\bibinfo {title} {Chapter four -
  {Quantum} limits in optical interferometry},\ }\href
  {https://doi.org/10.1016/bs.po.2015.02.003} {\bibfield  {journal} {\bibinfo
  {journal} {Prog. Optics}\ }\textbf {\bibinfo {volume} {60}},\ \bibinfo
  {pages} {345 } (\bibinfo {year} {2015})},\ \Eprint
  {https://arxiv.org/abs/arXiv:1405.7703} {arXiv:1405.7703} \BibitemShut
  {NoStop}%
\bibitem [{\citenamefont {Pezze}\ and\ \citenamefont
  {Smerzi}(2014)}]{Pezze2014Quantum}%
  \BibitemOpen
  \bibfield  {author} {\bibinfo {author} {\bibfnamefont {L.}~\bibnamefont
  {Pezze}}\ and\ \bibinfo {author} {\bibfnamefont {A.}~\bibnamefont {Smerzi}},\
  }\bibfield  {title} {\bibinfo {title} {Quantum theory of phase estimation},\
  }in\ \href@noop {} {\emph {\bibinfo {booktitle} {Atom Interferometry (Proc.
  Int. School of Physics 'Enrico Fermi', Course 188, Varenna)}}},\ \bibinfo
  {editor} {edited by\ \bibinfo {editor} {\bibfnamefont {G.}~\bibnamefont
  {Tino}}\ and\ \bibinfo {editor} {\bibfnamefont {M.}~\bibnamefont
  {Kasevich}}}\ (\bibinfo  {publisher} {IOS Press, Amsterdam},\ \bibinfo {year}
  {2014})\ pp.\ \bibinfo {pages} {691--741},\ \Eprint
  {https://arxiv.org/abs/arXiv:1411.5164} {arXiv:1411.5164} \BibitemShut
  {NoStop}%
\bibitem [{\citenamefont {T\'oth}\ and\ \citenamefont
  {Apellaniz}(2014)}]{Toth2014Quantum}%
  \BibitemOpen
  \bibfield  {author} {\bibinfo {author} {\bibfnamefont {G.}~\bibnamefont
  {T\'oth}}\ and\ \bibinfo {author} {\bibfnamefont {I.}~\bibnamefont
  {Apellaniz}},\ }\bibfield  {title} {\bibinfo {title} {Quantum metrology from
  a quantum information science perspective},\ }\href
  {https://doi.org/10.1088/1751-8113/47/42/424006} {\bibfield  {journal}
  {\bibinfo  {journal} {J. Phys. A: Math. Theor.}\ }\textbf {\bibinfo {volume}
  {47}},\ \bibinfo {pages} {424006} (\bibinfo {year} {2014})}\BibitemShut
  {NoStop}%
\bibitem [{\citenamefont {Pezz\`e}\ \emph {et~al.}(2018)\citenamefont
  {Pezz\`e}, \citenamefont {Smerzi}, \citenamefont {Oberthaler}, \citenamefont
  {Schmied},\ and\ \citenamefont {Treutlein}}]{Pezze2018Quantum}%
  \BibitemOpen
  \bibfield  {author} {\bibinfo {author} {\bibfnamefont {L.}~\bibnamefont
  {Pezz\`e}}, \bibinfo {author} {\bibfnamefont {A.}~\bibnamefont {Smerzi}},
  \bibinfo {author} {\bibfnamefont {M.~K.}\ \bibnamefont {Oberthaler}},
  \bibinfo {author} {\bibfnamefont {R.}~\bibnamefont {Schmied}},\ and\ \bibinfo
  {author} {\bibfnamefont {P.}~\bibnamefont {Treutlein}},\ }\bibfield  {title}
  {\bibinfo {title} {Quantum metrology with nonclassical states of atomic
  ensembles},\ }\href {https://doi.org/10.1103/RevModPhys.90.035005} {\bibfield
   {journal} {\bibinfo  {journal} {Rev. Mod. Phys.}\ }\textbf {\bibinfo
  {volume} {90}},\ \bibinfo {pages} {035005} (\bibinfo {year}
  {2018})}\BibitemShut {NoStop}%
\bibitem [{\citenamefont {Beatty}\ and\ \citenamefont
  {Stilck~Fran{\c{c}}a}(2025)}]{Beatty2025Order}%
  \BibitemOpen
  \bibfield  {author} {\bibinfo {author} {\bibfnamefont {E.}~\bibnamefont
  {Beatty}}\ and\ \bibinfo {author} {\bibfnamefont {D.}~\bibnamefont
  {Stilck~Fran{\c{c}}a}},\ }\bibfield  {title} {\bibinfo {title} {{Order p
  Quantum Wasserstein Distances from Couplings}},\ }\bibfield  {journal}
  {\bibinfo  {journal} {Annales Henri Poincar{\'e}}\ }\href
  {https://doi.org/10.1007/s00023-025-01557-z} {10.1007/s00023-025-01557-z}
  (\bibinfo {year} {2025})\BibitemShut {NoStop}%
\bibitem [{\citenamefont
  {Borsoni}(2026)}]{borsoni2026foldedoptimaltransportapplication}%
  \BibitemOpen
  \bibfield  {author} {\bibinfo {author} {\bibfnamefont {T.}~\bibnamefont
  {Borsoni}},\ }\href {https://arxiv.org/abs/2512.01722} {\bibinfo {title}
  {Folded optimal transport and its application to separable quantum optimal
  transport}} (\bibinfo {year} {2026}),\ \Eprint
  {https://arxiv.org/abs/2512.01722} {arXiv:2512.01722 [math.FA]} \BibitemShut
  {NoStop}%
\bibitem [{\citenamefont {{R. Bistron, M. Eckstein, and K.
  \.Zyczkowski}}(2022)}]{Bistron2022Monotonicity}%
  \BibitemOpen
  \bibfield  {author} {\bibinfo {author} {\bibnamefont {{R. Bistron, M.
  Eckstein, and K. \.Zyczkowski}}},\ }\bibfield  {title} {\bibinfo {title}
  {{Monotonicity of the quantum 2-Wasserstein distance}},\ }\href
  {https://arxiv.org/abs/2204.07405} {\bibfield  {journal} {\bibinfo  {journal}
  {arXiv:2204.07405}\ } (\bibinfo {year} {2022})}\BibitemShut {NoStop}%
\bibitem [{\citenamefont
  {Uhlmann}(1976)}]{Uhlmann1976TheTransitionProbability}%
  \BibitemOpen
  \bibfield  {author} {\bibinfo {author} {\bibfnamefont {A.}~\bibnamefont
  {Uhlmann}},\ }\bibfield  {title} {\bibinfo {title} {The "transition
  probability" in the state space of a *-algebra},\ }\href
  {https://doi.org/https://doi.org/10.1016/0034-4877(76)90060-4} {\bibfield
  {journal} {\bibinfo  {journal} {Rep. Math. Phys.}\ }\textbf {\bibinfo
  {volume} {9}},\ \bibinfo {pages} {273} (\bibinfo {year} {1976})}\BibitemShut
  {NoStop}%
\bibitem [{\citenamefont {Jozsa}(1994)}]{Jozsa1994Fidelity}%
  \BibitemOpen
  \bibfield  {author} {\bibinfo {author} {\bibfnamefont {R.}~\bibnamefont
  {Jozsa}},\ }\bibfield  {title} {\bibinfo {title} {Fidelity for mixed quantum
  states},\ }\href {https://doi.org/10.1080/09500349414552171} {\bibfield
  {journal} {\bibinfo  {journal} {J. Mod. Opt.}\ }\textbf {\bibinfo {volume}
  {41}},\ \bibinfo {pages} {2315} (\bibinfo {year} {1994})}\BibitemShut
  {NoStop}%
\bibitem [{\citenamefont {Werner}(1989)}]{Werner1989Quantum}%
  \BibitemOpen
  \bibfield  {author} {\bibinfo {author} {\bibfnamefont {R.~F.}\ \bibnamefont
  {Werner}},\ }\bibfield  {title} {\bibinfo {title} {Quantum states with
  {Einstein-Podolsky-Rosen} correlations admitting a hidden-variable model},\
  }\href {https://doi.org/10.1103/PhysRevA.40.4277} {\bibfield  {journal}
  {\bibinfo  {journal} {Phys. Rev. A}\ }\textbf {\bibinfo {volume} {40}},\
  \bibinfo {pages} {4277} (\bibinfo {year} {1989})}\BibitemShut {NoStop}%
\bibitem [{\citenamefont {Helstrom}(1976)}]{Helstrom1976Quantum}%
  \BibitemOpen
  \bibfield  {author} {\bibinfo {author} {\bibfnamefont {C.}~\bibnamefont
  {Helstrom}},\ }\href
  {https://www.elsevier.com/books/quantum-detection-and-estimation-theory/helstrom/978-0-12-340050-5}
  {\emph {\bibinfo {title} {Quantum Detection and Estimation Theory}}}\
  (\bibinfo  {publisher} {Academic Press, New York},\ \bibinfo {year}
  {1976})\BibitemShut {NoStop}%
\bibitem [{\citenamefont {Holevo}(1982)}]{Holevo1982Probabilistic}%
  \BibitemOpen
  \bibfield  {author} {\bibinfo {author} {\bibfnamefont {A.}~\bibnamefont
  {Holevo}},\ }\href@noop {} {\emph {\bibinfo {title} {Probabilistic and
  Statistical Aspects of Quantum Theory}}}\ (\bibinfo  {publisher}
  {North-Holland, Amsterdam},\ \bibinfo {year} {1982})\BibitemShut {NoStop}%
\bibitem [{\citenamefont {Braunstein}\ and\ \citenamefont
  {Caves}(1994)}]{Braunstein1994Statistical}%
  \BibitemOpen
  \bibfield  {author} {\bibinfo {author} {\bibfnamefont {S.~L.}\ \bibnamefont
  {Braunstein}}\ and\ \bibinfo {author} {\bibfnamefont {C.~M.}\ \bibnamefont
  {Caves}},\ }\bibfield  {title} {\bibinfo {title} {Statistical distance and
  the geometry of quantum states},\ }\href
  {https://doi.org/10.1103/PhysRevLett.72.3439} {\bibfield  {journal} {\bibinfo
   {journal} {Phys. Rev. Lett.}\ }\textbf {\bibinfo {volume} {72}},\ \bibinfo
  {pages} {3439} (\bibinfo {year} {1994})}\BibitemShut {NoStop}%
\bibitem [{\citenamefont {Braunstein}\ \emph {et~al.}(1996)\citenamefont
  {Braunstein}, \citenamefont {Caves},\ and\ \citenamefont
  {Milburn}}]{Braunstein1996Generalized}%
  \BibitemOpen
  \bibfield  {author} {\bibinfo {author} {\bibfnamefont {S.~L.}\ \bibnamefont
  {Braunstein}}, \bibinfo {author} {\bibfnamefont {C.~M.}\ \bibnamefont
  {Caves}},\ and\ \bibinfo {author} {\bibfnamefont {G.~J.}\ \bibnamefont
  {Milburn}},\ }\bibfield  {title} {\bibinfo {title} {Generalized uncertainty
  relations: Theory, examples, and {Lorentz} invariance},\ }\href
  {https://doi.org/10.1006/aphy.1996.0040} {\bibfield  {journal} {\bibinfo
  {journal} {Ann. Phys.}\ }\textbf {\bibinfo {volume} {247}},\ \bibinfo {pages}
  {135} (\bibinfo {year} {1996})}\BibitemShut {NoStop}%
\bibitem [{\citenamefont {Petz}(2008)}]{Petz2008Quantum}%
  \BibitemOpen
  \bibfield  {author} {\bibinfo {author} {\bibfnamefont {D.}~\bibnamefont
  {Petz}},\ }\href@noop {} {\emph {\bibinfo {title} {Quantum information theory
  and quantum statistics}}}\ (\bibinfo  {publisher} {Springer, Berlin,
  Heilderberg},\ \bibinfo {year} {2008})\BibitemShut {NoStop}%
\bibitem [{\citenamefont {T\'oth}\ and\ \citenamefont
  {Petz}(2013)}]{Toth2013Extremal}%
  \BibitemOpen
  \bibfield  {author} {\bibinfo {author} {\bibfnamefont {G.}~\bibnamefont
  {T\'oth}}\ and\ \bibinfo {author} {\bibfnamefont {D.}~\bibnamefont {Petz}},\
  }\bibfield  {title} {\bibinfo {title} {Extremal properties of the variance
  and the quantum fisher information},\ }\href
  {https://doi.org/10.1103/PhysRevA.87.032324} {\bibfield  {journal} {\bibinfo
  {journal} {Phys. Rev. A}\ }\textbf {\bibinfo {volume} {87}},\ \bibinfo
  {pages} {032324} (\bibinfo {year} {2013})}\BibitemShut {NoStop}%
\bibitem [{\citenamefont {Yu}(2013)}]{Yu2013Quantum}%
  \BibitemOpen
  \bibfield  {author} {\bibinfo {author} {\bibfnamefont {S.}~\bibnamefont
  {Yu}},\ }\bibfield  {title} {\bibinfo {title} {{Quantum Fisher Information as
  the Convex Roof of Variance}},\ }\href {http://arxiv.org/abs/1302.5311}
  {\bibfield  {journal} {\bibinfo  {journal} {arXiv:1302.5311}\ } (\bibinfo
  {year} {2013})}\BibitemShut {NoStop}%
\bibitem [{\citenamefont {T\'oth}\ \emph {et~al.}(2015)\citenamefont {T\'oth},
  \citenamefont {Moroder},\ and\ \citenamefont {G\"uhne}}]{Toth2015Evaluating}%
  \BibitemOpen
  \bibfield  {author} {\bibinfo {author} {\bibfnamefont {G.}~\bibnamefont
  {T\'oth}}, \bibinfo {author} {\bibfnamefont {T.}~\bibnamefont {Moroder}},\
  and\ \bibinfo {author} {\bibfnamefont {O.}~\bibnamefont {G\"uhne}},\
  }\bibfield  {title} {\bibinfo {title} {Evaluating convex roof entanglement
  measures},\ }\href {https://doi.org/10.1103/PhysRevLett.114.160501}
  {\bibfield  {journal} {\bibinfo  {journal} {Phys. Rev. Lett.}\ }\textbf
  {\bibinfo {volume} {114}},\ \bibinfo {pages} {160501} (\bibinfo {year}
  {2015})}\BibitemShut {NoStop}%
\bibitem [{\citenamefont {Apellaniz}\ \emph {et~al.}(2017)\citenamefont
  {Apellaniz}, \citenamefont {Kleinmann}, \citenamefont {G\"uhne},\ and\
  \citenamefont {T\'oth}}]{Apellaniz2017Optimal}%
  \BibitemOpen
  \bibfield  {author} {\bibinfo {author} {\bibfnamefont {I.}~\bibnamefont
  {Apellaniz}}, \bibinfo {author} {\bibfnamefont {M.}~\bibnamefont
  {Kleinmann}}, \bibinfo {author} {\bibfnamefont {O.}~\bibnamefont {G\"uhne}},\
  and\ \bibinfo {author} {\bibfnamefont {G.}~\bibnamefont {T\'oth}},\
  }\bibfield  {title} {\bibinfo {title} {Optimal witnessing of the quantum
  {Fisher} information with few measurements},\ }\href
  {https://doi.org/10.1103/PhysRevA.95.032330} {\bibfield  {journal} {\bibinfo
  {journal} {Phys. Rev. A}\ }\textbf {\bibinfo {volume} {95}},\ \bibinfo
  {pages} {032330} (\bibinfo {year} {2017})}\BibitemShut {NoStop}%
\bibitem [{\citenamefont {T\'oth}\ and\ \citenamefont
  {Fr\"owis}(2022)}]{Toth2022Uncertainty}%
  \BibitemOpen
  \bibfield  {author} {\bibinfo {author} {\bibfnamefont {G.}~\bibnamefont
  {T\'oth}}\ and\ \bibinfo {author} {\bibfnamefont {F.}~\bibnamefont
  {Fr\"owis}},\ }\bibfield  {title} {\bibinfo {title} {Uncertainty relations
  with the variance and the quantum {Fisher} information based on convex
  decompositions of density matrices},\ }\href
  {https://doi.org/10.1103/PhysRevResearch.4.013075} {\bibfield  {journal}
  {\bibinfo  {journal} {Phys. Rev. Research}\ }\textbf {\bibinfo {volume}
  {4}},\ \bibinfo {pages} {013075} (\bibinfo {year} {2022})}\BibitemShut
  {NoStop}%
\bibitem [{\citenamefont {Chiew}\ and\ \citenamefont
  {Gessner}(2022)}]{Chiew2022Improving}%
  \BibitemOpen
  \bibfield  {author} {\bibinfo {author} {\bibfnamefont {S.-H.}\ \bibnamefont
  {Chiew}}\ and\ \bibinfo {author} {\bibfnamefont {M.}~\bibnamefont
  {Gessner}},\ }\bibfield  {title} {\bibinfo {title} {Improving sum uncertainty
  relations with the quantum {Fisher} information},\ }\href
  {https://doi.org/10.1103/PhysRevResearch.4.013076} {\bibfield  {journal}
  {\bibinfo  {journal} {Phys. Rev. Research}\ }\textbf {\bibinfo {volume}
  {4}},\ \bibinfo {pages} {013076} (\bibinfo {year} {2022})}\BibitemShut
  {NoStop}%
\bibitem [{\citenamefont {M{\"{u}}ller-Rigat}\ \emph
  {et~al.}(2023)\citenamefont {M{\"{u}}ller-Rigat}, \citenamefont {Srivastava},
  \citenamefont {Kurdzia{\l{}}ek}, \citenamefont {Rajchel-Mieldzio{\'{c}}},
  \citenamefont {Lewenstein},\ and\ \citenamefont
  {Fr{\'{e}}rot}}]{MullerRigat2023CertifyingQuantum}%
  \BibitemOpen
  \bibfield  {author} {\bibinfo {author} {\bibfnamefont {G.}~\bibnamefont
  {M{\"{u}}ller-Rigat}}, \bibinfo {author} {\bibfnamefont {A.~K.}\ \bibnamefont
  {Srivastava}}, \bibinfo {author} {\bibfnamefont {S.}~\bibnamefont
  {Kurdzia{\l{}}ek}}, \bibinfo {author} {\bibfnamefont {G.}~\bibnamefont
  {Rajchel-Mieldzio{\'{c}}}}, \bibinfo {author} {\bibfnamefont
  {M.}~\bibnamefont {Lewenstein}},\ and\ \bibinfo {author} {\bibfnamefont
  {I.}~\bibnamefont {Fr{\'{e}}rot}},\ }\bibfield  {title} {\bibinfo {title}
  {Certifying the quantum {F}isher information from a given set of mean values:
  a semidefinite programming approach},\ }\href
  {https://doi.org/10.22331/q-2023-10-24-1152} {\bibfield  {journal} {\bibinfo
  {journal} {{Quantum}}\ }\textbf {\bibinfo {volume} {7}},\ \bibinfo {pages}
  {1152} (\bibinfo {year} {2023})}\BibitemShut {NoStop}%
\bibitem [{\citenamefont {Bhatia}\ \emph {et~al.}(2019)\citenamefont {Bhatia},
  \citenamefont {Jain},\ and\ \citenamefont {Lim}}]{BHATIA2019165}%
  \BibitemOpen
  \bibfield  {author} {\bibinfo {author} {\bibfnamefont {R.}~\bibnamefont
  {Bhatia}}, \bibinfo {author} {\bibfnamefont {T.}~\bibnamefont {Jain}},\ and\
  \bibinfo {author} {\bibfnamefont {Y.}~\bibnamefont {Lim}},\ }\bibfield
  {title} {\bibinfo {title} {On the {Bures-Wasserstein} distance between
  positive definite matrices},\ }\href
  {https://doi.org/https://doi.org/10.1016/j.exmath.2018.01.002} {\bibfield
  {journal} {\bibinfo  {journal} {Expositiones Mathematicae}\ }\textbf
  {\bibinfo {volume} {37}},\ \bibinfo {pages} {165} (\bibinfo {year}
  {2019})}\BibitemShut {NoStop}%
\bibitem [{\citenamefont {Cerezo}\ \emph {et~al.}(2020)\citenamefont {Cerezo},
  \citenamefont {Poremba}, \citenamefont {Cincio},\ and\ \citenamefont
  {Coles}}]{Cerezo2020variationalquantum}%
  \BibitemOpen
  \bibfield  {author} {\bibinfo {author} {\bibfnamefont {M.}~\bibnamefont
  {Cerezo}}, \bibinfo {author} {\bibfnamefont {A.}~\bibnamefont {Poremba}},
  \bibinfo {author} {\bibfnamefont {L.}~\bibnamefont {Cincio}},\ and\ \bibinfo
  {author} {\bibfnamefont {P.~J.}\ \bibnamefont {Coles}},\ }\bibfield  {title}
  {\bibinfo {title} {Variational {Q}uantum {F}idelity {E}stimation},\ }\href
  {https://doi.org/10.22331/q-2020-03-26-248} {\bibfield  {journal} {\bibinfo
  {journal} {{Quantum}}\ }\textbf {\bibinfo {volume} {4}},\ \bibinfo {pages}
  {248} (\bibinfo {year} {2020})}\BibitemShut {NoStop}%
\bibitem [{\citenamefont {Afham}\ \emph {et~al.}(2022)\citenamefont {Afham},
  \citenamefont {Kueng},\ and\ \citenamefont
  {Ferrie}}]{afham2022quantummeanstatesnicer}%
  \BibitemOpen
  \bibfield  {author} {\bibinfo {author} {\bibfnamefont {A.}~\bibnamefont
  {Afham}}, \bibinfo {author} {\bibfnamefont {R.}~\bibnamefont {Kueng}},\ and\
  \bibinfo {author} {\bibfnamefont {C.}~\bibnamefont {Ferrie}},\ }\href
  {https://arxiv.org/abs/2206.08183} {\bibinfo {title} {Quantum mean states are
  nicer than you think: fast algorithms to compute states maximizing average
  fidelity}} (\bibinfo {year} {2022}),\ \Eprint
  {https://arxiv.org/abs/2206.08183} {arXiv:2206.08183 [quant-ph]} \BibitemShut
  {NoStop}%
\bibitem [{\citenamefont {Gily\'en}\ and\ \citenamefont
  {Poremba}(2022)}]{gilyen2022improvedquantumalgorithmsfidelity}%
  \BibitemOpen
  \bibfield  {author} {\bibinfo {author} {\bibfnamefont {A.}~\bibnamefont
  {Gily\'en}}\ and\ \bibinfo {author} {\bibfnamefont {A.}~\bibnamefont
  {Poremba}},\ }\href {https://arxiv.org/abs/2203.15993} {\bibinfo {title}
  {Improved quantum algorithms for fidelity estimation}} (\bibinfo {year}
  {2022}),\ \Eprint {https://arxiv.org/abs/2203.15993} {arXiv:2203.15993
  [quant-ph]} \BibitemShut {NoStop}%
\bibitem [{\citenamefont {Streltsov}\ \emph {et~al.}(2010)\citenamefont
  {Streltsov}, \citenamefont {Kampermann},\ and\ \citenamefont
  {Bru\ss}}]{Streltsov2010Linking}%
  \BibitemOpen
  \bibfield  {author} {\bibinfo {author} {\bibfnamefont {A.}~\bibnamefont
  {Streltsov}}, \bibinfo {author} {\bibfnamefont {H.}~\bibnamefont
  {Kampermann}},\ and\ \bibinfo {author} {\bibfnamefont {D.}~\bibnamefont
  {Bru\ss}},\ }\bibfield  {title} {\bibinfo {title} {Linking a distance measure
  of entanglement to its convex roof},\ }\href
  {https://doi.org/10.1088/1367-2630/12/12/123004} {\bibfield  {journal}
  {\bibinfo  {journal} {New J. Phys.}\ }\textbf {\bibinfo {volume} {12}},\
  \bibinfo {pages} {123004} (\bibinfo {year} {2010})}\BibitemShut {NoStop}%
\bibitem [{\citenamefont {Horn}\ and\ \citenamefont
  {Johnson}(2012)}]{Horn_Johnson_2012}%
  \BibitemOpen
  \bibfield  {author} {\bibinfo {author} {\bibfnamefont {R.~A.}\ \bibnamefont
  {Horn}}\ and\ \bibinfo {author} {\bibfnamefont {C.~R.}\ \bibnamefont
  {Johnson}},\ }\href@noop {} {\emph {\bibinfo {title} {Matrix Analysis}}},\
  \bibinfo {edition} {2nd}\ ed.\ (\bibinfo  {publisher} {Cambridge University
  Press},\ \bibinfo {year} {2012})\BibitemShut {NoStop}%
\bibitem [{\citenamefont {Hill}\ and\ \citenamefont
  {Wootters}(1997)}]{Hill1997Entanglement}%
  \BibitemOpen
  \bibfield  {author} {\bibinfo {author} {\bibfnamefont {S.}~\bibnamefont
  {Hill}}\ and\ \bibinfo {author} {\bibfnamefont {W.~K.}\ \bibnamefont
  {Wootters}},\ }\bibfield  {title} {\bibinfo {title} {Entanglement of a pair
  of quantum bits},\ }\href {https://doi.org/10.1103/PhysRevLett.78.5022}
  {\bibfield  {journal} {\bibinfo  {journal} {Phys. Rev. Lett.}\ }\textbf
  {\bibinfo {volume} {78}},\ \bibinfo {pages} {5022} (\bibinfo {year}
  {1997})}\BibitemShut {NoStop}%
\bibitem [{\citenamefont {Wootters}(1998)}]{Wootters1998Entanglement}%
  \BibitemOpen
  \bibfield  {author} {\bibinfo {author} {\bibfnamefont {W.~K.}\ \bibnamefont
  {Wootters}},\ }\bibfield  {title} {\bibinfo {title} {Entanglement of
  formation of an arbitrary state of two qubits},\ }\href
  {https://doi.org/10.1103/PhysRevLett.80.2245} {\bibfield  {journal} {\bibinfo
   {journal} {Phys. Rev. Lett.}\ }\textbf {\bibinfo {volume} {80}},\ \bibinfo
  {pages} {2245} (\bibinfo {year} {1998})}\BibitemShut {NoStop}%
\bibitem [{\citenamefont {Bennett}\ \emph {et~al.}(1996)\citenamefont
  {Bennett}, \citenamefont {DiVincenzo}, \citenamefont {Smolin},\ and\
  \citenamefont {Wootters}}]{Bennett1996Mixed-state}%
  \BibitemOpen
  \bibfield  {author} {\bibinfo {author} {\bibfnamefont {C.~H.}\ \bibnamefont
  {Bennett}}, \bibinfo {author} {\bibfnamefont {D.~P.}\ \bibnamefont
  {DiVincenzo}}, \bibinfo {author} {\bibfnamefont {J.~A.}\ \bibnamefont
  {Smolin}},\ and\ \bibinfo {author} {\bibfnamefont {W.~K.}\ \bibnamefont
  {Wootters}},\ }\bibfield  {title} {\bibinfo {title} {Mixed-state entanglement
  and quantum error correction},\ }\href
  {https://doi.org/10.1103/PhysRevA.54.3824} {\bibfield  {journal} {\bibinfo
  {journal} {Phys. Rev. A}\ }\textbf {\bibinfo {volume} {54}},\ \bibinfo
  {pages} {3824} (\bibinfo {year} {1996})}\BibitemShut {NoStop}%
\bibitem [{\citenamefont {Uhlmann}(2010)}]{Uhlmann2010Roofs}%
  \BibitemOpen
  \bibfield  {author} {\bibinfo {author} {\bibfnamefont {A.}~\bibnamefont
  {Uhlmann}},\ }\bibfield  {title} {\bibinfo {title} {Roofs and convexity},\
  }\href {https://doi.org/10.3390/e12071799} {\bibfield  {journal} {\bibinfo
  {journal} {Entropy}\ }\textbf {\bibinfo {volume} {12}},\ \bibinfo {pages}
  {1799} (\bibinfo {year} {2010})}\BibitemShut {NoStop}%
\bibitem [{\citenamefont {Uhlmann}(1995)}]{Uhlmann1995Geometric}%
  \BibitemOpen
  \bibfield  {author} {\bibinfo {author} {\bibfnamefont {A.}~\bibnamefont
  {Uhlmann}},\ }\bibfield  {title} {\bibinfo {title} {Geometric phases and
  related structures},\ }\href
  {https://doi.org/https://doi.org/10.1016/0034-4877(96)83640-8} {\bibfield
  {journal} {\bibinfo  {journal} {Reports on Mathematical Physics}\ }\textbf
  {\bibinfo {volume} {36}},\ \bibinfo {pages} {461} (\bibinfo {year} {1995})},\
  \bibinfo {note} {proceedings of the XXVI Symposium on Mathematical
  Physics}\BibitemShut {NoStop}%
\bibitem [{\citenamefont {T\'oth}\ and\ \citenamefont
  {G\"uhne}(2009)}]{Toth2009Entanglement}%
  \BibitemOpen
  \bibfield  {author} {\bibinfo {author} {\bibfnamefont {G.}~\bibnamefont
  {T\'oth}}\ and\ \bibinfo {author} {\bibfnamefont {O.}~\bibnamefont
  {G\"uhne}},\ }\bibfield  {title} {\bibinfo {title} {Entanglement and
  permutational symmetry},\ }\href
  {https://doi.org/10.1103/PhysRevLett.102.170503} {\bibfield  {journal}
  {\bibinfo  {journal} {Phys. Rev. Lett.}\ }\textbf {\bibinfo {volume} {102}},\
  \bibinfo {pages} {170503} (\bibinfo {year} {2009})}\BibitemShut {NoStop}%
\bibitem [{\citenamefont {Miszczak}\ \emph {et~al.}(2009)\citenamefont
  {Miszczak}, \citenamefont {Puchala}, \citenamefont {Horodecki}, \citenamefont
  {Uhlmann},\ and\ \citenamefont {Zyczkowski}}]{Miszczak2009Sub}%
  \BibitemOpen
  \bibfield  {author} {\bibinfo {author} {\bibfnamefont {J.~A.}\ \bibnamefont
  {Miszczak}}, \bibinfo {author} {\bibfnamefont {Z.}~\bibnamefont {Puchala}},
  \bibinfo {author} {\bibfnamefont {P.}~\bibnamefont {Horodecki}}, \bibinfo
  {author} {\bibfnamefont {A.}~\bibnamefont {Uhlmann}},\ and\ \bibinfo {author}
  {\bibfnamefont {K.}~\bibnamefont {Zyczkowski}},\ }\bibfield  {title}
  {\bibinfo {title} {Sub- and super-fidelity as bounds for quantum fidelity},\
  }\href {https://doi.org/10.26421/QIC9.1-2-7} {\bibfield  {journal} {\bibinfo
  {journal} {Quantum Inf. Comput.}\ }\textbf {\bibinfo {volume} {9}},\ \bibinfo
  {pages} {103} (\bibinfo {year} {2009})}\BibitemShut {NoStop}%
\bibitem [{\citenamefont {H\"ubner}(1992)}]{Hubner1992Explicit}%
  \BibitemOpen
  \bibfield  {author} {\bibinfo {author} {\bibfnamefont {M.}~\bibnamefont
  {H\"ubner}},\ }\bibfield  {title} {\bibinfo {title} {Explicit computation of
  the bures distance for density matrices},\ }\href
  {https://doi.org/https://doi.org/10.1016/0375-9601(92)91004-B} {\bibfield
  {journal} {\bibinfo  {journal} {Phys. Lett. A}\ }\textbf {\bibinfo {volume}
  {163}},\ \bibinfo {pages} {239} (\bibinfo {year} {1992})}\BibitemShut
  {NoStop}%
\bibitem [{\citenamefont {Sommers}\ and\ \citenamefont
  {\.Zyczkowski}(2004)}]{Sommers2004Statistical}%
  \BibitemOpen
  \bibfield  {author} {\bibinfo {author} {\bibfnamefont {H.-J.}\ \bibnamefont
  {Sommers}}\ and\ \bibinfo {author} {\bibfnamefont {K.}~\bibnamefont
  {\.Zyczkowski}},\ }\bibfield  {title} {\bibinfo {title} {Statistical
  properties of random density matrices},\ }\href
  {http://stacks.iop.org/0305-4470/37/i=35/a=004} {\bibfield  {journal}
  {\bibinfo  {journal} {J. Phys. A: Math. Gen.}\ }\textbf {\bibinfo {volume}
  {37}},\ \bibinfo {pages} {8457} (\bibinfo {year} {2004})}\BibitemShut
  {NoStop}%
\bibitem [{\citenamefont {L{\'e}ka}\ and\ \citenamefont
  {Petz}(2013)}]{Leka2013Some}%
  \BibitemOpen
  \bibfield  {author} {\bibinfo {author} {\bibfnamefont {Z.}~\bibnamefont
  {L{\'e}ka}}\ and\ \bibinfo {author} {\bibfnamefont {D.}~\bibnamefont
  {Petz}},\ }\bibfield  {title} {\bibinfo {title} {Some decompositions of
  matrix variances},\ }\href
  {http://www.math.uni.wroc.pl/~pms/publicationsArticle.php?nr=33.2&nrA=1&ppB=191&ppE=199}
  {\bibfield  {journal} {\bibinfo  {journal} {Probab. Math. Statist.}\ }\textbf
  {\bibinfo {volume} {33}},\ \bibinfo {pages} {191} (\bibinfo {year}
  {2013})}\BibitemShut {NoStop}%
\bibitem [{\citenamefont {Petz}\ and\ \citenamefont
  {Virosztek}(2014)}]{Petz2014}%
  \BibitemOpen
  \bibfield  {author} {\bibinfo {author} {\bibfnamefont {D.}~\bibnamefont
  {Petz}}\ and\ \bibinfo {author} {\bibfnamefont {D.}~\bibnamefont
  {Virosztek}},\ }\bibfield  {title} {\bibinfo {title} {A characterization
  theorem for matrix variances},\ }\href
  {https://doi.org/10.14232/actasm-013-789-z} {\bibfield  {journal} {\bibinfo
  {journal} {Acta Sci. Math. (Szeged)}\ }\textbf {\bibinfo {volume} {80}},\
  \bibinfo {pages} {681} (\bibinfo {year} {2014})}\BibitemShut {NoStop}%
\bibitem [{\citenamefont {Vitagliano}\ \emph {et~al.}(2011)\citenamefont
  {Vitagliano}, \citenamefont {Hyllus}, \citenamefont {Egusquiza},\ and\
  \citenamefont {T\'oth}}]{Vitagliano2011Spin}%
  \BibitemOpen
  \bibfield  {author} {\bibinfo {author} {\bibfnamefont {G.}~\bibnamefont
  {Vitagliano}}, \bibinfo {author} {\bibfnamefont {P.}~\bibnamefont {Hyllus}},
  \bibinfo {author} {\bibfnamefont {I.~L.}\ \bibnamefont {Egusquiza}},\ and\
  \bibinfo {author} {\bibfnamefont {G.}~\bibnamefont {T\'oth}},\ }\bibfield
  {title} {\bibinfo {title} {Spin squeezing inequalities for arbitrary spin},\
  }\href {https://doi.org/10.1103/PhysRevLett.107.240502} {\bibfield  {journal}
  {\bibinfo  {journal} {Phys. Rev. Lett.}\ }\textbf {\bibinfo {volume} {107}},\
  \bibinfo {pages} {240502} (\bibinfo {year} {2011})}\BibitemShut {NoStop}%
\bibitem [{\citenamefont {Doherty}\ \emph {et~al.}(2002)\citenamefont
  {Doherty}, \citenamefont {Parrilo},\ and\ \citenamefont
  {Spedalieri}}]{Doherty2002Distinguishing}%
  \BibitemOpen
  \bibfield  {author} {\bibinfo {author} {\bibfnamefont {A.~C.}\ \bibnamefont
  {Doherty}}, \bibinfo {author} {\bibfnamefont {P.~A.}\ \bibnamefont
  {Parrilo}},\ and\ \bibinfo {author} {\bibfnamefont {F.~M.}\ \bibnamefont
  {Spedalieri}},\ }\bibfield  {title} {\bibinfo {title} {Distinguishing
  separable and entangled states},\ }\href
  {https://doi.org/10.1103/PhysRevLett.88.187904} {\bibfield  {journal}
  {\bibinfo  {journal} {Phys. Rev. Lett.}\ }\textbf {\bibinfo {volume} {88}},\
  \bibinfo {pages} {187904} (\bibinfo {year} {2002})}\BibitemShut {NoStop}%
\bibitem [{\citenamefont {Doherty}\ \emph {et~al.}(2004)\citenamefont
  {Doherty}, \citenamefont {Parrilo},\ and\ \citenamefont
  {Spedalieri}}]{Doherty2004Complete}%
  \BibitemOpen
  \bibfield  {author} {\bibinfo {author} {\bibfnamefont {A.~C.}\ \bibnamefont
  {Doherty}}, \bibinfo {author} {\bibfnamefont {P.~A.}\ \bibnamefont
  {Parrilo}},\ and\ \bibinfo {author} {\bibfnamefont {F.~M.}\ \bibnamefont
  {Spedalieri}},\ }\bibfield  {title} {\bibinfo {title} {Complete family of
  separability criteria},\ }\href {https://doi.org/10.1103/PhysRevA.69.022308}
  {\bibfield  {journal} {\bibinfo  {journal} {Phys. Rev. A}\ }\textbf {\bibinfo
  {volume} {69}},\ \bibinfo {pages} {022308} (\bibinfo {year}
  {2004})}\BibitemShut {NoStop}%
\bibitem [{\citenamefont {Doherty}\ \emph {et~al.}(2005)\citenamefont
  {Doherty}, \citenamefont {Parrilo},\ and\ \citenamefont
  {Spedalieri}}]{Doherty2005Detecting}%
  \BibitemOpen
  \bibfield  {author} {\bibinfo {author} {\bibfnamefont {A.~C.}\ \bibnamefont
  {Doherty}}, \bibinfo {author} {\bibfnamefont {P.~A.}\ \bibnamefont
  {Parrilo}},\ and\ \bibinfo {author} {\bibfnamefont {F.~M.}\ \bibnamefont
  {Spedalieri}},\ }\bibfield  {title} {\bibinfo {title} {Detecting multipartite
  entanglement},\ }\href {https://doi.org/10.1103/PhysRevA.71.032333}
  {\bibfield  {journal} {\bibinfo  {journal} {Phys. Rev. A}\ }\textbf {\bibinfo
  {volume} {71}},\ \bibinfo {pages} {032333} (\bibinfo {year}
  {2005})}\BibitemShut {NoStop}%
\bibitem [{\citenamefont {Virosztek}(2019)}]{VIROSZTEK201967}%
  \BibitemOpen
  \bibfield  {author} {\bibinfo {author} {\bibfnamefont {D.}~\bibnamefont
  {Virosztek}},\ }\bibfield  {title} {\bibinfo {title} {Jointly convex quantum
  {Jensen} divergences},\ }\href
  {https://doi.org/https://doi.org/10.1016/j.laa.2018.03.002} {\bibfield
  {journal} {\bibinfo  {journal} {Linear Algebra Appl.}\ }\textbf {\bibinfo
  {volume} {576}},\ \bibinfo {pages} {67} (\bibinfo {year} {2019})},\ \bibinfo
  {note} {proceedings of the ILAS 2017 Conference in Ames, Iowa}\BibitemShut
  {NoStop}%
\bibitem [{\citenamefont {Virosztek}(2021)}]{VIROSZTEK2021107595}%
  \BibitemOpen
  \bibfield  {author} {\bibinfo {author} {\bibfnamefont {D.}~\bibnamefont
  {Virosztek}},\ }\bibfield  {title} {\bibinfo {title} {The metric property of
  the quantum {Jensen-Shannon} divergence},\ }\href
  {https://doi.org/https://doi.org/10.1016/j.aim.2021.107595} {\bibfield
  {journal} {\bibinfo  {journal} {Adv. Math.}\ }\textbf {\bibinfo {volume}
  {380}},\ \bibinfo {pages} {107595} (\bibinfo {year} {2021})}\BibitemShut
  {NoStop}%
\bibitem [{\citenamefont {Pitrik}\ and\ \citenamefont
  {Virosztek}(2015)}]{Pitrik2015JointConvexityBregmanDivergence}%
  \BibitemOpen
  \bibfield  {author} {\bibinfo {author} {\bibfnamefont {J.}~\bibnamefont
  {Pitrik}}\ and\ \bibinfo {author} {\bibfnamefont {D.}~\bibnamefont
  {Virosztek}},\ }\bibfield  {title} {\bibinfo {title} {On the joint convexity
  of the {Bregman} divergence of matrices},\ }\href
  {https://doi.org/10.1007/s11005-015-0757-y} {\bibfield  {journal} {\bibinfo
  {journal} {Lett. Math. Phys.}\ }\textbf {\bibinfo {volume} {105}},\ \bibinfo
  {pages} {675} (\bibinfo {year} {2015})}\BibitemShut {NoStop}%
\bibitem [{\citenamefont {MATLAB}(2020)}]{MATLAB2020}%
  \BibitemOpen
  \bibfield  {author} {\bibinfo {author} {\bibnamefont {MATLAB}},\ }\href@noop
  {} {\emph {\bibinfo {title} {9.9.0.1524771(R2020b)}}}\ (\bibinfo  {publisher}
  {The MathWorks Inc.},\ \bibinfo {address} {Natick, Massachusetts},\ \bibinfo
  {year} {2020})\BibitemShut {NoStop}%
\bibitem [{\citenamefont {{\relax MOSEK ApS}}(2019)}]{MOSEK}%
  \BibitemOpen
  \bibfield  {author} {\bibinfo {author} {\bibnamefont {{\relax MOSEK ApS}}},\
  }\href {http://docs.mosek.com/9.0/toolbox/index.html} {\emph {\bibinfo
  {title} {The MOSEK optimization toolbox for MATLAB manual. Version 9.0.}}}
  (\bibinfo {year} {2019})\BibitemShut {NoStop}%
\bibitem [{\citenamefont {L{\"{o}}fberg}(2004)}]{Lofberg2004Yalmip}%
  \BibitemOpen
  \bibfield  {author} {\bibinfo {author} {\bibfnamefont {J.}~\bibnamefont
  {L{\"{o}}fberg}},\ }\bibfield  {title} {\bibinfo {title} {Yalmip : A toolbox
  for modeling and optimization in matlab},\ }in\ \href@noop {} {\emph
  {\bibinfo {booktitle} {In Proceedings of the CACSD Conference}}}\ (\bibinfo
  {address} {Taipei, Taiwan},\ \bibinfo {year} {2004})\BibitemShut {NoStop}%
\bibitem [{\citenamefont {{T{\'o}th}}(2008)}]{Toth2008QUBIT4MATLAB}%
  \BibitemOpen
  \bibfield  {author} {\bibinfo {author} {\bibfnamefont {G.}~\bibnamefont
  {{T{\'o}th}}},\ }\bibfield  {title} {\bibinfo {title} {{QUBIT4MATLAB V3.0: A
  program package for quantum information science and quantum optics for
  MATLAB}},\ }\href {https://doi.org/10.1016/j.cpc.2008.03.007} {\bibfield
  {journal} {\bibinfo  {journal} {Comput. Phys. Commun.}\ }\textbf {\bibinfo
  {volume} {179}},\ \bibinfo {pages} {430} (\bibinfo {year}
  {2008})}\BibitemShut {NoStop}%
\bibitem [{QUB()}]{QUBIT4MATLAB_actual_note_href}%
  \BibitemOpen
  \href@noop {} {}\bibinfo {note} {The package QUBIT4MATLAB is available at
  \href{https://www.mathworks.com/matlabcentral/fileexchange/8433}{https://www.mathworks.com/matlabcentral/fileexchange/8433},
  and at
  \href{https://gtoth.eu/qubit4matlab.html}{https://gtoth.eu/qubit4matlab.html}.}\BibitemShut
  {Stop}%
\end{thebibliography}%

\end{document}